\documentclass[preprint,12pt]{elsarticle}

\usepackage{graphicx}
\usepackage{amssymb}
\usepackage{amsmath}
\usepackage{hyperref}

\usepackage{lineno}
\usepackage{colortbl}

\newcommand{\CH}[1]{{\color{black} #1}}
\newcommand{\AI}[1]{{\color{black} #1}}
\usepackage{hyperref}

\def\apjs{ApJ}

\def\aap{A\&A}
\def\apjl{ApJ}

\journal{Journal Name}

\begin{document}

\begin{frontmatter}


\title{A review of quasi-periodic oscillations from black hole X-ray binaries: observation and theory}



\author{Adam R. Ingram and Sara E. Motta}

\address{Department of Physics, Astrophysics, University of Oxford, Denys Wilkinson Building, Keble Road, Oxford OX1 3RH, UK}

\begin{abstract}

Black hole and neutron star X-ray binary systems routinely show quasi-periodic oscillations (QPOs) in their X-ray flux. Despite being strong, easily measurable signals, their physical origin has long remained elusive. However, recent observational and theoretical work has greatly improved our understanding. Here, we briefly review the basic phenomenology of the different varieties of QPO in both black hole and neutron star systems before focusing mainly on low frequency QPOs in black hole systems, for which much of the recent progress has been made. We describe the detailed statistical properties of these QPOs and review the physical models proposed in the literature, with particular attention to those based on Lense-Thirring precession. This is a relativistic effect whereby a spinning massive object twists up the surrounding spacetime, inducing nodal precession in inclined orbits. We review the theory describing how an accretion flow reacts to the Lense-Thirring effect, including analytic theory and recent numerical simulations. We then describe recent observational tests that provide very strong evidence that at least a certain type of low frequency QPOs are a geometric effect, and good evidence that they are the result of precession. We discuss the possibility of the spin axis of the compact object being misaligned with the binary rotation axis for a large fraction of X-ray binaries, as is required for QPOs to be driven specifically by Lense-Thirring precession, as well as some outstanding gaps in our understanding and future opportunities provided by X-ray polarimeters and/or high throughput X-ray detectors.

\end{abstract}

\begin{keyword}
Black Holes \sep Neutron Stars \sep Accretion disc \sep Frame-dragging
% https://www.overleaf.com/project/5c363648ec26d84917a3b15b

\end{keyword}

\end{frontmatter}

%\linenumbers

\section{Introduction}
\label{S:1}

\AI{X-ray binary systems (XRBs) consist of a black hole (BH) or a neutron star (NS) accreting material from a stellar companion. Close to the compact object, the accreting material is heated to such high temperatures that it glows brightly in X-rays -- providing a means to observe the relativistic motion of matter in strong gravitational fields, and probe the ultra-dense matter that forms NSs. These systems cannot be directly imaged as their angular size is typically sub-nano-arcseconds, but the observed rapid X-ray variability provides a means to indirectly map the accretion flow. In particular, quasi-periodic oscillations (QPOs) are commonly observed in the X-ray flux. Upon their discovery more than 30 years ago, their potential as a powerful diagnostic was immediately recognised, but unambiguous determination of their physical origin has since proved challenging. Here we review the extensive observational phenomenology of QPOs, the prevailing theories for their physical origin, and progress made through observational tests over the last $\sim$decade.}

Although our main focus here will be on the BH systems, we will frequently refer to the NS systems as a basis for comparison. We will almost exclusively discuss low mass XRBs, in which the companion is less massive than the compact object, and mass transfer takes place via Roche-Lobe overflow\footnote{The only exception in this review is Cygnus X-1, for which the companion is a massive O-type star, and mass transfer takes place via a focused wind.}. Based on their long-term behaviour, low mass XRBs can be classified into persistent and transient sources. The former are always active, showing typical X-ray luminosities exceeding L$_X$ $\sim$ 10$^{36}$ erg/s. The latter spend most of their life in quiescence (L$_X$ $\sim$ 10$^{30}$--10$^{34}$ erg/s) with occasional bright outbursts \CH{(typically reaching luminosities of L$_X$ $\sim$ 10$^{37}$--10$^{38}$ erg/s)} that last $\sim$weeks to months, more rarely years, and are spaced by $\sim$months to decades. Apart from a few exceptions, BH systems are typically transients and most NS systems are persistent sources \cite{Done2007}.

QPOs are best studied in the Fourier domain. They appear in the power spectrum -- which is the modulus squared of the Fourier transform of the light curve \cite{vanderKlis1989} -- as narrow (width less than $\sim 1/2$ the centroid frequency) peaks. In BH systems, QPOs are generally split into low frequency (LF) QPOs, with centroid frequency $\lesssim 30$ Hz, and high frequency (HF) QPOs, with centroid frequency $\gtrsim 60$ Hz \cite{Belloni2010}. A similar distinction applies in NS systems, although the nomenclature is more complex with extra classes of QPO that are not observed in BHs, such as `hectohertz' QPOs, and with the highest frequency QPOs in NSs generally referred to as kHz QPOs rather than HF QPOs. The earliest reference in the literature to what we now recognise as a LF QPO is likely in the 1979 paper by \citet{Samimi1979}, in which they discuss the `sporadic quasi-periodic behaviour' of the X-ray light curve of GX 339-4 as observed by \textit{HEAO}. However, it is not clear if this quasi-periodicity is statistically significant in their data. The first robust detection was reported in 1983 by \citet{Motch1983} for the same source. An X-ray QPO with a centroid frequency of $\sim 0.1$ Hz was detected in data from the \emph{Ariel 6} rocket, and an optical QPO was simultaneously observed at about half that frequency in data from a 1.5 m ESO telesope located in La Silla. Only a couple of years later, QPOs were detected in X-ray data from the \emph{EXOSAT} satellite, from the NS source GX 5-1 \cite{vanderKlis1985}. The unprecedentedly high frequency ($\sim 20-40$ Hz) and persistence of the oscillations in GX 5-1, coupled with the correlation between QPO frequency and X-ray flux, suggested that they could be a geometric effect driven by a characteristic frequency of the inner accretion flow such as the orbital frequency \cite{VdK1989}. Over the following $\sim 5$ years, thanks mainly to observations by the \emph{EXOSAT} and \emph{GINGA} satellites, the existence of distinct QPO \textit{types} became clear for both NS \cite{Middleditch1986,VdK1989} and BH systems \cite{Miyamoto1991a}.
The \textit{Rossi X-ray Timing Explorer} (\textit{RXTE}), operational between 1996 and 2012, provided the richest database for the study of QPOs in XRBs, with thousands of observations in which hundreds of LF QPOs have been detected, contributing greatly to the current rich phenomenological picture.

\textit{RXTE} also enabled the first detections of kHz QPOs in NS systems \cite{vanderKlis1996} and HF QPOs in BH systems \cite{Morgan1997,Remillard1999}. Since HF QPOs have frequencies commensurate with the epicyclic frequencies of particle motion close to the innermost stable circular orbit (ISCO) of the BH, they have been the subject of great theoretical interest \cite{Abramowicz2001,Kato2004}. However, whereas kHz QPOs from NS XRBs are fairly strong and common features, HF QPOs from BH XRBs are very rare and weak features. LF QPOs, on the other hand, are very commonly observed from BH XRBs with very high signal to noise \cite{Motta2015}. This has allowed a detailed picture of their observational properties to be built over the past few decades, and a great amount of progress as to their physical interpretation to be made over the past decade. LF QPOs in BH XRBs are therefore the primary focus of this review. 

\begin{figure}
\centering
\includegraphics[width=0.99\textwidth]{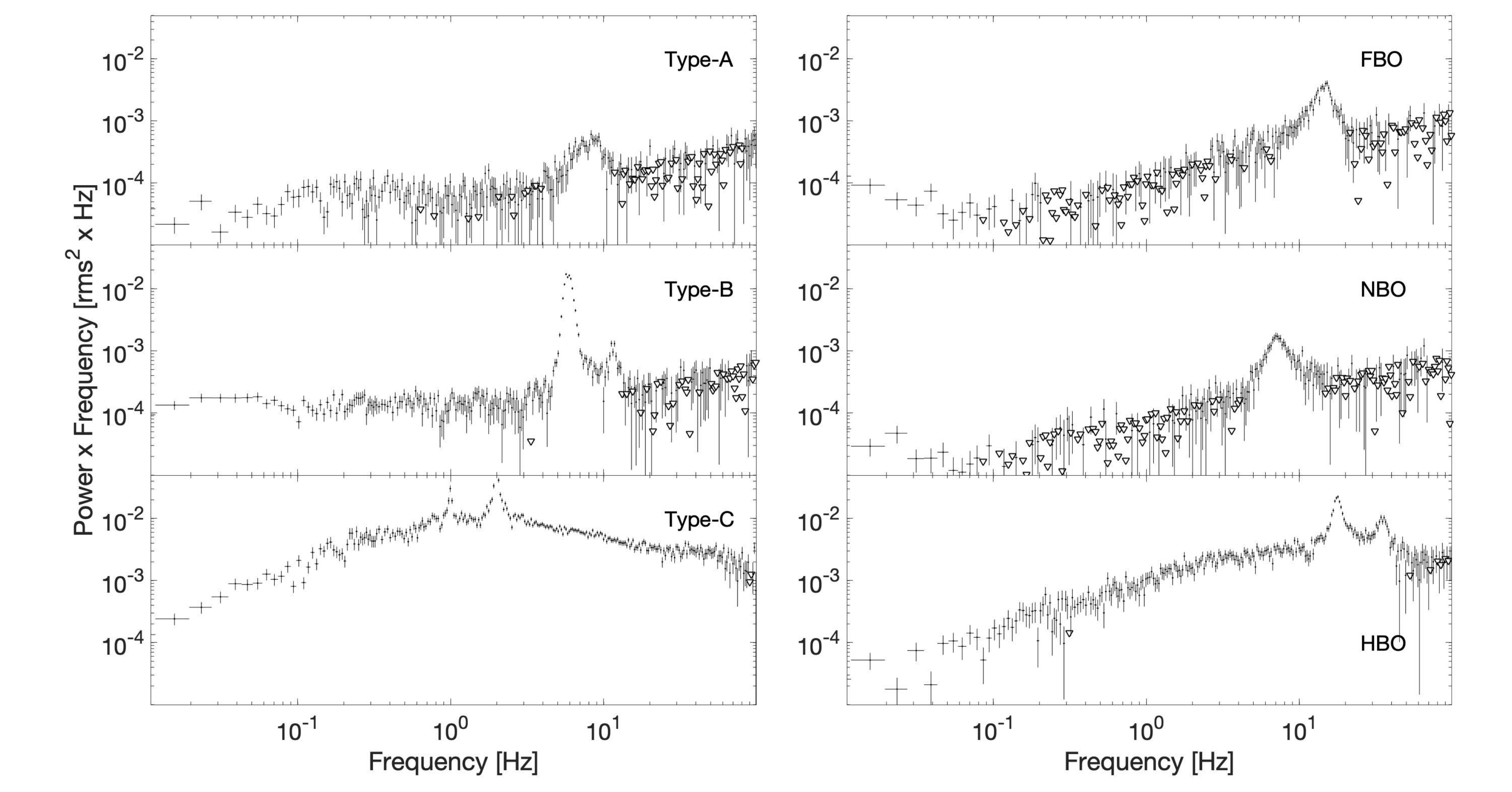}
\caption{\textit{Left  panels}: examples of LF QPOs from BH XRBs. From top to bottom, QPOs are taken from  XTE J1859+226, GX 339-4 and again GX 339-4. \textit{Right panels}: examples of QPOs from NS XRBs. From top to bottom, QPOs are taken from  GX17+2, again GX17+2 and Cyg X-2. \CH{Power spectra in the power$\times$frequency versus frequency form, and have been normalised in fractional rms$^2$.} The contribution of the Poisson noise has been subtracted. Arrows represent $3\sigma$ upper limits.}\label{fig:QPO_sel}
\end{figure}

LF QPOs in BH XRBs have been classified into three types: A, B and C \cite{Wijnands1999,Sobczak2000a,Casella2005,Motta2015}. Representative power spectra of these three types are shown in Fig \ref{fig:QPO_sel} (left). Type-A QPOs are weak and broad features, whereas Type-B and Type-C QPOs are strong and relatively narrow features. Type-C QPOs are typically observed together with broad band, flat-topped noise, which has been suggested to be caused by propagating fluctuations in the mass accretion rate \cite{Lyubarskii1997,Ingram2013}. For NS systems, the nomenclature is slightly different, and largely dependent on historical reasons: LF QPOs are classed as flaring branch oscillations (FBOs), normal branch oscillations (NBOs) and horizontal brand oscillations (HBOs), based on the location of the source in a colour-colour diagram \cite{Lewin1988,Hasinger1989} at the time of the QPO detection. Fig \ref{fig:QPO_sel} (right) shows typical examples of these three types, which have been suggested to be NS analogues of the Type-A, -B and -C, respectively, in BH systems \cite{Casella2005,Motta2017}. NBOs, FBOs and HBOs have been observed in the so-called Z-sources, low-mass NS XRBs accreting near or above the Eddington luminosity, $L_{\rm Edd}$ (see \cite{Hasinger1989}). In NS XRBs accreting at lower rates, the so-called Atoll sources, similar LF QPOs have been identified, and called HBO-like and FBO-like oscillations based on the similarities of their properties to the LF QPOs observed in Z-sources (see \cite{Motta2017} for details).

QPO properties evolve throughout an outburst and are tightly correlated with the evolution of the spectrum through a number of spectral states: the \textit{hard state}, the \textit{intermediate states} and the \textit{soft state}. In the hard state the X-ray spectrum is dominated by a hard power-law (photon-index $\approx$1.4-2) with a high energy cut-off around $\sim100$ keV, while in the soft state the spectrum is dominated by a thermal component peaking at $\sim 1$ keV. Both components are present in the intermediate states. The thermal component is well understood as originating from a geometrically thin, optically thick accretion disc, in which turbulent stress transports angular momentum outwards and heats the disc material \cite{Shakura1973,Novikov1973}. Since the disc is optically thick, it locally emits a blackbody spectrum, with the blackbody temperature increasing with proximity to the BH. The power-law component is thought to originate from Compton up-scattering of soft X-ray photons from the disc (or a jet, in the low-luminosity hard states) by a cloud of hot electrons located close to the BH \cite{Thorne1975,Sunyaev1979}, with the high energy cut-off determined by the characteristic electron temperature. There is still no consensus in the literature on the exact geometry of this cloud, which is often termed the \textit{corona}. The disc may evaporate inside of some truncation radius into a large scale height accretion flow that plays the role of the corona (\textit{truncated disc model} \cite{Eardley1975,Ichimaru1977}). Or, the corona may be located above the disc, and possibly magnetically supported \cite{Galeev1979,Haardt1991} (analogous to the solar corona). Alternatively it may be located at the base of the out-flowing jet, either being fairly compact \cite{Miyamoto1991,Fender1999} or extended \cite{Kylafis2015}.

The spectrum also displays a \textit{reflection} component that results from irradiation of the disc by the corona. The irradiated flux is reprocessed in the disc's upper atmosphere to emit characteristic features such as an iron K$\alpha$ fluorescence line at $\sim 6.4$ keV and a broad bump peaking at $\sim 20-30$ keV referred to as the Compton hump \cite{Lightman1980,Garcia2010}. The observed reflection spectrum is distorted by relativistic motion of the disc material (Doppler shifts) and the gravitational pull of the BH (gravitational redshift), leading to the observed iron line profile being asymmetrically broadened \cite{Fabian1989}. The observed iron line profile therefore encodes information about orbital motion in the disc, and provides a valuable diagnostic as to the physical origin of QPOs.

We will first summarise the detailed phenomenology of all QPO classes and their relation to spectral states and transitions (Section \ref{sec:phen}). We will then focus on details of LF QPOs in BH XRBs, investigating what makes them quasi-periodic as opposed to purely periodic signals  (Section \ref{sec:quasi}). We will discuss theoretical models of LF QPOs, many of which consider epicyclic frequencies in general relativity (GR) and their effect on the dynamics of the accretion flow. We will therefore first review some of the theory behind this (Section \ref{sec:LT}) before summarising the most prominent LF QPO models from the literature (Section \ref{sec:models}). We will then present the most constraining observational tests of the physical mechanism behind LF QPOs (Section \ref{sec:tests}), discussing the current state of the art (Section \ref{sec:discussion}) and finally concluding with a number of concluding remarks (Section \ref{sec:conclusions}). 

\section{Phenomenology of QPOs}
\label{sec:phen}

QPOs of all classes have been studied in great detail by fitting the Poisson noise subtracted power spectrum of many observations with various empirical models, the most commonly used consisting of a sum of Lorentzian functions (e.g. \cite{Belloni2002}). The Lorentzian function is given by \cite{vanstraaten2002}
\begin{equation}
    L(\nu) = \frac{a_0^2}{\pi/2+\arctan(\nu_0/\Delta)} \frac{\Delta}{\Delta^2+(\nu-\nu_0)^2},
    \label{eqn:lore}
\end{equation}
where $\nu$ is Fourier frequency, $\nu_0$ is the Lorentzian peak frequency (corresponding to the QPO centroid frequency), $\Delta$ is the half width at half maximum (HWHM), and $a_0^2$ is equal to the integral of $L(\nu)$ from $\nu=0$ to $\nu=\infty$. From Parseval's theorem, the power spectrum can be normalised such that its integral over all positive frequencies is equal to the square of the fractional root mean square deviation (or rms) of the corresponding time series (\textit{fractional rms normalisation}; \cite{Lewin1988,Belloni1990}). 
The rms essentially quantifies how variable a given time series is in a given Fourier frequency range.
% the emission for a given source is, in a certain frequency range, and in the energy range considered to produce the time series or its corresponding power-density spectrum. 
If the rms normalisation is used, then $a_0$ represents the fractional rms of the Lorentzian component. A \textit{quality factor} $Q=\nu_0/(2\Delta)$ is typically defined, which describes how narrowly peaked a given component is. It is also useful to define the frequency at which the Lorentzian component contributes most of its variance per logarithmic frequency interval, $\nu_{\rm max} = (\nu_0^2 + \Delta^2)^{1/2}$, since this allows the characteristic frequencies of broad and narrow components to be compared. This frequency is sometimes referred to as the \textit{characteristic frequency} \cite{Belloni2002}.  

QPOs often appear as a series of harmonically related peaks, which indicates that the oscillation in the time domain is more complex than purely sinusoidal. In this review, with the term `QPO frequency' we will refer to the centroid frequency of the fundamental peak, $\nu_{\rm qpo}$, and we employ the nomenclature standard in physics in which the centroid frequency of the $n^{\rm th}$ harmonic is $n \nu_{\rm qpo}$. Therefore, the first harmonic is the fundamental. In contrast, the centroid frequency of the $n^{\rm th}$ \textit{overtone} is $(n+1) \nu_{\rm qpo}$. We note that in the QPO literature the second harmonic is (erroneously) often referred to as `the harmonic' or `the first harmonic'.

%In this section, we review the observational phenomenology of QPOs. 
Since the QPO properties are tightly correlated with spectral state, we  review the phenomenology of states and states transitions in BHs before discussing LF QPOs and HF QPOs and their observed properties. %We then compare with NS systems.

\begin{figure}
\centering
\includegraphics[width=0.98\textwidth]{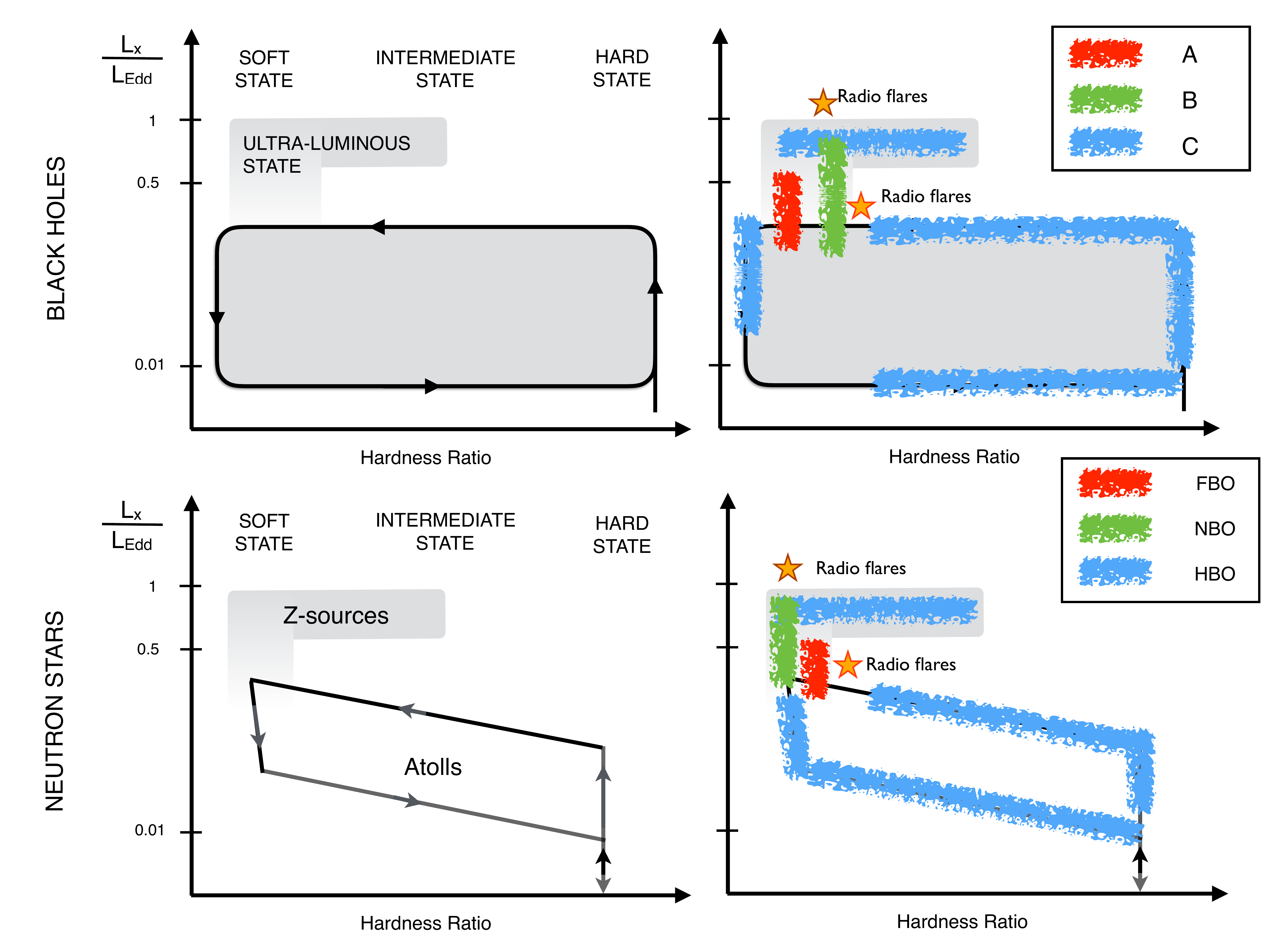}
\caption{Schematic description of the behaviour of BH (upper panels) and NS (lower panels) systems in a Hardness-Intensity diagram. On the left panels we show the typical behavior for NS and BH systems, indicating the relevant spectral states. On the right panels, we indicate where LF QPOs are found in both systems. Figures adapted from \cite{Munoz-Darias2014}.}\label{fig:sketch}
\end{figure}

\subsection{States and transitions} 
\label{sec:states}

During an outburst, BH XRBs typically cycle through several spectral-timing states - believed to be connected with different accretion rates onto the BH (e.g. \cite{Miyamoto1992}, \cite{Belloni2016}) -  which can be tracked using a hardness-intensity diagram (HID) \cite{Homan2001a,Belloni2016}. Fig \ref{fig:sketch} (top left) shows a sketch of the `q'-shaped loop traced out on the HID by a typical BH XRB in outburst. The hardness is defined as the ratio of the counts in a harder X-ray band to that in a softer X-ray band. Typical bounds for the hard and soft bands are $\sim 6-10$ keV and $4-6$ keV \cite{Belloni2005a}, but being rather arbitrarily chosen, the ranges used vary quite a lot across different studies and  different instruments. 
Outbursts begin in the hard state (bottom right of the HID), during which the time variability amplitude of the X-ray emission is high, and  the rms can be as high as $20-30\%$. Sources then transition to the soft state via two intermediate states: the \text{hard intermediate state} (HIMS) and the \text{soft intermediate state} (SIMS). This transition occurs at a roughly constant luminosity ($\sim {\rm tens}\% ~L_{\rm Edd}$). Since the hardness and the fractional rms variability amplitude are tightly correlated, the rms decreases during the state transition and is very low (sometimes consistent with zero) in the soft state \cite{Belloni2010}.

%\footnote{Note that some sources show extremely fast outburst rises, and the hard state is sometimes only observed very briefly (a few days), or missed completely}, 
%The HIMS and SIMS show characteristics of both the hard and soft states (both hard and soft emission play a significant role in the energy spectra), but differ significantly in their fast-time variability properties. During the major state transition from the hard to the soft state, very distinctive changes in the fast-time variability occur, contrasting sharply with rather smooth changes in the energy spectra, which evolve from Compton emission-dominated to disc emission-dominated in a rather continuous fashion.

Shortly before or just after the transition to the soft state, a few sources enter the so called ultra-luminous state (ULS; e.g. \cite{Motta2012}, top of the HID). GRO J1655-40 is the BH XRB that showed the clearest example of a ULS in the \textit{RXTE} era. Other sources have shown short excursions to this state, among these GX 339-4, XTE J1550-564, H1743-322, 4U 1630-47 (see e.g. \cite{Dunn2010}) and GRS 1915+105
% , which is almost always observed in such a state
\cite{Munoz-Darias2014}. The ULS exhibits fairly low variability amplitude (around 5-10\%), but large colour variations. Compared to the standard intermediate states, it extends to much higher luminosities, reaching and in some cases  exceeding the Eddington luminosity \cite{Uttley2015}. After the soft state is reached, the luminosity slowly decreases (in weeks to months) to a few percent of L$_{Edd}$ until a backward transition occurs, taking a source once more through the intermediate states and then to the hard state. After this second hard state, the outburst phase normally comes to an end, even though it is not unusual to witness a temporary re-brightening, which takes the sources up in the hard state for a time (typically a few weeks to a few months), and finally back to quiescence \cite{Chen1993, Bailyn1995, Chen1997}. Outburst phases in BH XRBs are characterized by hysteresis, i.e. the hard to soft transition always occurs at higher luminosity than the soft to hard transition \cite{Miyamoto1995,Maccarone2005}. This cyclic pattern translates into hysteresis loops in the HID
%hardness-intensity diagram (HID, \cite{Homan2001})
that are always observed to develop in an anti-clockwise manner. Similar loops are seen in an rms-intensity diagram \cite{Munoz-Darias2011a}, as expected from the known correlation between rms and hardness. 

\subsection{Low frequency QPOs}

As we alluded to in the Introduction, LF QPOs in BH XRBs have traditionally been classified into three types: A, B and C. Fig \ref{fig:sketch} (top right) shows where in the hysteresis loop on the HID each type of QPO is observed. We see that Type-C QPOs are first observed in the hard state, whereas Type-B and Type-A QPOs are only seen during the transition to the soft state. Below, we discuss these three QPO classes in the order they appear in an outburst.

\begin{figure}
\centering
\includegraphics[width=0.98\textwidth,trim=0.0cm 2.0cm 0.0cm 4.5cm,clip=true]{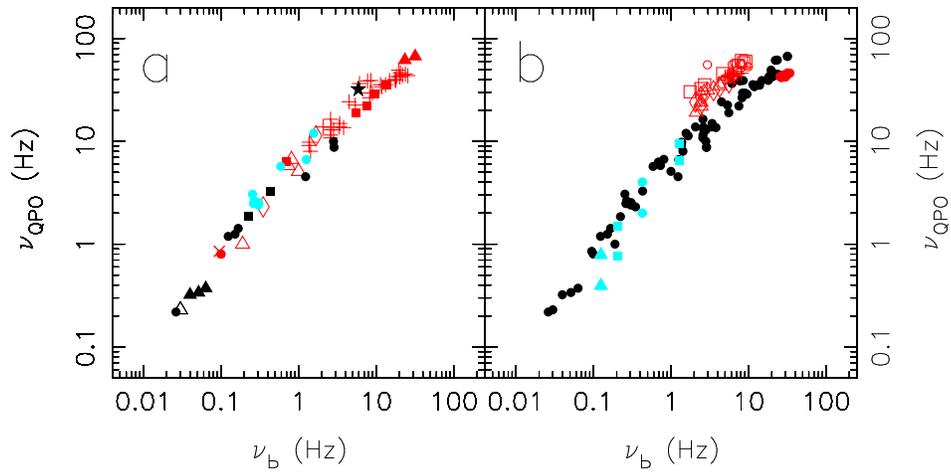}
\caption{Type-C QPO frequency versus the low frequency break frequency (reproduced from \citet{Wijnands1999}). \textit{Left:} black symbols represent BH systems, whereas red and blue points represent NS systems. \textit{Right:} Black symbols now represent all points from the left panel and red symbols are Z-sources (high accretion rate NS systems). Blue symbols are sources for which there are two QPOs and it is unclear which one should be classified as Type-C (both are plotted).}
\label{fig:wijnands}
\end{figure}
%Left, Bottom, Right, Top

\subsubsection{Type-C QPOs}

Type-C QPOs (Fig. \ref{fig:QPO_sel} bottom) are by far the most common type of QPO in BH systems. Type-C QPOs can be detected essentially in all \CH{accretion states} (Fig. \ref{fig:sketch}, top right), \CH{even though they are particularly prominent in the hard-intermediate state and at the bright end of the low-hard state.} The centroid frequency is tightly correlated with spectral state, rising from a few mHz in the hard state at low luminosities to $\sim~10$ Hz in the intermediate states. They are also occasionally observed in the soft state and the ULS (see, e.g., \cite{Motta2012}), where they can reach $\sim$30 Hz. Type-C QPOs are characterised by a high-amplitude (up to 20\% rms), narrow ($Q \gtrsim 8$) peak in the power spectrum, coincident with `flat-top' noise characterised by a low and high frequency break. A number of harmonics are generally detected, of which the one with the highest rms amplitude is usually (but not exclusively) identified with the fundamental. A second harmonic and sometimes even higher-order harmonics are often seen in addition to the fundamental, as well as a so-called sub-harmonic peak (centered at half the fundamental frequency). As shown in Fig. \ref{fig:wijnands}, the Type-C QPO frequency correlates with the low frequency break in the power spectrum \cite{Wijnands1999a}. The QPO frequency also correlates tightly with spectral parameters such as the photon index $\Gamma$ \cite{Vignarca2003} and modelling the flat-top noise as a sum of broad Lorentzians reveals that it correlates with the characteristic frequency of all these broad components \cite{Psaltis1999}. Type-C QPOs have also been observed at optical (e.g. \cite{Motch1983,Imamura1990,Gandhi2010}), ultraviolet \cite{Hynes2003a} and infrared \cite{Kalamkar2016} wavelengths. Simultaneous multi-wavelength observations have revealed that the UV/optical/IR QPO centroid frequency is sometimes coincident (within errors) with the X-ray QPO fundamental frequency \cite{Hynes2003a,Durant2009,Gandhi2010} and sometimes (again, within errors) with half of the X-ray fundamental frequency \cite{Motch1983,Kalamkar2016}. Due to the quality and quantity of Type-C QPO observations, we will concentrate on these features more than any other throughout this review.

\subsubsection{Type-B QPOs}

Type-B QPOs (see Fig. \ref{fig:QPO_sel}) have been detected in a large number of BH XRBs and they appear during the SIMS (Fig. \ref{fig:sketch}, top right). In fact, the SIMS is \textit{defined} by the presence of a Type-B QPO \cite{Belloni2016}. They are characterised by a relatively high amplitude (up to $\sim$5\% rms) and narrow ($Q \gtrsim 6$) peak, with a centroid frequency at 5-6 Hz (but see \cite{Motta2011}, where type-B QPOs were found at $\approx$1-3 Hz). Type-B QPOs generally appear in the power spectrum coincident with weak red noise (few percent rms or less) that increases in amplitude at low frequencies ($\leq$0.1 Hz). A weak second harmonic is often present, sometimes together with a sub-harmonic peak. In a few cases, the sub-harmonic and fundamental have comparable amplitude, resulting in a \textit{cathedral-like} QPO shape \cite{Casella2004}. Rapid transitions in which Type-B QPOs appear and disappear on very short time scales are sometimes observed in some sources (e.g. \cite{Nespoli2003} and \cite{Casella2004} for the cases of GX 339-4 and XTE J1859+226). These transitions are difficult to resolve at present, as they take place on a timescale shorter than a few seconds. Type-B QPOs occur at a similar time to discrete jet ejections evidenced by transient radio flares \cite{Fender2004} and transient jets resolved in high angular resolution radio images \cite{Corbel2005}. This property led to speculation that they are causally connected with the jet ejection, perhaps indicating that the inner regions of the accretion flow are ejected during the Type-C to Type-B transition. However,  \citet{Fender2009} showed that the association is not so clean, with the Type-B QPO sometimes occurring slightly before and sometimes slightly after the inferred ejection.

\subsubsection{Type-A QPOs}

Type-A QPOs (see Fig. \ref{fig:QPO_sel}) are the least common type of \CH{LF} QPO in BH XRBs. The entire \textit{RXTE} archive only contains $\sim 10$ significant Type-A QPO detections. Figure  \ref{fig:sketch} (top right) shows that Type-A QPOs normally appear in the soft state, just after the hard to soft transition has taken place. They appear as a weak (few percent rms) and broad ($Q \lesssim 3$) peak with centroid frequency of approximately $6-8$ Hz. Neither a sub-harmonic nor a second harmonic are usually detected, possibly because of the broadness of the fundamental peak, or because of the intrinsic low-amplitude of this type of QPO. Type-A QPOs are associated with a very low amplitude red noise. Originally, these LF QPOs were dubbed \textit{Type A-II} \cite{Homan2001}. LF QPOs dubbed \textit{Type A-I} \cite{Wijnands1999} were strong, broad and appeared together with a very low-amplitude red noise, and a `shoulder' on the right-hand side of this QPO was clearly visible and interpreted as a very broadened second harmonic peak. \citet{Casella2005} showed that these \textit{Type A-I} LF QPOs should instead be classified as Type-B QPOs. Since Type-A QPOs are very weak, rare features, we hereafter will not discuss them in much detail.

\subsection{High-frequency QPOs}

%\begin{figure}
%\centering
%\includegraphics[width=0.55\textwidth]{HFQPOs.pdf}
%\caption{Power spectra showing the HF QPOs observed in \textit{RXTE} data from the BH XRB GRO J1655-40, adapted from \cite{Motta2014a}. The QPO at $\approx$500 Hz is detected in a harder energy band than that at $\approx$350 Hz. Note that the power spectrum containing the $\approx$500 Hz QPO has been shifted upward for clarity. The red line marks the best fitting empirical model. }\label{fig:HFQPO}
%\end{figure}

HF QPOs are rare features in BH systems, with all detections to date being made by \textit{RXTE}. The first reported detection, in 1997, was from GRS 1915+105 at $\sim 67$ Hz \cite{Morgan1997}. Since then, sixteen years of \textit{RXTE} observations have yielded only a handful of detections in other sources (XTE J1550-564, GRO J1655-40, XTE J1859+226, H 1743-322 , GX 339-4, XTE J1752-223, 4U 1630-47, IGR J17091-3624), but all at a few hundred Hz, i.e., at significantly higher frequencies than the $\sim 67$ Hz QPO from GRS 1915+105. \CH{It is worth noticing that some of these detections have been later proven to be not statistically significant, leaving only two source a few solid detections from two sources (XTE J1550-564, GRO J1655-40, see \cite{Belloni2012}).}
The $\sim 67$ Hz QPO has emerged as a somewhat special case, in that a QPO at similar frequency has only been seen from one other source (IGR J17091-3624 \cite{Altamirano2012a}), and there are many detections of this feature from GRS 1915+105 \cite{Belloni2013}, \CH{including a number of cases found in the ASTROSAT data, where the 67Hz QPO varies significantly in frquency over short time scales \cite{Belloni2019}. The above contrasts sharply with} the handful of statistically significant detections  of $\gtrsim 100$ Hz HF QPOs from other sources \cite{Belloni2012}.  

\begin{figure}
\centering
\includegraphics[width=0.8\textwidth]{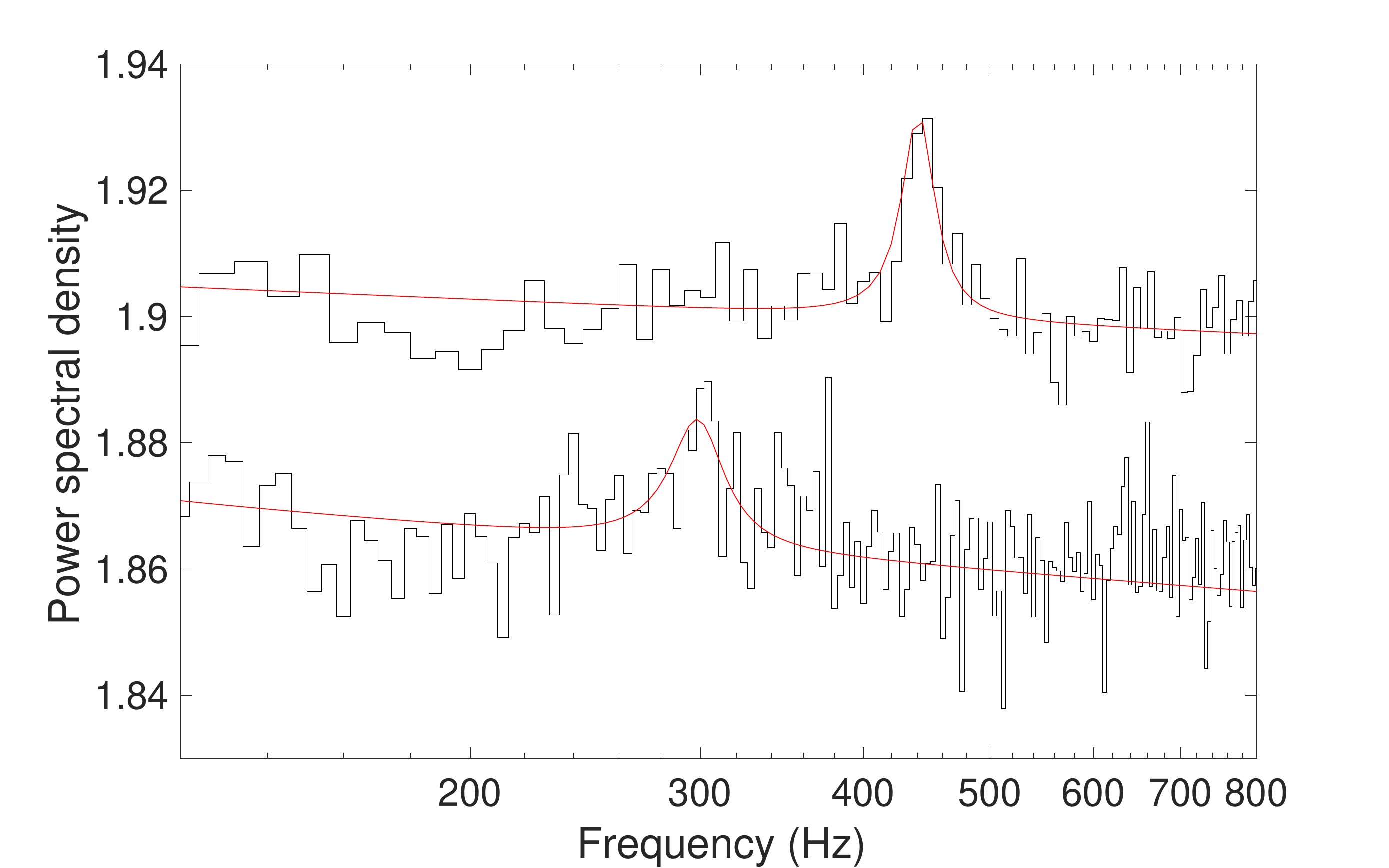}
\caption{\textit{Left  panels}: examples of HF QPOs from BH XRBs (from GROJ1655-40, adapted from \cite{Motta2014a}). The QPO at $\approx$500 Hz is detected in a harder energy band than that at $\approx$350 Hz. Note that the power spectrum containing the $\approx$500 Hz QPO has been shifted upward for clarity. The red line marks the best fitting empirical model.  \textit{Right panels}: examples of HF QPOs from NS XRBs (from Sco X-1). The two (strong) kHz QPOs are both detected in the entire energy range (2-30 keV).}\label{fig:HFQPOs}
\end{figure}

HF QPOs appear only in high flux observations, at a fairly specific hardness ratio \cite{Belloni2012}. Although this may be a selection effect given that the noise level decreases with count rate, there are many high flux observations without HF QPO detections, which suggests that a high count rate is not the only parameter required for the detection of an HF QPO. HF QPOs can be observed as single or double peaks (in which case they are called the lower and upper HF QPOs). Only one source, GRO J1655-40 (see Fig. \ref{fig:HFQPOs}), showed two clear simultaneous peaks \cite{Strohmayer2001,Motta2014}, which however appear in two different energy bands (the lower peak is visible at low energies, i.e. $E \lesssim 10$ keV, the higher peak at higher energies, i.e. $E \gtrsim 10$ keV), while all the other systems only showed single peaks, which appear to move in frequency with time \cite{Belloni2014}. A claim of two simultaneous HF QPOs from XTE J1550-564 (with the lower one at a $2.3 \sigma$ significance level \cite{Remillard2002}) was later shown to be an effect of averaging a large number of observations \cite{Mendez2013}.  H1743-322 also showed a significant HF QPO, and a weak tentative simultaneous peak that may correspond to a second HF QPO \cite{Homan2005}.

The fractional rms of a typical HF QPO is $\sim 0.5-6\%$ in the full \textit{RXTE} band, increasing steeply with energy \cite{Belloni2012}. This energy dependence explains why \textit{RXTE} was best suited for detecting HF QPOs, having a high effective area above $\sim 5$ keV. In the case of GRS 1915+105, the HF QPO rms amplitude can reach $\approx$20\% in the 20-40 keV  energy band \cite{Morgan1997}. The typical quality factor is $Q \sim 5$ and $Q \sim 10$ for the lower and upper HF QPO, respectively. For the $\sim 67$ Hz QPO, $Q\sim 20$ typically, but covers a range $Q\sim 5-30$. In both GRO J1655-40 and H1743-322, the frequencies of the simultaneous HF QPOs are close to being in a 3:2 ratio \cite{Strohmayer2001,Remillard2002,Remillard2006}. \CH{The same is true for the two peaks reported for XTE J1550-564, but as stated above, these peaks are very likely not real simultaneous HF QPOs.} This motivated a family of models suggesting that these HF QPOs result from some kind of resonance \cite{Abramowicz2001}. However, since HF QPOs are very rare and only detected in a specific state, this frequency ratio could well be a coincidence.

\subsection{Neutron star systems}
\label{sec:NSQPOs}

NS XRBs show states similar to those seen in BH systems except the phenomenology is somewhat more complex, presumably due to the solid surface and anchored magnetic field \cite{VDK2006}. State classification historically used a colour-colour diagram (CCD) \cite{Hasinger1989} instead of an HID \cite{VDK2006}, with sources being classified as either \textit{Atoll} or \textit{Z} sources, based on the shape of the pattern traced by these systems in a CCD. Both Atoll and Z-sources show three main states that roughly correspond to the hard, intermediate and soft state seen in BH systems. In Atoll sources, the states are traditionally referred to as the \textit{island}, \textit{lower banana} and \textit{upper banana} branches, whereas in Z-sources they are referred to as the \textit{horizontal}, \textit{normal} and \textit{flaring} branches. A few sources (e.g. XTE J1701-462 \cite{Lin2009}), have shown full evolution between Z and Atoll like spectra. This clarified that the main difference between these two classes of sources is their average accretion rate, significantly higher in Z-sources than in Atoll sources. Recently, \citet{Munoz-Darias2014} and \citet{Motta2017} showed that the spectral, timing and multi-wavelength properties of BH and NS systems are remarkably similar (see Fig \ref{fig:sketch}, bottom). These authors demonstrated that NS and BH systems cross three main accretion states that can be determined purely by the fractional variability amplitude: hard state (rms$\gtrsim 20\%$), intermediate states ($5\% \lesssim {\rm rms} \lesssim 20$) and soft state (rms$\lesssim 5\%$). The type of evolution of a NS or BH across states and transitions appears to be mostly determined by the average accretion rate on the compact object, rather than by the nature of the compact object itself. Below $\sim$50\% L$_{\rm Edd}$ (typical of Atoll sources and the majority of BH low-mass XRBs) hysteresis is the natural form that state transitions take. Above $\sim$50\% L$_{\rm Edd}$ (typical of Z-sources and highly-accreting BHs, such as GRS 1915+105) hysteresis does not take place, and instead fast transitions from and to the soft state occur at a roughly constant luminosity. Transient flaring behaviour is also observed in such sources, but the variability level is typically quite low \cite{Munoz-Darias2014}. %Both when hysteresis happens and when it does not, 
The main difference that remains between BH and NS low mass XRBs resides in the transition velocity between states: the fastest transitions are seen in Z-sources and the slowest in BH systems, and Atoll sources sit somewhere between the two.

Classification of LF QPOs in NS systems is more difficult than for BH systems due to the somewhat richer phenomenology, but three main types of LF QPOs have been identified for Z-sources \cite{VDK2006}: horizontal branch oscillations (HBOs), normal branch oscillations (NBOs) and flaring branch oscillations (FBOs). These are thought to correspond respectively to Type -C, -B and -A QPOs in BH systems (see Fig \ref{fig:QPO_sel}, right panels) \cite{Casella2005,Motta2017}, partly because they occur when the source showing them is in the same region of the HID (see Fig \ref{fig:sketch}, bottom right), and partly because HBOs follow the same correlation between QPO frequency and break frequency as BHs (see Fig \ref{fig:wijnands}). \citet{Motta2017} recently showed that the same classification can also be applied to Atoll systems, which however do not appear to show NBO-like QPOs.

QPOs are also seen at high frequencies in NSs, where they are traditionally called kHz QPOs instead of HF QPOs. The first two sources in which kHz QPO were detected were 4U 1728--34 and Sco X-1. Peaks at $\approx$700 Hz and $\approx$1100 Hz, apparently not directly connected with the NS spin period, were detected in the power spectrum of these two systems \cite{vanderKlis1996,Strohmayer1996}. Since then, many kHz QPO detections have been made. These features are notable for having a much larger amplitude and quality factor than HF QPOs, explaining why they are observed so much more commonly.

NS systems also display QPOs that do not have a BH equivalent. Some examples are: (i) $\sim$1 Hz QPOs sometimes seen in dipping sources \cite{Jonker1999}, (ii) hectohertz QPOs seen in Atoll sources \cite{Altamirano2008},  (iii) a $\sim$ 26 Hz QPO only seen in the dipping flaring branch of Cyg X-2 \cite{Kuulkers1995}.

\subsection{mHz QPOs}

In 2001, \citet{Revnivtsev2001} reported the discovery of a new class of low
frequency quasi-periodic variations of the X-ray flux in
three NS X-ray binaries (4U1608-52, 4U1636-536 and Aql X-1), which were later observed in more systems,
\CH{\cite{Strohmayer2011,Trudolyubov2001, Strohmayer2018,Mancuso2019}. }
These QPOs are typically referred to as to mHz QPOs, named after the frequency range at which they appear. They exhibit properties that differ from other QPO types, in that they occur only in a narrow range of X-ray luminosity, their fractional rms amplitude decreases with energy and they disappear at the onset of thermonuclear (type I) X-ray bursts \cite{Revnivtsev2001}. Based on their observed properties, these QPOs are generally interpreted as a the result of a special regime of the nuclear burning on the NS surface, which generates a quasi-periodic flaring in the emission from the  surface. 

\citet{Altamirano2018} reported on the discovery of mHz QPOs in the a BH candidate H1743-322. Clearly, surface-bound processes should be excluded in this source (unless it turns out to harbour a NS instead of a BH). The authors suggested that these QPOs are not analogues of the NS mHz QPOs, but may instead be the BH equivalent of the so-called 1-Hz QPOs seen in dipping NS systems \cite{Jonker1999,Homan2001a} (except with a much lower centroid frequency). This particular class of QPOs has only been reported for H 1743-322, but we note that most \textit{RXTE} observations are fairly short and, moreover, are typically analysed by averaging the power spectrum over many even shorter ($\lesssim 100$s) segments. It is therefore entirely possible that many mHz QPOs are still to be found in the \textit{RXTE} archive. Indeed, QPOs with frequency below the range typically associated with LF QPOs have been reported for Cygnus X-1 \cite{Rapisarda2017a} and V404 Cyg \cite{Huppenkothen2017}.
%The nature of this type of QPOs, observed only in H1743-322 in the hard state remains, however, dubious, and it has been speculated that they might be even connected with the compact, relativistic jets observed in the same state \cite{Altamirano2018}. 

\subsection{Other objects}

\subsubsection{AGN}
QPOs are extremely common features in accreting BH X-ray binaries and, based on the scale-invariance of the accretion process, one may expect that QPOs should be detected around the accreting super-massive ($M\sim 10^{5-10}~ M_\odot$) BHs powering active galactic nuclei (AGN).
All timescales are expected to scale with BH mass, and so any AGN QPOs are expected to be at frequencies orders of magnitude lower than their X-ray binary counterparts. This makes them very difficult to detect with the comparatively short and irregularly spaced observations currently available. In particular, \citet{Vaughan2005} showed that strong Type-C QPOs typical of BH X-ray binaries with their frequencies scaled to AGN BH masses would not be detectable with existing AGN monitoring campaigns. It is therefore not surprising that there have been very few convincing detections of AGN QPOs to date.

A number of early claims of AGN QPO detections were later disfavoured. Most commonly, the QPO is originally claimed to be highly statistically significant, then a later re-analysis demonstrates that the inferred significance reduces dramatically when the broad band noise is (correctly) modelled as red noise instead of white noise (e.g. see discussion in \cite{Benlloch2001}). Or in the case of the Seyfert galaxy NGC 6814, a strong $\sim 12$ ks periodicity reported by \citet{Mittaz1989} was later shown to originate from another (probably Galactic) source 37 arcmin from the AGN core \cite{Madejski1993}. The first robust detection of a QPO in an AGN was a $\sim$1hr X-ray modulation from the Seyfert galaxy RE J1034+396 \cite{Gierlinski2008}. \citet{Alston2014} later reported on another detection of this $\sim$1 hr QPO in RE J1034+396, and \citet{Alston2015} dicovered a $\sim 2$ hr QPO in another AGN, MS 2254.9–3712. Both QPOs show a relatively high amplitude ($\sim$5-6$\%$) and Q-factor ($>$ 8), a soft lag at the QPO frequency and a relatively hard spectrum (they are both detected above 1 keV). These QPOs have been suggested to be the AGN equivalent of HF QPOs \cite{Middleton2010}, but more detections are required before strong conclusions can be made.

%\cite{Gonzales-Martin2012})
% More, robust detections of QPOs in AGN are needed to better determine the nature of quasi-periodic signals around super-massive BHs, and to confirm or exclude the association with HFQPOs in stellar-mass BH XRBs.

\subsubsection{Ultra-luminous X-ray sources (ULXs)}

ULXs have inferred intrinsic X-ray luminosities well above the Eddington limit of a stellar-mass BH and are spatially resolved from the nucleus of their host galaxy. They were therefore long thought to be powered either by intermediate mass ($M\sim 10^{2-4}~M_\odot$) BHs accreting at sub-Eddington rates, or stellar-mass BHs accreting at super-Eddington rates. mHz QPOs are often detected from ULXs (e.g. \CH{ \cite{Mucciarelli2006, Strohmayer2009}}). It has been argued that these mHz QPOs are ULX analogues of Type-C QPOs, and thus that the QPO frequencies are $\sim$2 orders of magnitude lower because the BHs in ULXs are $\sim 2$ orders of magnitude heavier. Moreover, \citet{Pasham2014} reported on the discovery of a pair of 3:2 frequency ratio QPOs in the ULX M82 X-1, which they argued are the ULX analogue of HF QPOs. From this association, they derived a BH mass of $M \sim 400~M_\odot$. However, the field was dramatically changed by the discovery of pulsations from the ULX M82 X-2 \cite{Bachetti2014}. A number of subsequent detections of pulsations from other ULXs \cite{Fuerst2016,Israel2017,Israel2017a} have confirmed that at least a fraction of ULXs are powered by super-Eddington accreting NSs and have cast doubt on the simple QPO mass scaling arguments for the intermediate mass BH hypothesis -- particularly since M82 X-2 itself appears to exhibit an $\sim 8$ mHz QPO in its power-spectrum that was previously suggested to be a Type-C QPO analogue \cite{Caballero-Garcia2013}. It therefore seems more likely either that these mHz QPOs are simply analogues of the mHz QPOs observed in Galactic XRBs, or are entirely another class of QPO, perhaps associated with super-Eddington accretion rates. As is the case for AGN QPOs, a larger number of detections is needed to allow any further progress to be made in interpreting ULX QPOs.

\section{What makes QPOs quasi-periodic?}
\label{sec:quasi}

We now delve into the detailed observational properties of LF QPOs in BH XRBs, asking the question: \textit{what makes them quasi-periodic as opposed to purely periodic?}.

\subsection{Frequency and amplitude modulation}
\label{sec:freqamp}

\begin{figure}
%\includegraphics[width=8cm,trim=2.0cm 0.5cm
%        2.0cm 0.0cm,clip=true]
	\includegraphics[width=0.48\textwidth,trim=1.8cm 0.5cm
        2.8cm 0.5cm,clip=true]{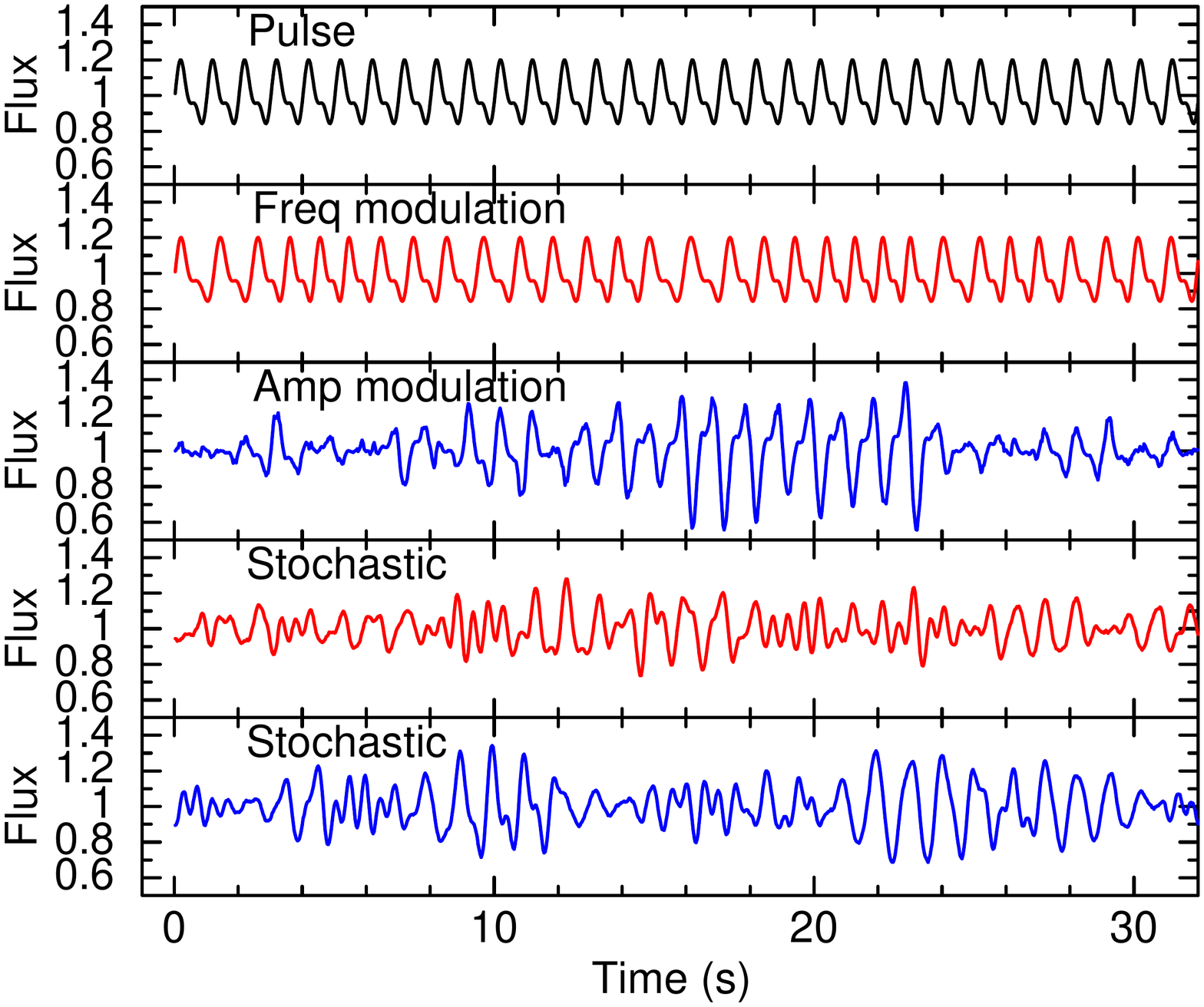} ~
	\includegraphics[width=0.48\textwidth,trim=1.8cm 2.0cm
        2.8cm 11.0cm,clip=true]{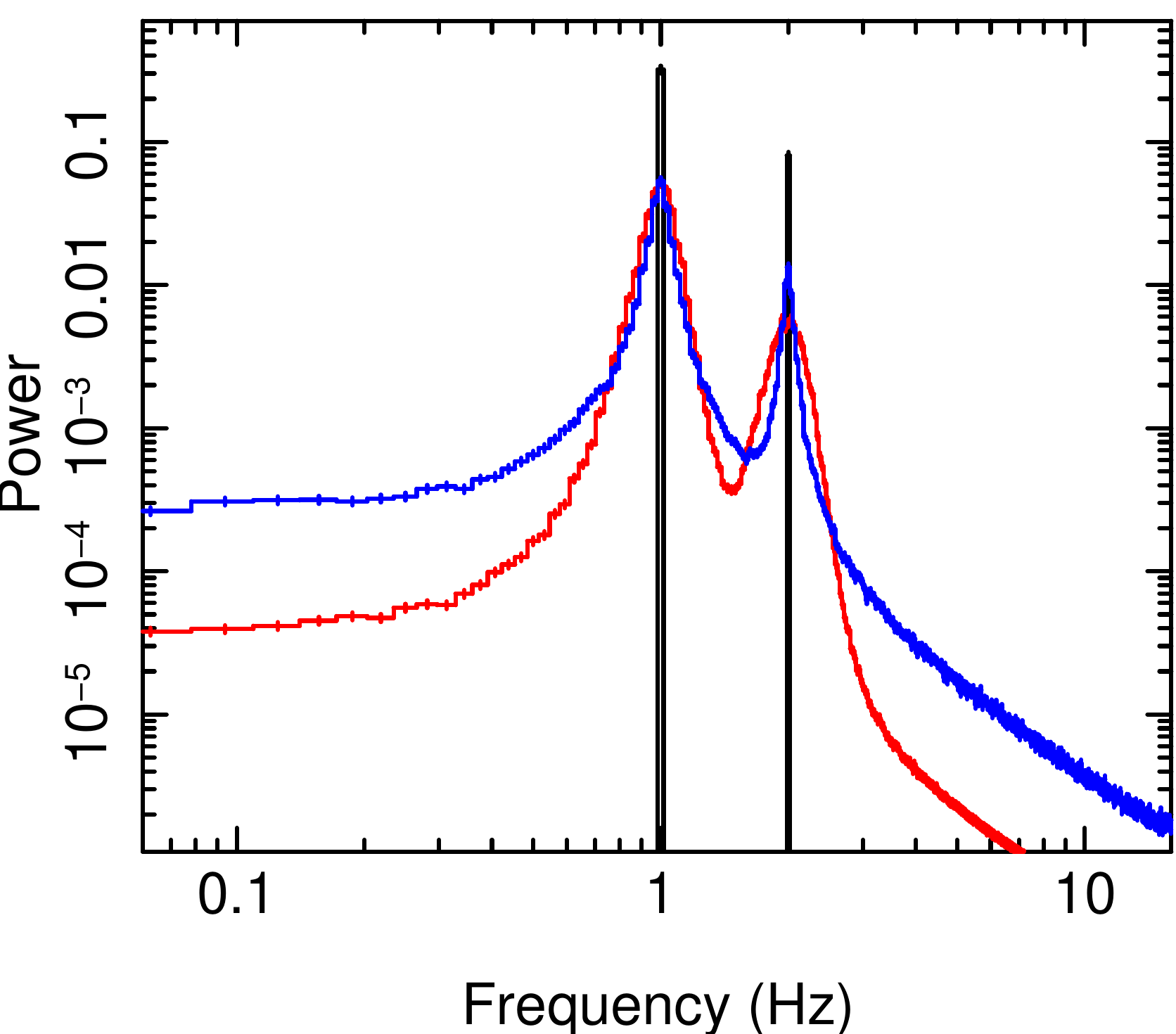}
    \centering
    \caption{Time series (left) and their corresponding power spectra (right). The periodic time series (top left) has a power spectrum consisting of two $\delta$-functions (black). Introducing frequency modulation (second from top) leads to the two harmonics in the power spectra having the same quality factor (red). Introducing amplitude modulation (middle) leads to the two harmonics in the power spectra having the same width (blue). However, the maximally stochastic time series in the second from bottom and bottom panels respectively have \textit{identical} power spectra to the red and blue lines plotted on the right.  }
    \label{fig:modmechs}
\end{figure}

First, let us discuss the effects that may make an oscillation quasi-periodic instead of periodic. As a starting point, we can represent a QPO as a sum of two harmonic components:
\begin{equation}
    f(t) = 1 + a_1(t) \sin[ \varphi(t) ] + a_2(t) \sin[ 2 ( \varphi(t) - \psi(t) ) ].
    \label{eqn:simpleQPO}
\end{equation}
Here, $a_1$ and $a_2$ are the amplitudes of the two harmonic components and the QPO phase is
\begin{equation}
    \varphi(t) = \varphi_0 + 2\pi \int_{0}^{t} \nu_{\rm qpo}(t') dt',
\end{equation}
where $\varphi_0$ is the QPO phase at $t=0$, $\nu_{\rm qpo}$ is the centroid frequency of the fundamental and $\psi$ is the phase difference between the harmonics.
%We will discuss this phase difference further in the following sub-section, but from equation \ref{eqn:simpleQPO} we can see that a peak in the fundamental component will be followed by a peak in the second harmonic a time $\sim \psi/(\pi \nu_{\rm qpo})$ later.
For the simplest case whereby the frequency, phase difference and amplitudes are constant, we would have a purely periodic function, whose power spectrum would be a sum of $\delta$-functions (i.e. a pulsation). This case is demonstrated by the black lines in Fig \ref{fig:modmechs}, with the time series on the top left ($\psi=0.1\pi$, $a_2=a_1/2$) and the corresponding power spectrum on the right.

Now let us explore the effect of \textit{frequency modulation} by allowing the QPO frequency to change with time. In order to generate a time series for $\nu_{\rm qpo}(t)$ with the desired properties, we first model its power spectrum (using fractional rms normalisation) as a Lorentzian function (Equation \ref{eqn:lore}). We use $\nu_0=0$ (i.e. a zero-centered Lorentzian), $a_0=0.1$ and $\Delta=0.1$ Hz. We then use the algorithm of \citet{Timmer1995} to generate a maximally stochastic time series from this input power spectrum. Essentially, the Fourier components of this time series have amplitudes corresponding to the target power spectrum and completely random phases (see e.g. \cite{Ingram2013} for an expanded discussion). The time series shown in the second panel from the top
in Fig \ref{fig:modmechs} (left) is calculated using this variable QPO frequency, with the amplitudes of both harmonics held constant (i.e. pure frequency modulation). The corresponding power spectrum is plotted on the right (red). We see that the harmonic peaks now have a finite width, and that both harmonics have the same \textit{fractional} width (i.e. they have the same $Q$, and therefore appear to have the same width on a logarithmic scale). This happens because the centroid frequency of the first and second harmonic at time $t$ is respectively $\nu_{\rm qpo}(t)$ and $2\nu_{\rm qpo}(t)$, and therefore the range of frequencies covered by the second harmonic is always twice that of the first.

%Left, Bottom, Right, Top

We can instead explore the effect of \textit{amplitude modulation} by holding the QPO frequency constant and allowing the harmonic amplitudes to vary. We use a Timmer and Koenig simulation to generate a time series for $a_1(t)$ from a zero-centered Lorentzian power spectrum with $a_0=0.9$ and $\Delta=0.05$ Hz, and set $a_2(t)=a_1(t)/2$.
%The mean value of $a_1(t)$ is set to zero.
The resulting QPO time series is in the middle panel of Fig \ref{fig:modmechs} (left), and the corresponding power spectrum is on the right (blue). This pure amplitude modulation gives the harmonic peaks the same \textit{absolute} width as each other, which happens because each harmonic component is multiplied by essentially the same stochastic time series.

\begin{figure}
%\includegraphics[width=8cm,trim=2.0cm 0.5cm
%        2.0cm 0.0cm,clip=true]
	\includegraphics[width=0.48\textwidth,trim=0.0cm 0.0cm
        0.0cm 0.5cm,clip=true]{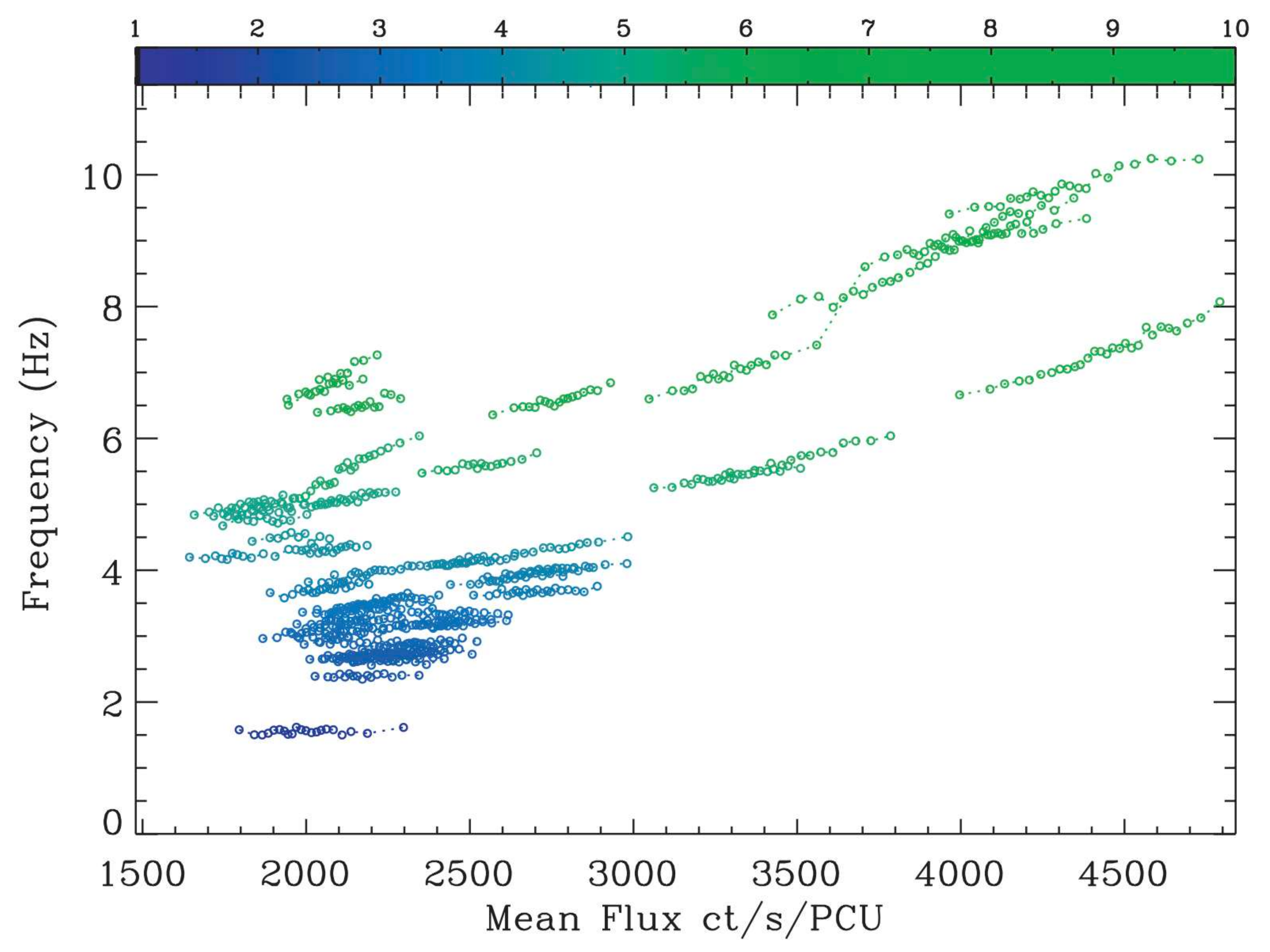} ~
	\includegraphics[width=0.48\textwidth,trim=0.0cm 0.0cm
        0.0cm 2.5cm,clip=true]{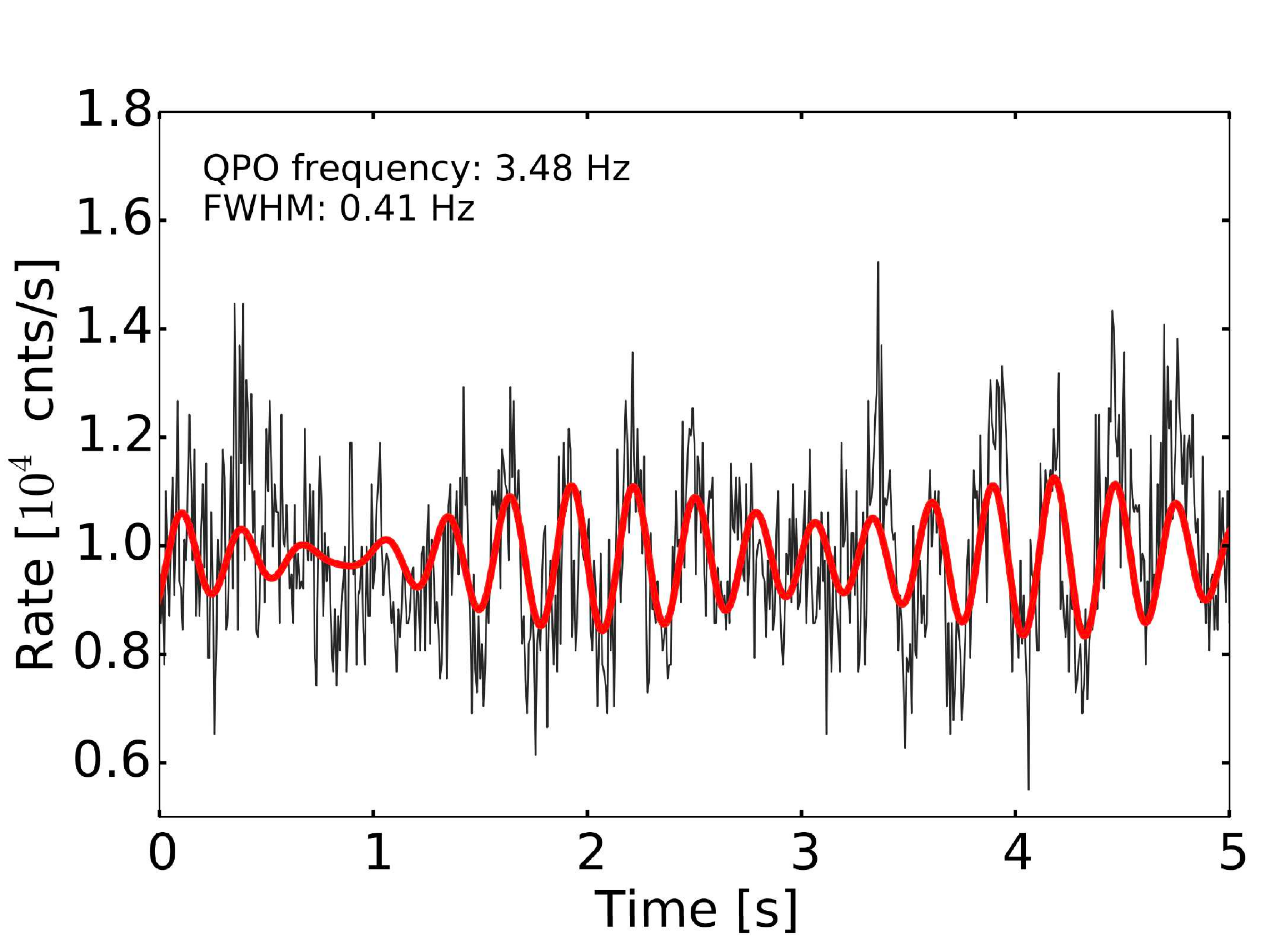} ~
    \centering
    \caption{\textit{Left:} QPO frequency measured in 3 s segments plotted against the mean count rate of those segments (adapted from \citet{Heil2011}). This demonstrates that short timescale fluctuations in the QPO frequency and mean count rate are correlated, with the gradient of the correlation increasing with mean QPO frequency (colour bar represents mean QPO frequency). \textit{Right:} Raw (black) and optimally filtered (red) X-ray light curve from GRS 1915+105 (reproduced from \citet{vandeneijnden2016}). The filtered light curve reveals `coherent intervals', between which the QPO signal loses coherence. Both plots are created with \textit{RXTE} data.}
    \label{fig:heil2011}
\end{figure}

Insight into the dominant de-cohering mechanism can therefore, in principle, be gained by modelling observed power spectra with a multi-Lorentzian model. If the harmonics have the same width, $\Delta$, we can conclude that amplitude modulation is the most important effect, and if they have the same quality factor, we can instead conclude that frequency modulation is the most important effect. A number of studies have shown that Type-C QPOs are consistent with the latter \cite{Rao2010,Ratti2012,Pawar2015}, even when higher harmonics than the second are detected. A complication is the sub-harmonic, which often has a lower quality factor. This could be explained by additional amplitude modulation. \citet{Heil2011} investigated frequency modulation of the Type-C QPO in XTE J1550-564 by splitting the light curve into $3$ s segments and stacking the power spectra of those segments based on count rate (flux binning). They found that the QPO frequency correlates with count rate (see Fig \ref{fig:heil2011}, left), and that the higher the mean QPO frequency, the steeper the gradient of this correlation. \citet{Nespoli2003} had earlier found similar results for the Type-B QPO in GX 339-4.

%Left, Bottom, Right, Top

However, the power spectrum \AI{includes only the amplitudes of the Fourier components, and discards the phases. Therefore much of the information in the time series is not represented by the power spectrum}. To demonstrate this, \AI{we take the two power spectra shown in Fig \ref{fig:modmechs} (right) and generate time series with these power spectra using the} Timmer and Koenig algorithm. \AI{The resulting time series are shown in the bottom two panels} of Fig \ref{fig:modmechs} (left). \AI{We see that the red `stochastic' time series (second from bottom) is dramatically different from the red `frequency modulation' time series (second from top), yet they have \textit{identical} power spectra to one another. Likewise, the blue `stochastic' time series is very different to the blue `amplitude modulation' time series (middle), eventhough the two have the same power spectrum.} This demonstrates that the relative widths of the QPO harmonics in the power spectrum do not uniquely determine the mechanism that makes the QPO quasi-periodic. For this information, we must use statistics more sophisticated that the power spectrum that preserve phase. Fig \ref{fig:heil2011} (right) shows a result from \citet{vandeneijnden2016}, who applied an optimal (Weiner) filter (see e.g \cite{Press1992}) to X-ray light curves from GRS 1915+105 in an attempt to isolate the Type-C QPO from the associated broad band noise. The black line is the raw data, and the red line is the filtered data. We see both frequency and amplitude modulation. In particular, we see what were termed \textit{coherent intervals} whereby the QPO signal remains coherent for $\sim Q$ cycles, then loses coherence before the next coherent interval begins (there are $\sim 2.5$ coherent intervals in the plot). The amplitude reaches a minimum between coherent intervals and a maximum in the middle of the intervals. Interestingly, \citet{vandeneijnden2016} used the same filtering technique to study observations in which the QPO frequency depends on energy band to find that during each coherent interval, the QPO phase of the energy band with higher frequency accelerates ahead of the slower band, only for the phases to reset at the start of the next coherent interval. Evidence of these coherent intervals can also be found using wavelet transforms \cite{Lachowicz2010} or the Hilbert-Huang transform \cite{Su2015}.

\subsection{QPO waveform}
\label{sec:qpowaveform}

If we wish to measure the waveform of the QPO -- the count rate as a function of QPO phase (what pulsar astronomers would call the `pulse profile') -- we require more information than can be determined from the power spectrum alone. In fact, from the power spectrum alone, we do not know if a well defined QPO waveform even \textit{exists}. The QPO could instead be purely stochastic noise with the variability amplitude peaking at harmonically related frequencies -- as is the case for the Timmer and Koenig simulations plotted in the bottom two panels of Fig \ref{fig:modmechs} (left). The difference between these maximally stochastic time series and the other quasi-periodic time series plotted in Fig \ref{fig:modmechs} (left), that \textit{do} have well defined waveforms, is that the phases of the QPO harmonic components are correlated with one another in the latter case and not in the former case. We can think of this in terms of the phase difference between harmonics, $\psi$ (see equation \ref{eqn:simpleQPO}). $\psi(t)$ varying randomly with a uniform distribution results in a maximally stochastic time series whereas $\psi(t)$ varying around a well-defined mean value instead leads to the harmonic components being correlated -- in which case it makes sense to try to constrain a QPO waveform.

\begin{figure}
%\includegraphics[width=8cm,trim=2.0cm 0.5cm
%        2.0cm 0.0cm,clip=true]
	\includegraphics[width=0.48\textwidth,trim=1.0cm 1.0cm
        2.5cm 10.0cm,clip=true]{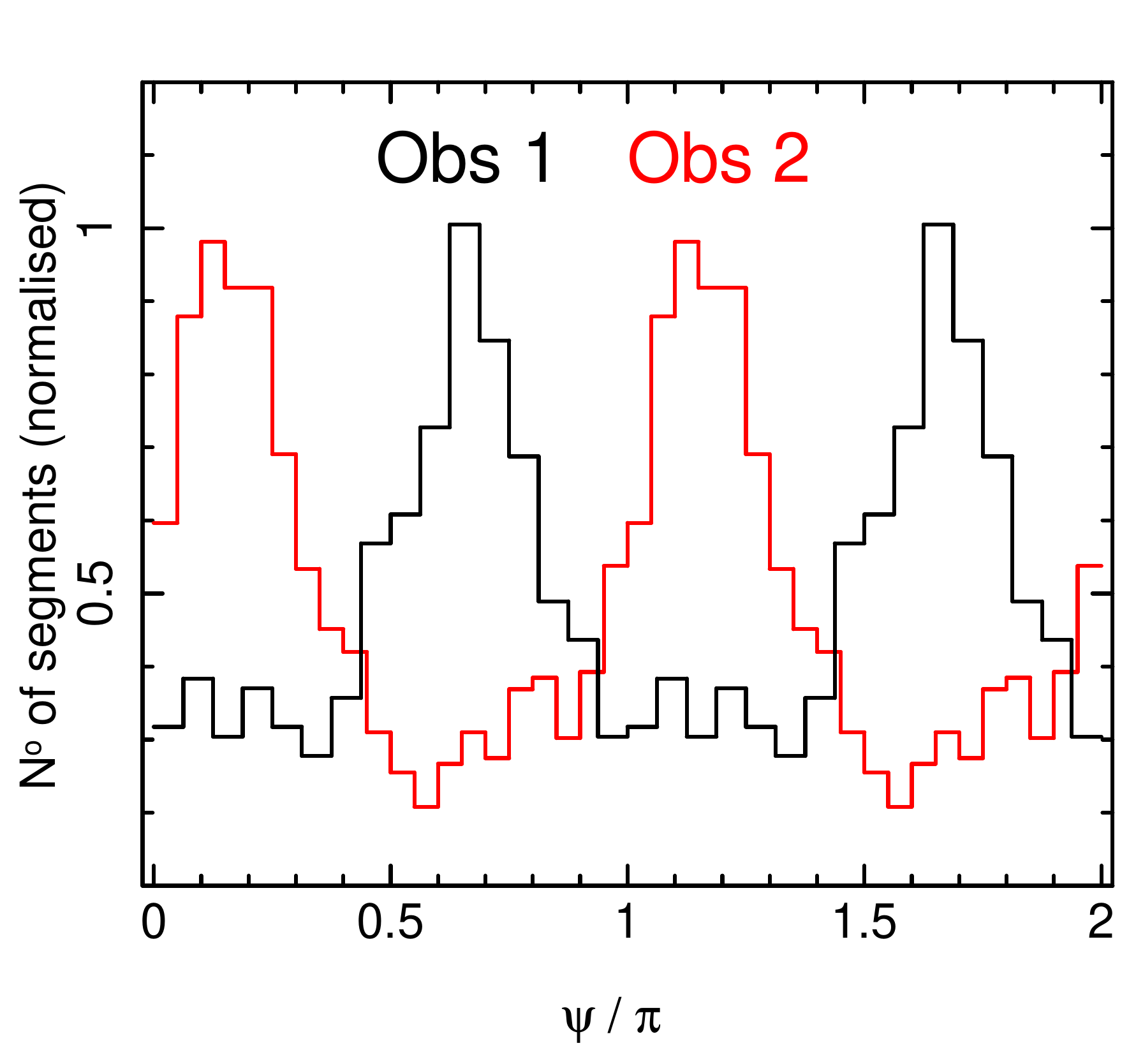} ~
	\includegraphics[width=0.48\textwidth,trim=1.0cm 1.0cm
        2.5cm 10.0cm,clip=true]{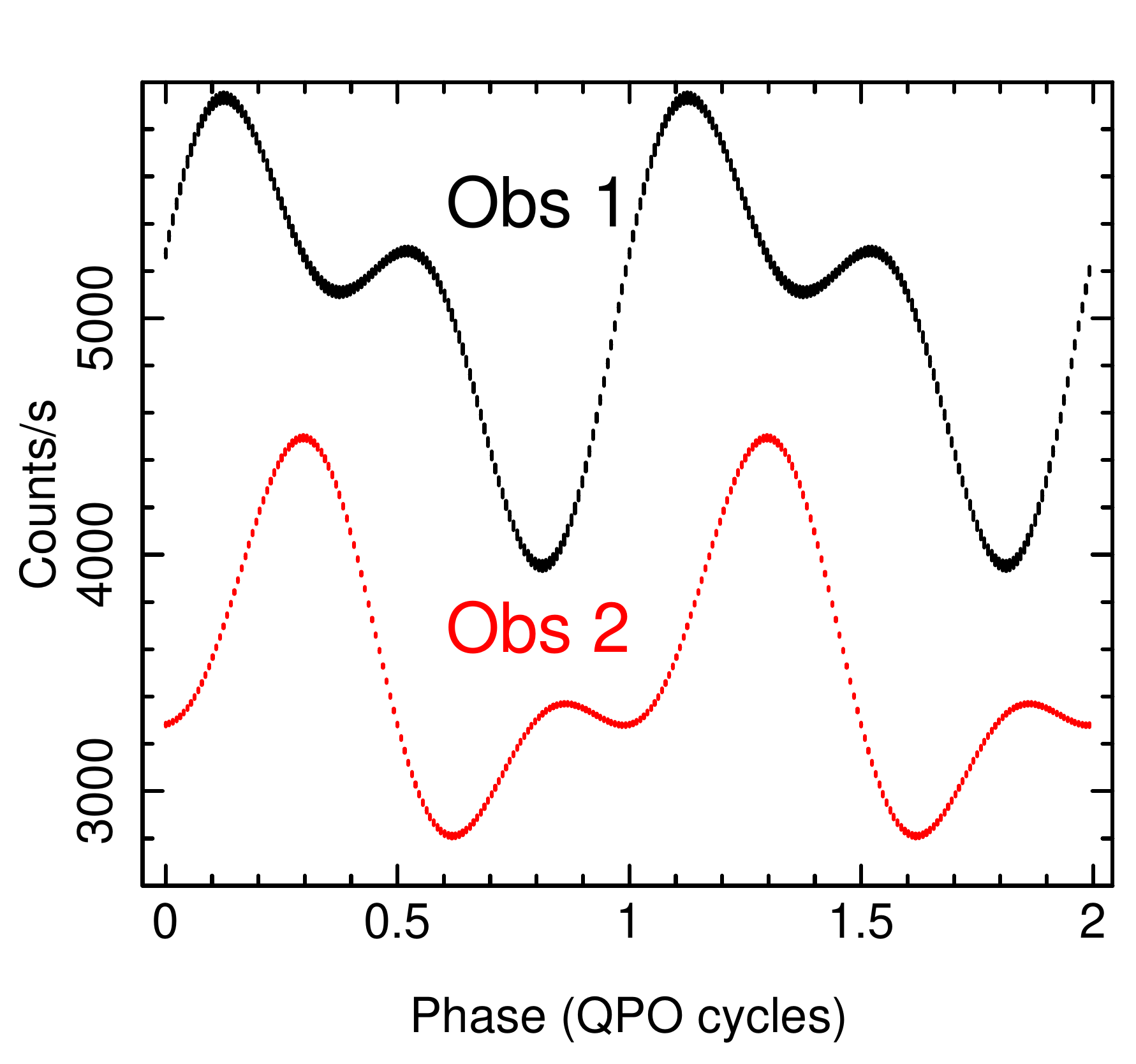} ~
    \centering
    \vspace{-1cm}
    \caption{\textit{Left:} Distribution of the phase difference between the first and second QPO harmonics, $\psi$, for two \textit{RXTE} observations of GRS 1915+105. Observations 1 (black) and 2 (red) have QPO frequencies $\nu_{\rm qpo}\sim 0.5$ Hz and $\nu_{\rm qpo}\sim 2.25$ Hz respectively. We see that $\psi$, which is defined on the interval $0\rightarrow\pi$, varies around a well-defined mean ($\psi\approx 0.67 \pi$ and $\psi \approx 0.13\pi$ for obs 1 and 2 respectively). \textit{Right:} QPO waveforms reconstructed using the measured $\psi$ values. Both plots are reproduced from \citet{Ingram2015}}
    \label{fig:waveform}
\end{figure}
%Left, Bottom, Right, Top

\citet{Ingram2015} investigated variations of this phase difference between harmonics for two \textit{RXTE} observations of Type-C QPOs in GRS 1915+105. They split the X-ray light curves up into short segments (with each segment containing $\sim Q$ QPO cycles) and for each segment calculated $\psi$ from the Fourier transform of the light curve. Fig \ref{fig:waveform} (left) shows histograms of these many measurements of $\psi$ for the two observations, which have QPO fundamental frequencies of $\sim 0.5$ Hz (obs 1, black) and $\sim 2.25$ Hz (obs 2, red). We see that these histograms each have a very clear peak, meaning that $\psi(t)$ does indeed vary around a well-defined mean value. The mean value of $\psi$ for each observation can be measured by determining the peak of the corresponding histogram, and an `average' QPO waveform can be reconstructed from combining this with measurements of the harmonic amplitudes (which can be determined from the power spectrum). Fig \ref{fig:waveform} (right) shows the waveforms re-constructed for the same two observations of GRS 1915+105. QPO waveforms can alternatively be constrained using phase-folding algorithms. \citet{Tomsick2001} filtered \textit{RXTE} light curves of GRS 1915+105 in order to isolate the QPO from the broadband noise (similar to our Fig \ref{fig:heil2011}, right). They then defined peaks in the filtered light curve as zero points of QPO phase in order to phase fold. That their results are similar to those of \citet{Ingram2015} provides confidence in the reconstruction technique. It also, in hindsight, tells us that we already knew in 2001 that the phase difference between harmonics varies around a well-defined mean. This is because the \citet{Tomsick2001} algorithm tracks the phase of the QPO fundamental, which would be uncorrelated with the phase of the second harmonic if $\psi(t)$ followed a uniform distribution. This would have resulted in the \citet{Tomsick2001} waveforms all being purely sinusoidal due to the second harmonic component cancelling completely in the phase-folded waveforms. Therefore, the very fact that the \citet{Tomsick2001} waveforms have harmonic content shows that $\psi$ is not uniformly random. If $\psi$ were constant, then the amplitude of the second harmonic in the phase-folded waveform would be the same as that in the Fourier reconstructed waveform. Instead, $\psi(t)$ varies around a mean value, meaning that the phase-folding method always predicts smaller second harmonic amplitudes than the reconstruction method due to destructive interference. Furthermore, any error in assigning instantaneous QPO phase leads to a reduction of the overall amplitude of the phase-folded waveform for exactly the same reason. This provides a good test of any phase-folding algorithm: if the QPO waveform ends up with a very small amplitude, as would be the case if one simply folded on a constant QPO period, one can conclude that the algorithm has not worked, and that the results have little meaning.

\begin{figure}
%\includegraphics[width=8cm,trim=2.0cm 0.5cm
%        2.0cm 0.0cm,clip=true]
	\includegraphics[width=0.6\textwidth,trim=0.5cm 1.0cm
        1.5cm 10.0cm,clip=true]{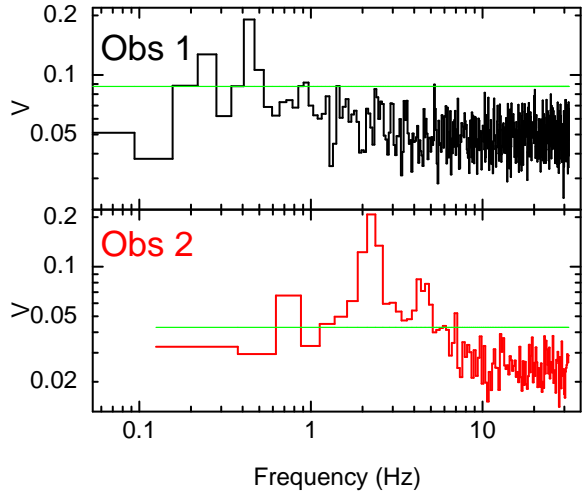}
    \centering
    \caption{Kuiper's statistic $V$, which is a measure of phase correlation between the Fourier component at a given frequency and twice that frequency, plotted against the lower frequency of the pair for the two GRS 1915+105 observations featured in Fig \ref{fig:waveform}. The $\psi$ distribution is $3\sigma$ inconsistent with uniform for $V$ above the green line. We see that the phases of the QPO harmonics couple strongly to each other, in contrast to pairs of broadband noise frequencies. Reproduced from \citet{Ingram2015}.}
    \label{fig:KPspec}
\end{figure}
%Left, Bottom, Right, Top

\citet{Ingram2015} further investigated the GRS 1915+105 data by measuring the phase difference between the Fourier components at each frequency $\nu$ and at $2\nu$. For each frequency pair, they calculated \textit{Kuiper's statistic}, $V$, to determine whether or not the distribution of $\psi$ is consistent with a uniform distribution (high $V$ means the distribution is \textit{not} consistent with uniform). Fig \ref{fig:KPspec} shows Kuiper's statistic plotted against $\nu$ for the two GRS 1915+105 observations featured in the previous figure. Values of $V$ above the green line correspond to a non-uniform distribution with $>3\sigma$ confidence. The peaks at the QPO fundamental frequency ($\sim 0.5$ Hz and $\sim 2.25$ Hz for obs 1 and 2 respectively) indicate coupling between the first and second harmonics, and the peaks at twice and thrice the QPO fundamental in the bottom plot indicate respectively coupling between the second and fourth harmonics and the third and sixth harmonics! The peak at half the QPO frequency for observation 1 indicates coupling between the sub-harmonic and the fundamental, and the peak at a third of the QPO frequency in observation 2 turns out to indicate coupling between the QPO and the broadband noise (see the following section). 
%It will be a recurring theme throughout this review that the sub-harmonic is somewhat of a mystery.
%Fig \ref{fig:KPspec} throws some doubt on the simplest explanation of the sub-harmonic: that it is actually the fundamental. But why in this case would the phases between the second and fourth harmonics be correlated so much better than the phases of the first and second harmonics?
The other feature of note in Fig \ref{fig:modmechs} is the difference between the QPO and the broadband noise: there is no evidence of strong phase coupling between pairs of frequencies both dominated by the broadband noise (although we will see in the next sub-section that there are some weak correlations).

\begin{figure}
%\includegraphics[width=8cm,trim=2.0cm 0.5cm
%        2.0cm 0.0cm,clip=true]
	\includegraphics[width=\textwidth,trim=0.0cm 0.0cm
        0.0cm 0.0cm,clip=true]{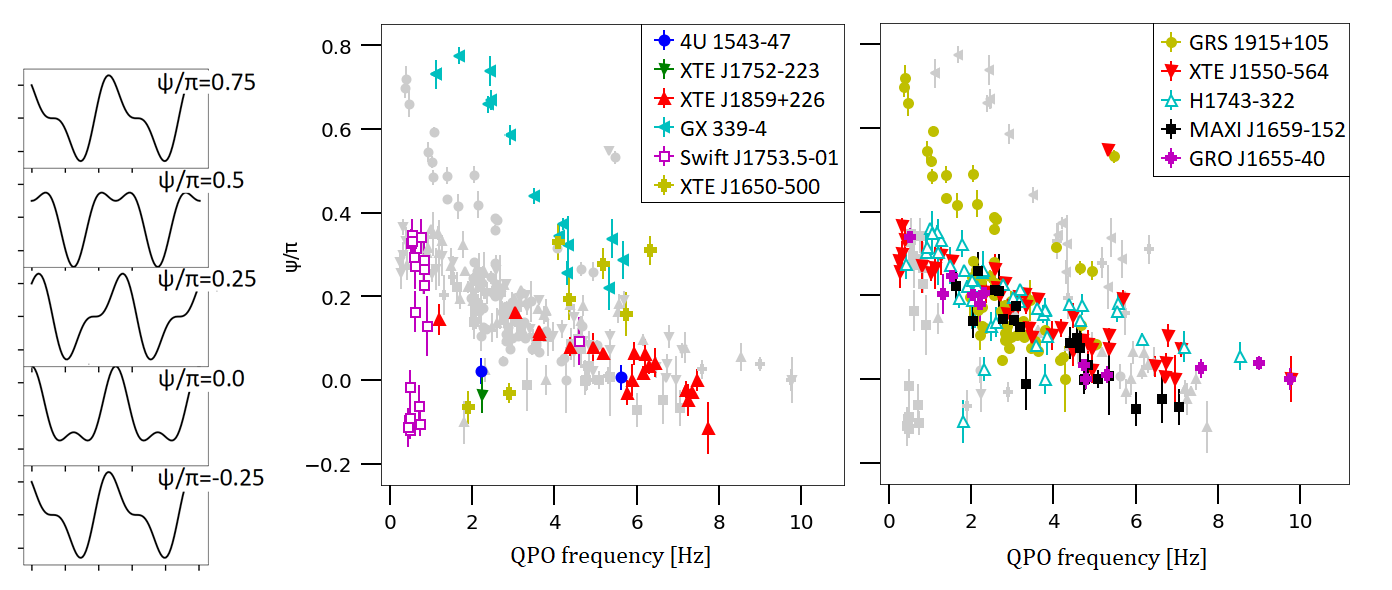}
    \centering
    \caption{Phase difference between QPO harmonics, $\psi$, plotted against QPO fundamental frequency for many \textit{RTXE} observations of Type-C QPOs. The coloured points on the left and right panels correspond respectively to low and high inclination sources (sources as labelled), and the grey points correspond to the coloured points from the other plot. Examples of QPO waveforms for given values of $\psi$ are displayed on the left. Reproduced from \citet{DeRuiter2019}.}
    \label{fig:iris}
\end{figure}

Recently, \citet{DeRuiter2019} measured $\psi$ for many observations of QPOs in the \textit{RXTE} archive. They found that $\psi$ is always measured to vary around a well defined mean whenever the data are good enough to do so. Fig \ref{fig:iris} shows their measurements of $\psi$ plotted against QPO frequency for many observations of Type-C QPOs, with the coloured points in the left and right panels corresponding respectively to low (i.e. the binary system is viewed more face-on) and high (i.e. the binary system is viewed more edge-on) inclination sources. We see that $\psi$ generally reduces from $\sim 0.4\pi$ to $\sim 0$ as the QPO frequency evolves from $\sim 0.1$ to $\sim 10$ Hz, with seemingly less scatter for the high inclination sources. This confirms the hints from Fig \ref{fig:waveform} that the waveform evolves systematically with QPO frequency (see the example waveforms for different values of $\psi$ to the left of Fig \ref{fig:iris}). \citet{DeRuiter2019} also found that Type-B QPOs have well defined waveforms, and that the measured $\psi$ value evolves with QPO frequency in a manner that differs from the Type-C case (see their Fig 6). This implies that these two QPO types have a genuinely different origin, with the difference between the two being more than simply the associated broadband noise.

%Left, Bottom, Right, Top

\subsection{The bi-spectrum: coupling of the QPO and broadband noise}
\label{sec:bispec}

Further insight can be gained from the bi-spectrum, which is defined as
\begin{equation}
    B(\nu_j,\nu_k) = \langle X(\nu_j) X(\nu_k) X^*(\nu_{j+k}) \rangle,
\end{equation}
where $X(\nu_j)$ is the $j^{\rm th}$ frequency of the Fourier transform of the X-ray light curve \cite{Kim1979,Maccarone2011}, and the averaging indicated by the angle brackets is over different light curve segments. Essentially, the bi-spectrum describes the correlations between the phases of Fourier components at three frequencies: $\nu_j$, $\nu_k$ and $\nu_{j+k}$, and is only defined for $\nu_{j+k}$ less than the Nyquist frequency (and only non-trivially defined for $\nu_j<\nu_k$). The bi-spectrum informs on how well correlated the square of a time series is with the time series itself. Defining $z(t_k)=x^2(t_k)$, it can be shown using the convolution theorem that
\begin{equation}
    \langle X(\nu_j) Z^*(\nu_j) \rangle = \sum_{k=-N/2+1}^{N/2} B(\nu_j,\nu_k),
\end{equation}
where $N$ is the number of time intervals in a light curve segment. Therefore a time series with zero bi-spectrum -- as is generated by the \citet{Timmer1995} algorithm -- is uncorrelated with its own square. It is possible to simulate a time series with a specified bi-spectrum using the algorithm of \citet{Vanhoff1997}, which was developed for the application of modelling waves in shallow water.

\begin{figure}
%\includegraphics[width=8cm,trim=2.0cm 0.5cm
%        2.0cm 0.0cm,clip=true]
	\includegraphics[width=\textwidth,trim=0.0cm 3.0cm
        0.0cm 6.0cm,clip=true]{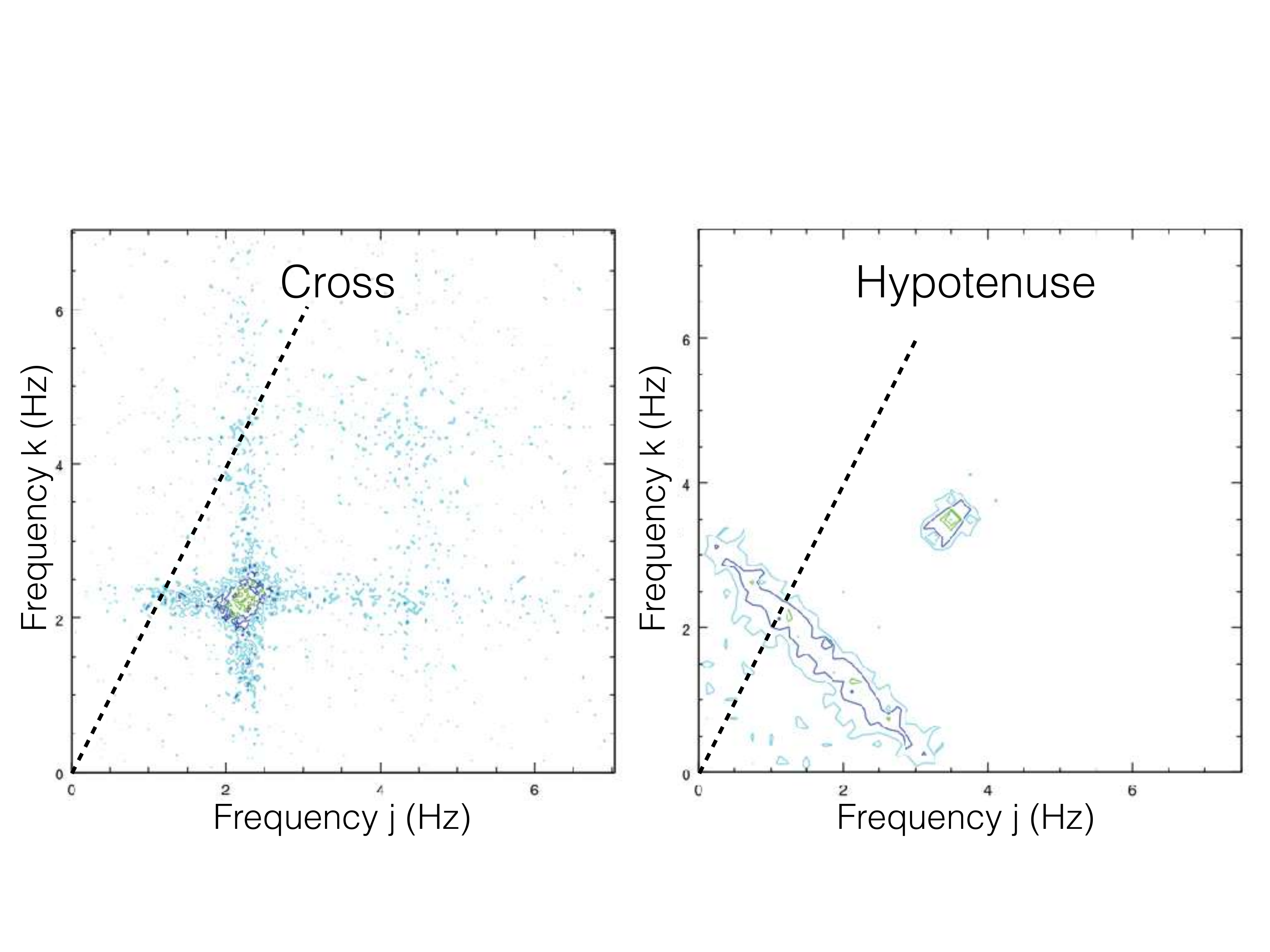}
    \centering
    \caption{Two examples of the bi-coherence of GRS 1915+105, adapted from \citet{Maccarone2011}. In general, three patterns are observed: the `cross' pattern (left), the `hypotenuse pattern' (right), and the `web' pattern is a combination of the cross and the hypotenuse. The dashed lines mark $\nu_j=2\nu_k$.}
    \label{fig:bicoh}
\end{figure}
%Left, Bottom, Right, Top

It is convenient to define the squared bi-coherence
\begin{equation}
    b^2(\nu_j,\nu_k) = \frac{|B(\nu_j,\nu_k)|^2}{ \langle |X(\nu_j)X(\nu_k)|^2 \rangle \langle |X(\nu_{j+k})|^2 \rangle},
\end{equation}
which can take values between $b^2=0$ and $b^2=1$ \cite{Kim1979}. Fig \ref{fig:bicoh} shows some contour plots of the bi-coherence of GRS 1915+105, reproduced from \citet{Maccarone2011}. They identified characteristic patterns displayed by Type-C QPOs in the source: the `cross' pattern (left), the `hypotenuse' pattern (right), and the `web' pattern, which is a combination of the other two.
%The `cross' pattern (left) includes peaks where both $\nu_1$ and $\nu_2$ are at a QPO harmonic peak (with the strongest peak at $\nu_1=\nu_2=\nu_{\rm qpo}$) and lines of high bi-coherence for $\nu_1=\nu_{\rm qpo}$ and $\nu_2=\nu_{\rm qpo}$. The `hypotenuse' pattern (right) also includes a peak at $\nu_1=\nu_2=\nu_{\rm qpo}$, alongside a diagonal line at $\nu_1+\nu_2=\nu_{\rm qpo}$. Note that the symmetry around $\nu_1=\nu_2$ in these plots is trivial.
The pattern exhibited by the source was found to correlate with the radio brightness and the QPO frequency. Recently, \citet{Arur2019} conducted a systematic analysis of GX 339-4 to find that the bi-spectrum transitions smoothly from the web to the hypotenuse pattern as the Type-C QPO frequency increases (no observations displayed the cross pattern for GX 339-4). Type-B QPOs only have bi-coherence peaks at $\nu_j=\nu_k=\nu_{\rm qpo}$, and Type-A QPOs have low bi-coherence. The peaks at combinations of QPO harmonics indicate phase coupling between QPO harmonics, whereas the diagonal line of raised bi-coherence in the hypotenuse pattern indicates coupling between the QPO and the broadband noise. The broadband noise itself has a non-zero bi-coherence, albeit more than an order of magnitude lower than that of the QPO \cite{Uttley2005}. This is expected, since the broadband noise is not Gaussian distributed, but log-normal distributed, which naturally arises from propagation of fluctuations in mass accretion rate \cite{Uttley2005}.

The dashed line along $\nu_k=2\nu_j$ enables us to compare these bi-coherence plots with the Kuiper's statistic plot in Fig \ref{fig:KPspec} if we read the frequency in Fig \ref{fig:KPspec} as $\nu_j$. This is because high bi-coherence roughly corresponds to large $V$. We see that the cross pattern along the dashed line consists of peaks at $\nu_j=\nu_{\rm qpo}/2$ and $\nu_j=\nu_{\rm qpo}$ as is seen for observation 1 in Fig \ref{fig:KPspec}, whereas the hypotenuse pattern along this line instead consists of a peak at $\nu_j=\nu_{\rm qpo}/3$. Since observation 2 in Fig \ref{fig:KPspec} includes such a peak at $\nu_j=\nu_{\rm qpo}/3$ in addition to peaks at each QPO harmonic frequency, we can conclude that observation 2 from \citet{Ingram2015} corresponds to a web pattern, and observation 1 to a cross pattern. From this comparison, we can make two insights that were not appreciated by \citet{Ingram2015} at the time: i) the peak at half the QPO frequency for observation 1 corresponds to the $\nu_k=\nu_{\rm qpo}$ line in the cross pattern of the bi-coherence, and ii) the peak at a third of the QPO frequency in observation 2 indicates phase coupling between the QPO and the broadband noise. It is still rather unclear quite what this non-linear interaction between the QPO and broadband noise is, but some simple models are explored and discussed in \cite{Maccarone2011} and \cite{Arur2019}. The phase difference between harmonics, $\psi$, can also be related to the bi-spectrum, through the bi-phase \cite{Maccarone2013}. Indeed, \citet{Arur2019} recently showd that the bi-spectrum can be used to reconstruct the QPO waveform.

\section{Lense-Thirring precession and disc theory}
\label{sec:LT}

Before exploring specific models for LF QPOs in the following section, we will first lay some important theoretical foundations. Many of the models in the literature consider the relativistic effect of Lense-Thirring precession, and so we take some time to introduce the concept in this section. As we will see, Lense-Thirring precession can greatly affect the dynamics of accretion discs that are initially misaligned with the equatorial plane of a spinning BH. Here, we briefly summarise the analytic theory for such tilted discs and compare with the results of recent numerical simulations. Hereafter, we will employ the convention whereby lower case $r$ is radius expressed in units of a gravitational radius $R_g=GM/c^2$, such that $r=R/R_g$ (where $M$, $G$ and $c$ are respectively BH mass, Newton's gravitational constant and the speed of light in a vacuum).

\subsection{Test mass frequencies}

For a test mass orbiting a spinning BH in a nearly circular orbit slightly perturbed from the BH equatorial plane, the orbital, radial and vertical epicyclic angular frequencies in the Kerr metric are \cite{Bardeen1972,Merloni1999}
\begin{equation}
\begin{split}
\Omega_\phi &= \pm \frac{c}{R_g} \frac{1}{r^{3/2}\pm a} \\
\Omega_r   &= \Omega_\phi \sqrt{ 1 - \frac{6}{r} \pm \frac{8a}{r^{3/2}} - \frac{3a^2}{r^2} } \\
\Omega_z &= \Omega_\phi \sqrt{ 1 \mp \frac{4a}{r^{3/2}} + \frac{3a^2}{r^2} },
\end{split}
\label{eqn:epicyclic}
\end{equation}
where $a=J_{\rm BH}/(M c R_g)$ is the dimensionless spin parameter, $J_{\rm BH}$ is the angular momentum of the hole and top and bottom signs represent pro- and retro-grade spin respectively. The corresponding frequencies in units of cycles per unit time are $\nu_{\rm \phi,r,z}=\Omega_{\rm \phi,r,z} /(2\pi)$. In Newtonian gravity, all three of these frequencies are always equal, meaning that orbits always close. In the Kerr metric, however, the inequality between the orbital and radial epicyclic frequencies leads to \textit{periastron precession}. This is a rotation of the semi-major axis of an orbit with angular frequency $\Omega_{\rm per} = \Omega_\phi - \Omega_r$. At the time of Einstein proposing GR, periastron precession had already been observed, and had remained unexplained for more than 50 years, in Mercury's orbit around the Sun.

The inequality between the orbital and vertical epicyclic frequencies leads to \textit{Lense-Thirring precession}. This is a nodal precession (i.e. a vertical wobble) of orbits misaligned with the BH equatorial plane, with angular frequency $\Omega_{\rm LT} = \Omega_\phi - \Omega_z$. It results from the \textit{frame dragging effect}, which is the dragging of inertial frames about the BH spin axis. Lense-Thirring precession is named after the authors who first derived the effect, although the \citet{Lense1918} expression for the frequency was derived in the weak field limit ($a/r \ll 1$), this being decades before the derivation of the Kerr metric \cite{Kerr1963}. Their expression
\begin{equation}
\Omega_{LT} \approx \pm \frac{c}{R_g}\frac{2a}{r^3},
\label{eqn:nult}
\end{equation}
can be derived from equations \ref{eqn:epicyclic} via a first order Taylor expansion. The hierarchy of the precession frequencies is $\Omega_\phi > \Omega_{\rm per} > \Omega_{\rm LT}$.

\subsection{The diffusive and wave-like regimes}

If a disc is initially tilted with respect to the equatorial plane of a spinning BH, it will be warped by the differential nature of the Lense-Thirring precession frequency. The response of the disc to this turns out to depend on whether the \citet{Shakura1973} dimensionless viscosity parameter $\alpha$ is greater than or less than the disc scale height $H/R$. For $\alpha > H/R$, the warp is communicated by viscosity (the \textit{diffusive regime}) and for $\alpha < H/R$, the warp is instead communicated by pressure waves (the \textit{wave-like regime}). Here we will take some time to understand what these two regimes are and where this comparison between $\alpha$ and $H/R$ comes from. We can start by splitting the disc into rings and representing the angular momentum per unit surface area of a ring as the vector
\begin{equation}
\mathbf{L}= \Sigma R^2 \Omega_\phi \mathbf{\hat{l}} = \Sigma R^2 \Omega_\phi (\cos\gamma \sin\beta , \sin\gamma \sin\beta , \cos\beta ).
\end{equation}
Here, $\Sigma$ is the \textit{surface density} (the mass density integrated over the disc height), the z-axis $\mathbf{\hat{k}}$ aligns with the BH spin axis, $\beta$ is the (polar) \textit{tilt} angle and $\gamma$ is the (azimuthal) \textit{twist} angle. We take a hat to denote a unit vector throughout. We see that a disc is respectively \textit{misaligned}, \textit{warped} and \textit{twisted} if $\beta \ne 0$, $\partial \beta / dR \neq 0$ and $\partial \gamma / dR \neq 0$. Warps are communicated via the \textit{internal torque}, $\mathbf{G} = \mathbf{R} \times \mathbf{F}$, where $\mathbf{F}$ is the shear force acting at the boundary between the rings (see equations 1-4 in \cite{Nixon2012a}) caused by the rings moving past one another in the locally azimuthal direction due to differential rotation and in the locally vertical direction due to differential nodal precession.

In the diffusive regime, mass is assumed to pass between rings only via the boundaries between them. \citet{Pringle1992} showed that mass conservation gives
\begin{equation}
\frac{\partial \Sigma}{\partial t} + \frac{1}{R} \frac{\partial}{\partial R} \left( \Sigma R v_R \right) = 0,
\label{eqn:masscons}
\end{equation}
which turns out to be the same as for a planar disc \cite{Frank2002}, and angular momentum conservation gives
\begin{equation}
\frac{\partial \mathbf{L}}{\partial t} = \frac{1}{R} \frac{\partial }{\partial R}\left( v_R R \mathbf{L} \right) = \frac{1}{2\pi R} \frac{\partial \mathbf{G}}{\partial R} - \left( \frac{\Omega_\phi^2-\Omega_z^2}{2\Omega_\phi} \right) \mathbf{\hat{k}} \times \mathbf{L},
\label{eqn:dLdt}
\end{equation}
where $v_R$ is radial velocity and the final term of equation \ref{eqn:dLdt} describes the warp forced by differential Lense-Thirring precession. These equations can be solved by defining kinematic viscosity in the azimuthal and vertical directions, $\nu_1$ and $\nu_2$. Using the $\alpha$ prescription of \citet{Shakura1973}, we can set $\nu_1 = \alpha c_s H$, where $c_s$ is the sound speed. \citet{Papaloizou1983} and \citet{Ogilvie1999} found that $\nu_2=c_s H /(2\alpha)$, and so the internal torque in each direction can be calculated for a given value of $\alpha$ (see e.g. equations 21-23 in \cite{Nixon2016}).
%Using the $\alpha$ prescription of \cite{Shakura1973}, we can set $\nu_1 = \alpha c_s H = \alpha \Omega H^2$ (since the sound speed $c_s=\Omega H$ for a disc in vertical hydrostatic equilibrium). Similarly, $\nu_2 = \alpha_2 c_s H$. 
% \citet{Papaloizou1983} and \citet{Ogilvie1999} showed that $\alpha_2 = 1 / (2\alpha)$. Since $\alpha<1$ is expected, the vertical viscosity is larger than the usual azimuthal viscosity\footnote{There is also a third, more subtle, component to the internal torque, $\mathbf{G}_3$, that causes misaligned rings to precess \cite{Papaloizou1983,Ogilvie1999}, which we ignore for the purposes of this discussion.}. The internal torque in each direction can then be calculated for a given value of $\alpha$ (see e.g. equations 21-23 in \cite{Nixon2016}).
Solving equations \ref{eqn:masscons} and \ref{eqn:dLdt} then indicates that $\beta(R)$ tends to the initial tilt angle $\beta=\beta_0$ for large $R$ and to $\beta=0$ at small $R$, with the two regimes connected by a smooth warp (also see \cite{Chakraborty2017,Banerjee2019}). This result was first derived by \citet{Bardeen1975}, and therefore such a setup is often called the \textit{Bardeen-Patterson configuration}. We note, however, that their equations did not conserve angular momentum, but the result still held up to subsequent, more sophisticated treatments \cite{Papaloizou1983,Pringle1992}.

\begin{figure}
%\includegraphics[width=8cm,trim=2.0cm 0.5cm
%        2.0cm 0.0cm,clip=true]
	\includegraphics[width=0.7\textwidth,trim=0.0cm 0.0cm
        0.0cm 0.0cm,clip=true]{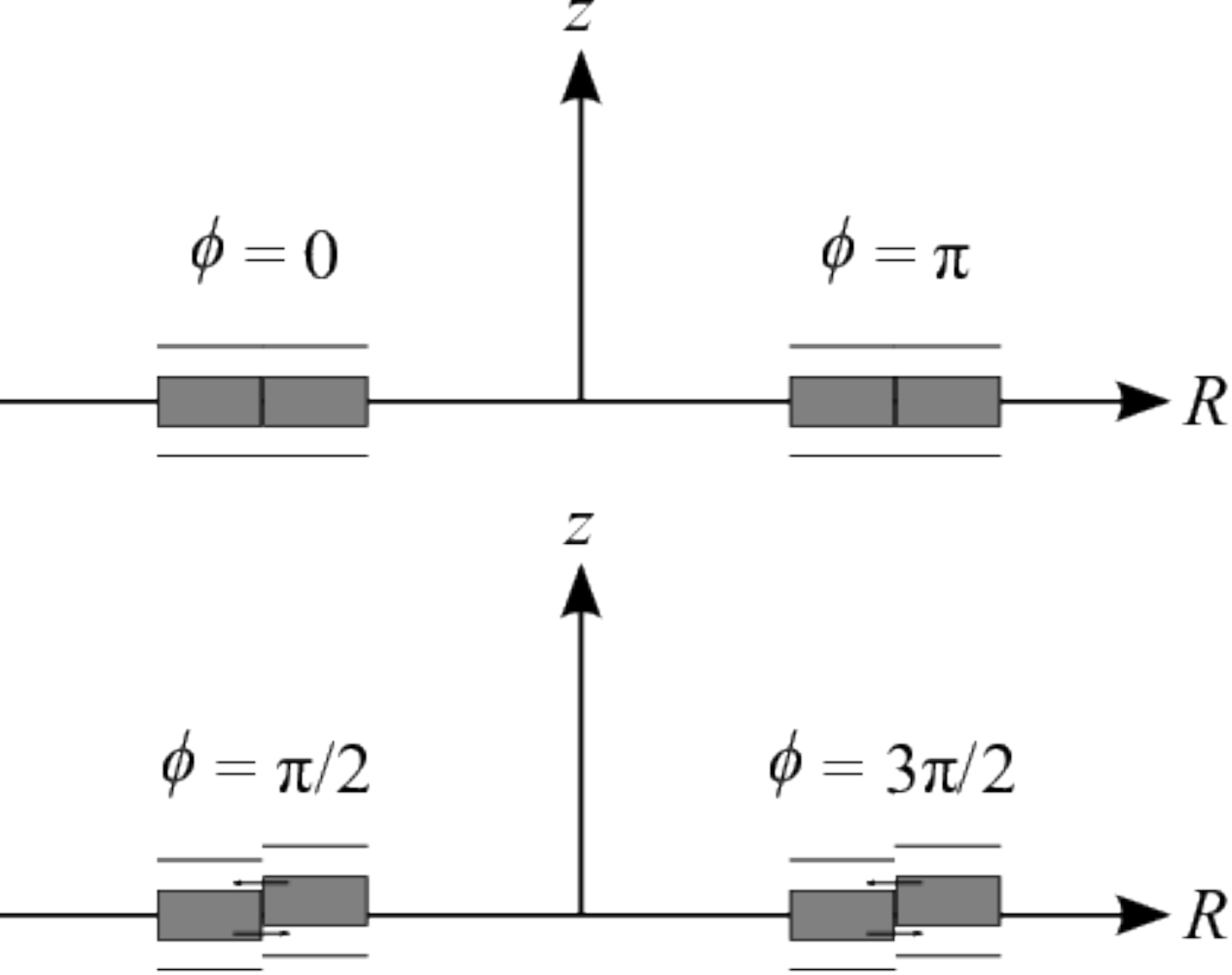}
    \centering
    \caption{Illustration of how a misalignment between two disc rings turns a vertical pressure gradient into an oscillating radial pressure gradient. Reproduced from \citet{Nixon2016}.}
    \label{fig:wavelike}
\end{figure}
%Left, Bottom, Right, Top

In the opposite, wave-like regime, the linearized equation for angular momentum conservation is essentially the same as the diffusive regime (equation \ref{eqn:dLdt}), whereas the internal torque obeys \cite{Lubow2000,Lubow2002,Nixon2016}
\begin{equation}
\frac{ \partial \mathbf{G}}{ \partial t}
+ \alpha \Omega_\phi \mathbf{G} =
\left( \frac{\Omega_\phi^2-\Omega_r^2}{2\Omega_\phi} \right)~\mathbf{\hat{l}}\times \mathbf{G}
+\Sigma~R^3~\frac{\Omega_z^2}{\Omega_\phi}~\frac{c_s^2}{4}~\frac{ \partial \mathbf{\hat{l}}}{ \partial R}.
\label{eqn:wavetorque}
\end{equation}
We can understand that these two equations have wave-like solutions for $\mathbf{\hat{l}}$, therefore explaining the name `wave-like regime', by exploring the invicid ($\alpha=0$), Keplerian ($\Omega_\phi=\Omega_r=\Omega_z$) limit. In this case, equations (\ref{eqn:dLdt}) and (\ref{eqn:wavetorque}) can be combined to find
\begin{equation}
\frac{\partial^2\mathbf{\hat{l}}}{\partial t^2} = \frac{1}{R L} \frac{\partial}{\partial R} \left( R L \frac{c_s^2}{4}\frac{ \partial \mathbf{\hat{l}}}{ \partial R} \right).
\end{equation}
Making a final illustrative assumption that $R L c_s^2$ is independent of $R$, we see that the above equation reduces to a classical wave equation with velocity $c_s/2$. We therefore see that a warp in an invicid disc launches a pressure wave with speed $c_s/2$. This is called a \textit{bending wave}, since it describes the disc shape.

This can be understood physically as warps in the thick disc launching pressure waves (note that the final term in equation \ref{eqn:wavetorque} is related to the vertically integrated pressure: $\Sigma c_s^2 [\Omega_z/\Omega_\phi]^2$ \cite{Lubow2002}). Fig \ref{fig:wavelike} illustrates that a misalignment between two rings lines up the high pressure mid-plane of one ring with the lower pressure atmosphere of its neighbour, creating a radial pressure gradient from the existing vertical pressure gradient. An orbiting gas parcel will therefore experience an oscillating pressure gradient, launching a pressure wave in the disc. The wavelength of the pressure wave in the invicid, Keplerian limit is approximately \cite{Lubow2002}
\begin{equation}
\lambda_{\rm bw} \approx \frac{\pi (H/R) }{(6 a)^{1/2}}~R^{9/4}.
\label{eqn:lambdabw}
\end{equation}
Since the bending wave wavelength is a fairly strong function of $R$, the stable warp, $\beta(R)\sim \sin\beta(R)~\propto \sin\lambda_{\rm bw}(R)$ is smooth far from the BH and oscillatory close to the BH. These oscillations at small $R$ are referred to in the literature as \textit{radial tilt oscillations} \cite{Ivanov1997,Lubow2002}.
% In a more realistic case with non-zero $\alpha$, the tilt oscillations still exist, but are much less dramatic since they are smoothed out somewhat by the viscous damping term.

In a more realistic $\alpha>0$ limit, the $\alpha \Omega_\phi \mathbf{G}$ term in equation \ref{eqn:wavetorque} introduces an exponential damping of the bending waves, on a characteristic timescale $t_{\rm damp} \sim 1/(\alpha\Omega_\phi)$ \cite{Pringle1999}. The waves therefore travel a typical distance $l_{\rm damp} \sim c_s~t_{\rm damp} = H/\alpha$ (since $c_s=\Omega_\phi H$ for a disc in vertical hydrostatic equilibrium) before they are damped by viscosity. If $l_{\rm damp} \lesssim R$, we can conclude that we are in the diffusive regime rather than a wave-like regime, since viscosity damps out the waves and governs the disc dynamics. Therefore, the diffusive regime is characterised by $H/R \lesssim \alpha$ and the wave-like regime by $H/R \gtrsim \alpha$.

\subsection{Solid body precession}
\label{sec:solidbody}

In the wave-like regime, solutions corresponding to solid-body precession are possible. That is, $\Omega_\phi(R)$ is still given by Equation (\ref{eqn:epicyclic}), $\beta(R)$ is independent of time and $\gamma(R,t) = \Omega_{\rm prec}~t + \gamma_0(R)$. We can calculate the angular frequency of solid body precession, $\Omega_{\rm prec}$, from the total angular momentum of the disc $\mathbf{J}_{\rm disc}$ and the Lense-Thirring torque on the total disc $\mathbf{T}_{\rm LT}$, which are related as $\mathbf{T}_{\rm LT} = \mathbf{\Omega}_{\rm prec} \times \mathbf{J}_{\rm disc}$. Thus $\Omega_{\rm prec}=T_{\rm LT}/(\sin\beta_0 J_{\rm disc})$,
% \begin{equation}
% \Omega_{prec} = \frac{T_{disc}}{\sin\beta_0 J_{disc}},
% \label{eqn:Omegaprec1}
% \end{equation}
where $\beta_0$ is the angle between $\mathbf{J}_{\rm disc}$ and BH spin axis. The angular momentum and torque acting on a disc annulus are respectively $d\mathbf{J}_{\rm disc}(R)=L(R)~2\pi R~\mathbf{l}(R)~dR$ and $d\mathbf{T}_{\rm LT}(R)=\mathbf{\Omega}_{\rm LT}(R) \times d\mathbf{J}_{\rm disc}(R)$. For a disc that is not too strongly warped ($\beta(R)\sim \beta_0$), we can set $|\mathbf{\hat{k}}\times \mathbf{\hat{l}}|=\sin\beta_0$ and integrate over all disc radii to get
\begin{equation}
\Omega_{\rm prec} = \frac{\int_{R_{in}}^{R_{out}} \Omega_{\rm LT}(R)~L(R)~R~dR}
{ \int_{R_{in}}^{R_{out}}~L(R)~R~dR }.
\label{eqn:omegaprec}
\end{equation}
This equation holds for an arbitrary disc twist, as long as that twist is frozen in time. Using the weak field formula for $\Omega_{\rm LT}$ (equation \ref{eqn:nult}) and specifying $\Sigma(r)\propto r^{-\zeta}$ then gives \cite{Liu2002,Fragile2007,Ingram2009}
\begin{equation}
\nu_{\rm prec} \approx \frac{(5-2\zeta)}{\pi(1+2\zeta)} \frac {a \left[ 1-(r_{\rm in}/r_{\rm out})^{1/2+\zeta} \right]} {r_{\rm out}^{5/2-\zeta} r_{\rm in}^{1/2+\zeta} \left[ 1 - (r_{\rm in}/r_{\rm out})^{5/2-\zeta} \right] } \frac{c}{R_g}.
\label{eqn:nuprec}
\end{equation}
A more correct description can of course be obtained by using the more accurate formula for $\Omega_{\rm LT}$ (equations \ref{eqn:epicyclic}) and integrating numerically (e.g. \cite{Ingram2011,Ingram2012,Rapisarda2014}), or using a polynomial approximation \cite{DeFalco2018}. Further accuracy can be achieved if the radial profile of the tilt angle is known.

\subsection{Numerical simulations}

\begin{figure}
	\includegraphics[width=0.8\textwidth,trim=0.0cm 0.0cm
        0.0cm 0.0cm,clip=true]{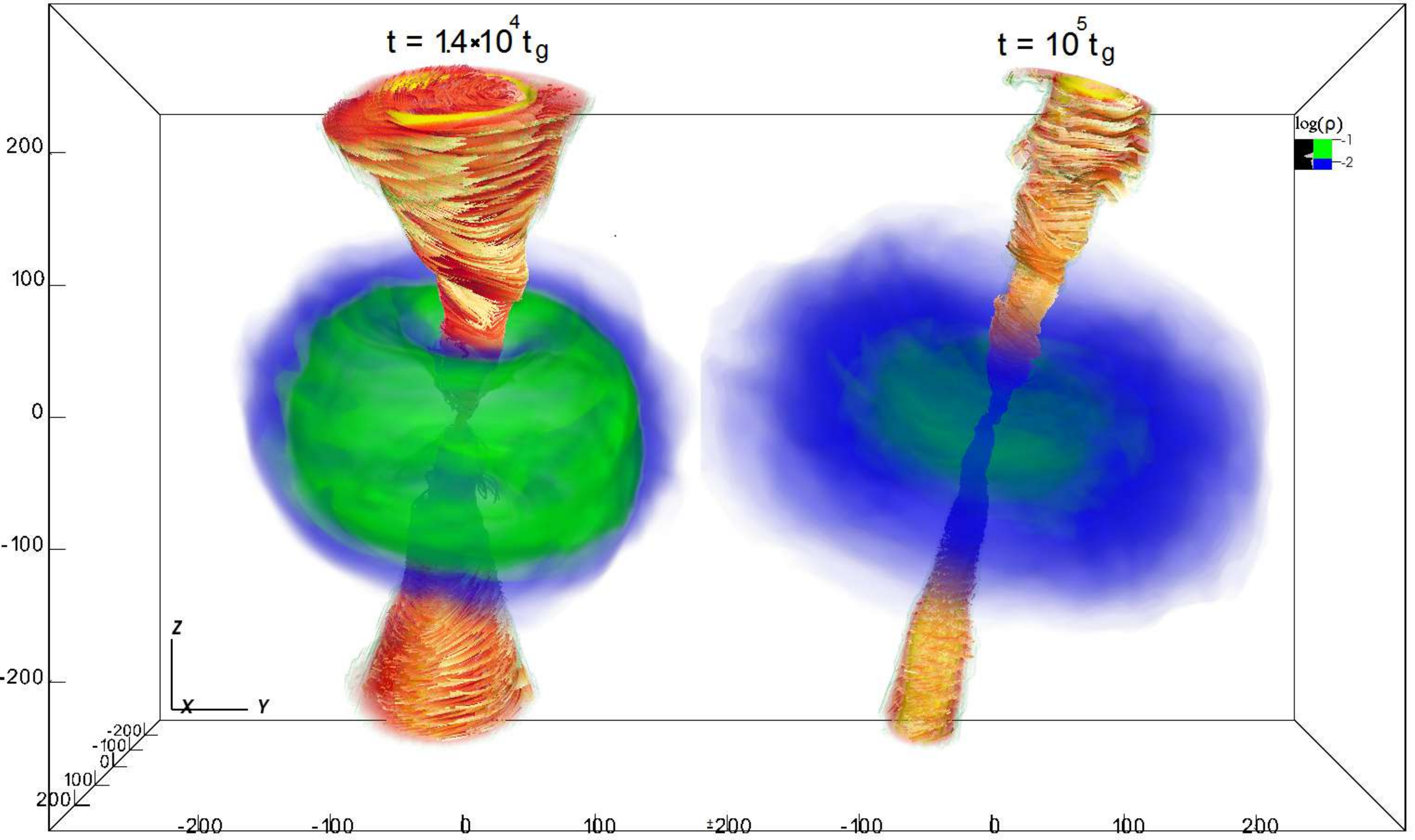}
    \centering
    \caption{Volume rendering of density (blue/green) and magnetic energy density (yellow/red) for one of the thick disc GRMHD simulations ran by \citet{Liska2018} (reproduced from said paper) at two times (as labelled; $t_g=R_g/c$). We see precession of both thick disc (blue/green) and jet (yellow/red).}
    \label{fig:liska1}
\end{figure}
%Left, Bottom, Right, Top

\begin{figure}
	\includegraphics[width=0.98\textwidth,trim=0.0cm 12.5cm
        0.0cm 4.0cm,clip=true]{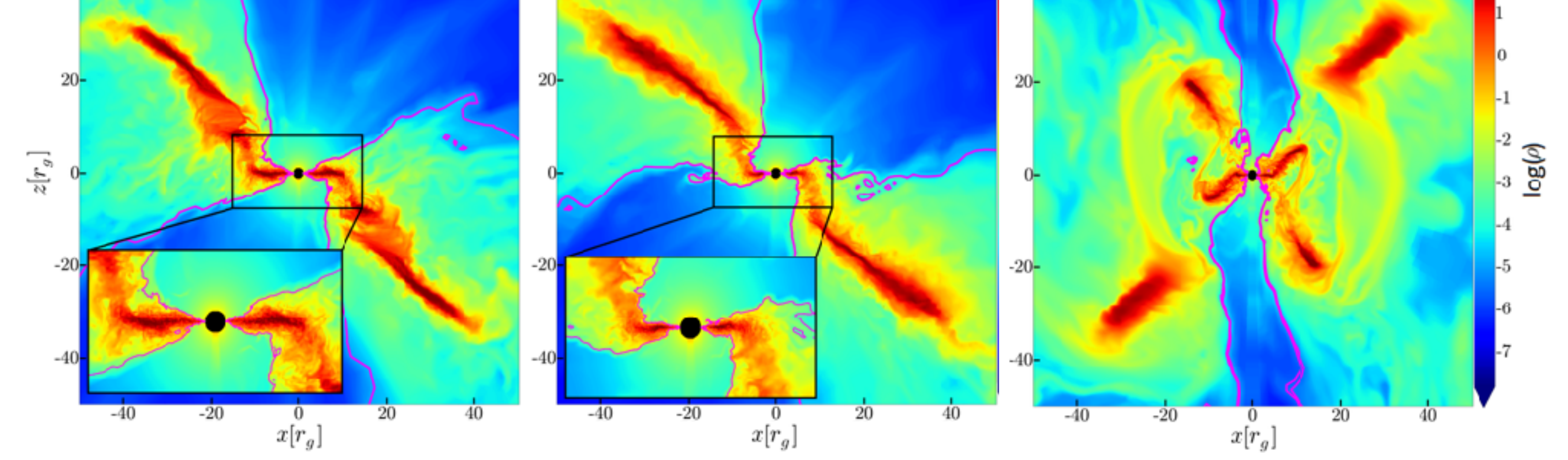}
    \centering
    \caption{Vertical slices through the grid of thin tilted disc GRMHD simulations, reproduced from \citet{Liska2019a}. Parameters are: $H/R=0.015$, $\beta_0=45^\circ$ (left); $H/R=0.03$, $\beta_0=45^\circ$ (center); $H/R=0.03$, $\beta_0=65^\circ$ (right). The tilt transitions sharply (disc breaking) at $r_{\rm bp}\sim 10$ (left) and $r_{\rm bp}\sim 5$ (centre), and the inner disc tears for the highest initial tilt (right).}
    \label{fig:liska2}
\end{figure}
%Left, Bottom, Right, Top

Qualitatively, the predictions of analytic theory have been reproduced by numerical simulations. Smoothed particle hydrodynamics (SPH) simulations consider many particles interacting in an effective potential, but still rely on an input viscosity prescription (i.e. values of $\nu_1$ and $\nu_2$). GR magneto-hydrodynamics (GRMHD) simulations instead solve the MHD equations on a grid given an initial configuration of mass and magnetic field density. The \textit{magneto-rotational instability} \cite{Balbus1991}, which forms due to the tangling of magnetic field lines by differential rotation, takes the place of a viscosity prescription. Such grid-based simulations therefore provide the best tests for analytic theory.

Thick discs ($H/R > \alpha$) were the first to be explored globally in GRMHD, being the least computationally expensive due to the comparatively few total cells required to properly resolve their scale height. \citet{Fragile2007} found that such a thick disc with an initial tilt angle of $\beta_0=15^\circ$ did indeed undergo solid-body precession as predicted by the analytic theory, at a precession frequency consistent with Equation (\ref{eqn:nuprec}). Recently \citet{Liska2018} repeated this setup, but were able to achieve higher resolution, particularly around the poles of the grid. This enabled them to resolve jets, which they found to precess with the thick disc (see Fig \ref{fig:liska1}).

Thin discs ($H/R < \alpha$) are rather harder to resolve in GRMHD, and so for a long time the Bardeen-Petterson configuration could only be studied in SPH (e.g. \cite{Nelson2000,Nealon2015}). \citet{Liska2019} recently achieved the required resolution for the first time to find that an $H/R=0.03$, $\beta_0=10^\circ$ disc around an $a=0.9375$ BH forms a Bardeen-Petterson configuration with transition radius $r_{\rm bp} \sim 5$, which is smaller than expected from analytic theory (e.g. \cite{Kumar1985}). For very large $\beta_0$ and/or very small $H/R$, the smooth transition of the Bardeen-Petterson configuration is predicted to become rather sharp (\textit{disc breaking} \cite{Lodato2010}), or even separate into discrete, individually precessing rings (\textit{disc tearing} \cite{Nixon2012a}). \citet{Nixon2012a} estimated that the disc tears at radius
\begin{equation}
r_{\rm tear} \lesssim \left( \frac{4}{3} |\sin\beta_0| \frac{a}{\alpha H/R} \right)^{2/3},
\end{equation}
and demonstrated that breaking and tearing occur in SPH simulations. \citet{Liska2019a} recently confirmed that this also happens in GRMHD (see Fig \ref{fig:liska2}).

\section{Models for low frequency QPOs}
\label{sec:models}

Many models have been proposed for LF QPOs, ranging from well-developed theories that have been confronted with the data to ideas briefly discussed in one or a few papers. In this section, we describe in detail the models that feature prominently in the literature and also summarise some other models that have received less attention. A high fraction of the models assume a truncated disc / hot inner flow geometry, whereby the thin disc truncates at some radius larger than the ISCO and the accretion flow inside of this takes the form of a hot, large scale height accretion flow (see Fig \ref{fig:precschem}c) that plays the role of the X-ray corona. Wherever this geometry is assumed we will refer to the inner radius of the flow as $r_i$ and the transition radius between disc and flow as $r_{\rm tr}$. All of these models were proposed to explain Type-C QPOs, although we note that they could all equally be candidates to explain Type-B QPOs. We have attempted to make this review of models fairly comprehensive, but fear that the shear size of the literature likely prevents this from being a completely exhaustive summary.

\subsection{The relativistic precession model}

The relativistic precession model (RPM) \cite{Stella1998,Stella1999} is perhaps the simplest of all the models. The LF QPO fundamental frequency is simply assumed to be the Lense-Thirring precession frequency at some characteristic radius, perhaps the disc truncation radius. Moreover, the lower and upper HF QPOs are postulated to be associated respectively with periastron precession and orbital motion. The model was originally proposed to explain the LF QPOs and pair of kHz QPOs observed from NSs \cite{Stella1998}, but was soon after extended to BH QPOs \cite{Stella1999}. These frequencies of geodesic motion may modulate the X-ray flux if there are extended bright clumps of material all located at the same radius of the accretion disc on slightly elliptical orbits (to ensure periastron precession) that are slightly titled with respect to the BH equatorial plane (to ensure Lense-Thirring precession). The flux would then be modulated through relativistic Doppler boosting \cite{Bakala2014}.
% , with the highest / lowest X-ray flux coinciding with times when a clump is approaching / receding from the observer.
It is plausible that such hot spots would form at the truncation radius, where turbulence is likely generated. As the truncation radius moves inwards and the spectrum consequently softens, all geodesic frequencies increase, reproducing the observed evolution of QPO frequency.

% Fig \ref{fig:bakala} shows a dynamical power spectrum predicted from the RPM. This figure is reproduced from the LOFT M3 Yellow Book and is based on the calculations of \citet{Bakala2014}. Here, a luminous region with azimuthal and radial extent of $315^\circ$ and $0.5~R_g$ orbits a BH with mass $M=7.1~M_\odot$ and spin $a=0.6$ on a slightly elliptical orbit. The orbital axis is inclined by $63^\circ$ to the observer's line of sight and the radius varies between $4$ and $4.5~R_g$ in a manner that correlates with flux (right hand panel). All GR effects are taken into account in order to calculate an X-ray light curve and eventually the power spectrum plotted in Fig \ref{fig:bakala}. The LF QPO, which is the strongest feature in the power spectrum, corresponds to the Lense-Thirring precession frequency, as expected. HF QPOs at the perisatron precession and orbital frequencies are also seen. The predicted QPO at the radial epicyclic frequency is too weak to have been discovered with existing X-ray observatories, but could potentially be detected in the future by large area X-ray detectors such as \textit{eXTP} and \textit{STROBE-X}. We see that the frequencies of the LF QPO and both HF QPOs positively correlate with the flux, whereas the radial epicyclic frequency anti-correlates with the flux.

An extension to this model was proposed by \citet{Schnittman2006}, who considered a precessing ring.
% Again, the location of this ring would need to move towards the BH to explain increases in QPO frequency, and therefore it would make most sense that the ring be located at the truncation radius.
Another mechanism for modulating the flux with the Lense-Thirring precession and orbital frequencies was suggested by \citet{Psaltis2000}. This model considers a disc with some narrow transition radius ($\delta r / r \sim 0.01$), perhaps coinciding with an abrupt change in disc properties (again, perhaps the truncation radius). A white noise of driving density perturbations are assumed to originate from outside the transition region, and the response of the disc in the transition region to these perturbations is calculated. The output power spectrum includes a zero-centered Lorentzian with HWHM $\sim (\delta r / r) \nu_{\rm visc}$ (where $\nu_{\rm visc}$ is the viscous frequency \cite{Frank2002}), and narrow peaks at the orbital and Lense-Thirring precession frequencies of the transition region. The latter peak describes a precessing, one-armed azimuthal density perturbation with a vertical tilt, similar to the precessing ring model of \citet{Schnittman2006}.

% The transition region acts as a low pass filter, resulting in an output power spectrum that is the sum of a number of modes. One mode resembles the broad band noise, its power spectrum consisting of a zero-centered Lorentzian with width $\sim (\delta r / r) \nu_{\rm visc}$, where $\nu_{\rm visc}$ is the viscous frequency \cite{Frank2002}. There is a mode with a narrow peak at the orbital frequency of the transition region and the mode that resembles the LF QPO is a narrow Lorentzian peaking at the Lense-Thirring precession frequency of the transition region. This mode describes a precessing, one-armed azimuthal density perturbation with a vertical tilt (i.e. similar to the precessing ring model of \citet{Schnittman2006}).

% \begin{figure}
% 	\includegraphics[width=0.8\textwidth,trim=0.0cm 0.5cm
%         0.0cm 0.0cm,clip=true]{BakalaQPOs.pdf}
%     \centering
%     \caption{Dynamical power spectrum generated from an extended luminous structure orbiting a BH with mass $M=7.1~M_\odot$ and spin $a=0.6$. The orbit is slightly elliptical and slightly tilted out of the BH equatorial plane and the radius varies between $4$ and $4.5~R_g$ in a manner that is correlated with flux (see right hand panel). The peaks (black), from lowest to highest frequency, correspond to Lense-Thirring precession, radial epicyclic motion, periastron precession and orbital motion. Plot reproduced from the LOFT M3 Yellow Book, using calculations from \cite{Bakala2014}.}
%     \label{fig:bakala}
% \end{figure}
% %Left, Bottom, Right, Top

\subsection{The precessing inner flow model}

\begin{figure}
	\includegraphics[width=\textwidth,trim=0.0cm 5.5cm
        0.0cm 1.0cm,clip=true]{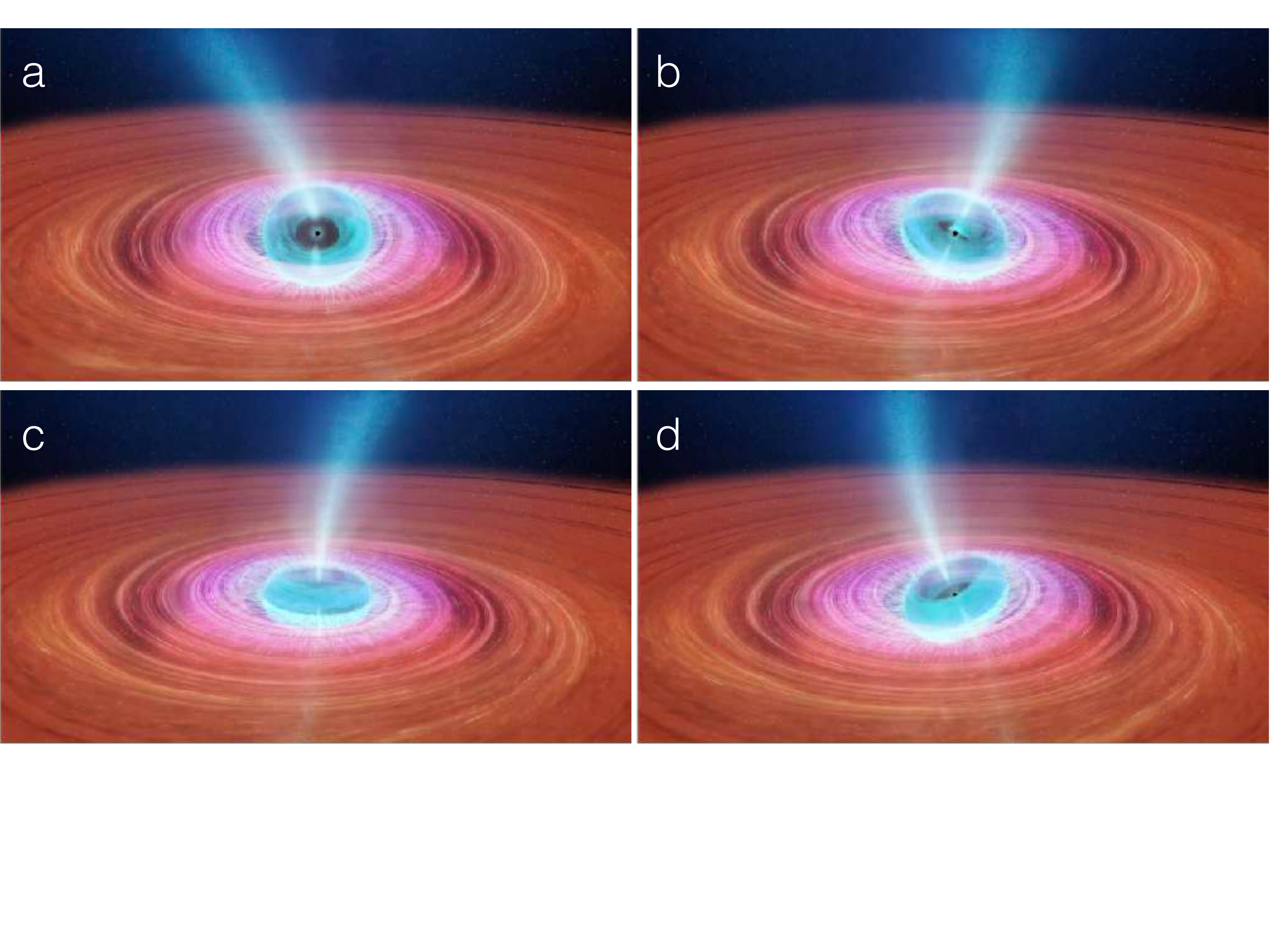}
    \centering
    \caption{Schematic illustration of the precessing inner flow model. The disc (orange) remains stationary, whilst the inner flow (blue torus) and the jet precess around the BH spin axis (which is pointing upwards and slightly towards the observer in this example). Here, precession and disc rotation are both anti-clockwise. Adapted from an animation made by ICRAR in association with \cite{Miller-Jones2019} (the original movie can be found at \url{https://vimeo.com/332582739}).}
    \label{fig:precschem}
\end{figure}
%Left, Bottom, Right, Top

The model of \citet{Ingram2009}, which assumes a truncated disc / hot inner flow geometry, is illustrated in Fig \ref{fig:precschem}. The BH spin axis is assumed to be moderately misaligned with the rotational axis of the binary system ($\sim 10-15^\circ$). In the soft state, the thin disc is in a Bardeen-Petterson configuration, but in the hard state the disc is assumed to be truncated outside of the Bardeen-Petterson transition radius, meaning that the disc feeds misaligned material to the inner flow. This then causes the entire inner flow to precess, as is discussed in Section \ref{sec:solidbody}, as well as the base of the jet \cite{Liska2018}.

% This model was originally inspired by the simulations of \citet{Fragile2007}, which demonstrated that a large scale height accretion flow can precess as a solid body with the precession frequency given by a surface density weighted average of the test mass precession frequency (equation \ref{eqn:omegaprec}). \citet{Liska2018} have since shown in a higher resolution simulation that the same configuration can launch a jet that precesses with the flow (see Fig Y), or lagging slightly behind it (\cite{Liska2019b}). 

As illustrated in Fig \ref{fig:precschem}, the flow precesses around the BH spin axis such that the misalignment angle between the flow and BH spin, $\beta$, remains constant. Because the disc is stationary and misaligned with the spin axis by the same angle $\beta$, the angle between the disc and flow rotation axes varies over each precession cycle between $0$ (panel c) and $2\beta$ (panel a). Therefore viewers at different azimuths will see different things. For the reader, maximum misalignment in Fig \ref{fig:precschem} (panel a) occurs when the flow faces maximally towards us, but the flow would face maximally away from an observer positioned behind the page at maximum misalignment. The X-ray flux is then modulated by a combination of effects. The most obvious are Doppler boosting and projected area effects. Maximal Doppler boosting occurs when the flow is viewed most edge-on, since blue shifts from the approaching material dominate over red shifts from receding material. The projected area, on the other hand, is largest when the flow is viewed most face-on. The predicted flux as a function of precession phase is therefore a trade-off between these two considerations \cite{Veledina2013,Ingram2015a}.
% It turns out that Doppler boosting is most important when the truncation radius is small, because the rotational velocity in the entire flow is very high, and projected area effects are most important when the truncation radius is large.
It is also important to consider the angular dependence of Compton scattering, which is not isotropic for a non-spherical scattering medium such as the inner flow \cite{Sunyaev1985}. Finally, the intrinsic luminosity and temperature of the inner flow should vary with precession phase as the angle between disc and flow varies and consequently so too does the luminosity of disc photons irradiating the flow \cite{Zycki2016,You2018}.

\citet{Ingram2009} showed that the spin dependence of the flow precession frequency is partially cancelled out by the expected spin dependence of the surface density profile, $\Sigma(r)$. Analytic theory and GRMHD simulations predict that $\Sigma$ is $\sim$constant outside of the so-called \textit{bending wave radius} but drops-off steeply inside of this in the region dominated by radial tilt oscillations \cite{Fragile2007,Fragile2009a}. The bending wave radius can be estimated as the largest radius where the bending waves turn over, $r_{\rm bw} \sim \lambda_{\rm bw}(r_{\rm bw})/4$, giving (from Equation \ref{eqn:lambdabw}) $r_{\rm bw} \approx 2.5 (h/r)^{-4/5} a^{2/5}$. The increase of this radius with increasing spin can explain why different sources display similar (but not identical; e.g. \cite{Shaposhnikov2006}) evolution in QPO frequency \cite{Stiele2013,Franchini2017} without having to make the uncomfortably fine-tuned assumption that all BH XRBs have the same spin. The highest possible precession frequency, on the other hand, \textit{should} depend on BH spin, since this is when the truncation radius is only just larger than the ISCO (and therefore smaller than $r_{\rm bw}$), and the highest observed QPO frequency does indeed vary from one object to the other \cite{Motta2014,Ingram2014,Franchini2017}.

The dependence of the precession frequency on the surface density profile also provides a natural explanation for the observed short term variations in QPO frequency (see Section \ref{sec:freqamp}) if the broad band noise is assumed to originate from propagating fluctuations of the mass accretion rate in the inner flow \cite{Lyubarskii1997,Arevalo2006,Ingram2011}. This is because a local increase in accretion rate will cause a local over-density in the surface density. Over-densities in the inner / outer regions of the flow will speed up / slow down precession, and increases in the accretion rate close to the hole will lead to greater increases in X-ray flux than those far from the hole due to the centrally peaked gravitational emissivity \cite{Ingram2011}. Therefore, the precession model predicts the short timescale correlations between QPO frequency and X-ray flux seen by \citet{Heil2011} (see Fig \ref{fig:heil2011}, left).

This model makes a number of testable predictions, most notably regarding the reflection signature. First of all, the changing angle between disc and flow should result in the reflection fraction changing over the course of each precession cycle. Most characteristically, the inner flow will predominantly illuminate different disc azimuths as it goes through each precession cycle, meaning that the iron line should be blue shifted when the flow shines on the approaching disc material (panel d) and red shifted when the flow shines on the receding disc material (panel b) \cite{Schnittman2006,Ingram2012a}. \citet{Fragile2016} extended this model to also attribute HF QPOs to the breathing mode and vertical epicyclic frequency of the inner flow. The model has also been considered as a basis to interpret multi-wavelength observations of QPOs. \citet{Kalamkar2016} suggested that the IR QPO they detected from GX 339-4 is due to jet precession. \citet{Veledina2013} modelled optical QPOs by considering optical photons produced in the inner precessing flow by the cyclo-synchrotron process, and \citet{Veledina2015} instead considered reprocessing of X-ray photons from a precessing inner flow into optical photons by the outer accretion disc.

\subsection{Corrugation modes}

Corrugation modes (c modes) are transverse standing waves in the disc height with some resonant angular frequency $\omega_c$, that are trapped in a region between the disc inner radius $r_{\rm in}$ and another radius called the \textit{inner vertical resonance} (IVR) $r_{\rm ivr}$, and excited by perturbations in an intrinsic disc quantity such as density (e.g. \cite{Kato1980,Wagoner1999,Kato2001}). Visualisations of such oscillations can be found in Fig 2 of \citet{Tsang2013}. The study of c modes is part of the wider field of discoseismology, which encapsulates a number of different modes of oscillation (e.g. gravity modes, pressure modes). The c modes are the discoseismic mode with resonant frequencies in the region of the observed LF QPO frequencies (gravity modes and pressure modes oscillate at higher frequencies). The resonant frequency is
\begin{equation}
\omega_c = m \Omega_\phi(r_{\rm ivr}) - j^{1/2} \Omega_z(r_{\rm ivr}),
\label{eqn:omegac}
\end{equation}
where $m$ and $j$ are integers,
\begin{equation}
    r_{\rm ivr} = r_{\rm in} + C ( a^{-1/3} - 1 ),
\end{equation}
and the constant $C$ depends on the sound speed in the disc (typical value $C=0.17$; \cite{Wagoner1999}). We see from equation (\ref{eqn:omegac}) that, in the case of $j=m=1$, the resonant frequency is nothing other than the Lense-Thirring precession frequency at the IVR, $\omega_c = \Omega_\phi(r_{\rm ivr}) - \Omega_z(r_{\rm ivr})$, making this essentially yet another model that associates the LF QPO with Lense-Thirring precession. The resonant frequency is therefore set by the mass and spin of the hole, the sound speed and the disc inner radius. The model can therefore reproduce the observed changes in LF QPO frequency if the disc inner radius moves towards the hole as the QPO frequency is observed to increase (changes in QPO frequency can also be driven, to a lesser extent, by changes in the sound speed).

In order to understand the origin of the standing wave, let us explore the result of introducing a plane wave perturbation in the disc density with angular frequency $\omega$. This causes a plane wave in the disc height with dispersion relation \cite{Tsang2009,Okazaki1987}
\begin{equation}
c_s^2 k^2 = \frac{ (\Omega_r^2 - \tilde{\omega}^2) (j \Omega_z^2 - \tilde{\omega}^2) }{\tilde{\omega}^2},
\label{eqn:cdisp}
\end{equation}
where $\tilde{\omega}\equiv \omega - m\Omega_\phi$, and $k=2\pi/\lambda$ is the radial wave number. 
Waves can only propagate in regions of the disc for which the RHS of equation (\ref{eqn:cdisp}) is $>0$, thus in the regions where the following conditions are fulfilled
\begin{eqnarray}
\tilde{\omega}^2 &<& \Omega_r^2 < j \Omega_z^2 \\
\tilde{\omega}^2 &>& j \Omega_z^2 > \Omega_r^2.
\label{eqn:cmode}
\end{eqnarray}
The latter of these two expressions (equation \ref{eqn:cmode}) corresponds to the propagation region of c modes (the former corresponds to gravity modes), which extends from $r_{\rm in}$ out to $r_{\rm ivr}$, defined as $\tilde{\omega}(r_{\rm ivr})=-j^{1/2}\Omega_z(r_{\rm ivr})$. It is therefore possible for a wave with angular frequency $\omega$ to propagate in the region $r_{\rm in}$ to $r_{\rm ivr}$. If these two boundaries are reflective (see \cite{Tsang2009}), then a standing wave trapped in the region $r_{\rm in}$ to $r_{\rm ivr}$ will result only for eigen-frequencies, $\omega_c$, that correspond to waves with an integer number of nodes in the region. Therefore, similarly to the \citet{Psaltis2000} model, a white noise of input perturbations will result in the excitation of resonant modes.

\subsection{Accretion ejection instability}

\begin{figure}
	\includegraphics[width=0.8\textwidth]{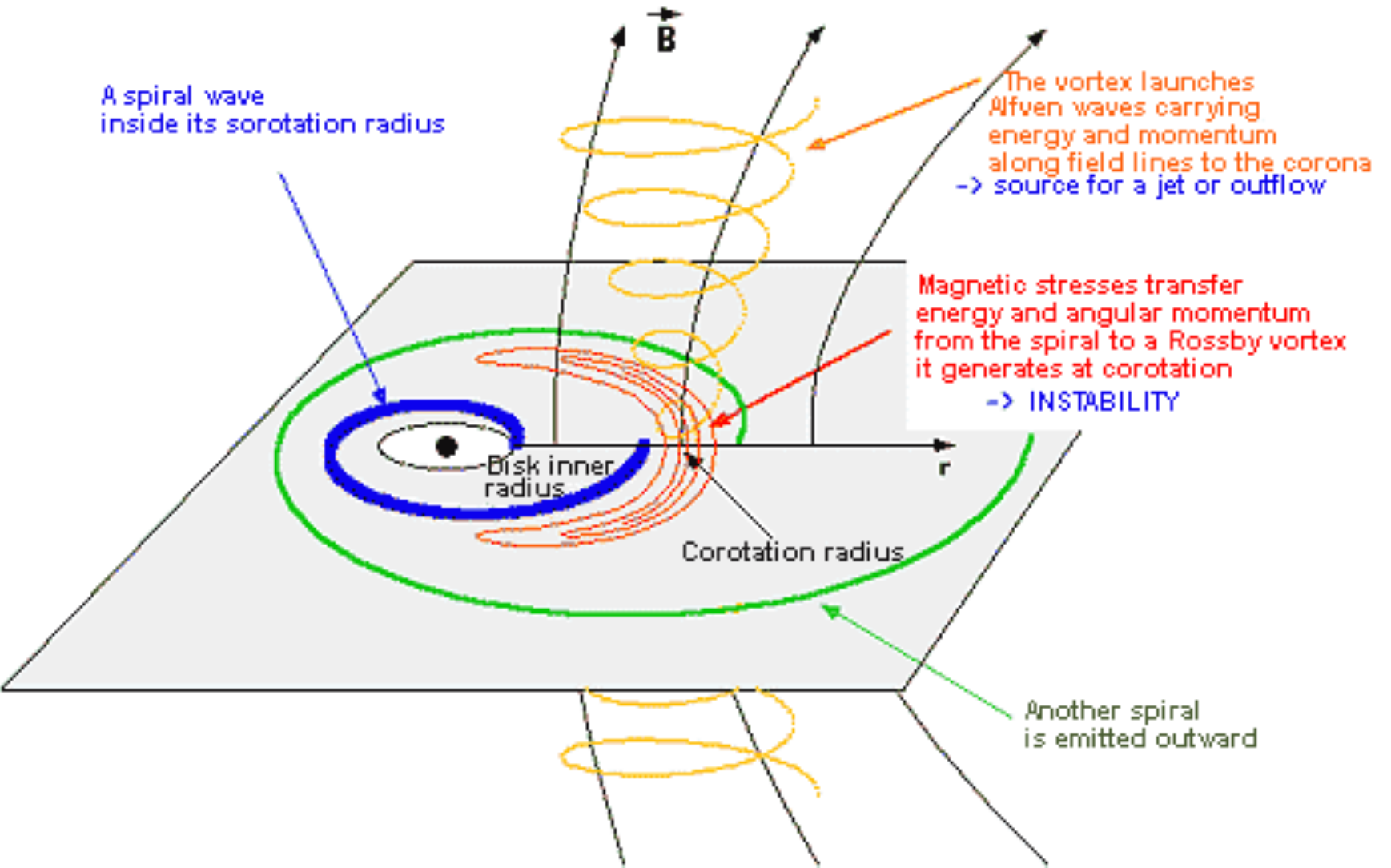}
    \centering
    \caption{Schematic of the accretion ejection instability model. Perturbations in poloidal magnetic field strength launch spiral pressure waves in the disc (blue) that are reflected at the inner Lindblad resonance. This causes an over-density to form at the co-rotation radius (red contours), which is sheared by differential rotation to create a Rossby vortex and ultimately launch Alfv'en waves along the poliodal magnetic field lines threading the disc (yellow). Schematic reproduced from \citet{Tagger2007}.}
    \label{fig:aei}
\end{figure}

The Accretion Ejection Instability (AEI; \cite{Tagger1999}) is a spiral wave instability in the density and scale height of a thin disc threaded with a strong vertical (poloidal) magnetic field (strong as in gas and magnetic pressure are roughly equal). Fig \ref{fig:aei} shows a schematic of the instability. The spiral structures each have an integer number of turns between the disc inner radius and the inner Lindblad resonance and rotate about the disc surface with angular eigen-frequency, $\omega_i$. The spiral arms result from perturbations to the strength of the vertical magnetic field threading the disc.

A plane wave perturbation in field strength with angular frequency $\omega$ results in spiral waves in disc density with approximately the following dispersion relation
\begin{equation}
    \frac{2 B_0^2}{\Sigma} \frac{|k|}{r} + \frac{k^2}{r}c_s^2 = \tilde{\omega}^2 - \Omega_r^2,
    \label{eqn:aeidisp}
\end{equation}
where $B_0$ is the unperturbed magnetic field strength (see \cite{Varniere2002} for a more accurate expression). This is identical to the dispersion relation of self-gravity driven spiral density waves in galaxies, except here magnetic terms act as negative self-gravity. The waves can only propagate in regions for which the RHS of equation (\ref{eqn:aeidisp}) is $>0$, and are therefore damped between the inner and outer Lindblad resonances, defined as the two radii where $\tilde{\omega}^2 - \Omega_r^2=0$. Waves from the inner disc propagate towards the co-rotation radius, where $\tilde{\omega}=0$, which is located between the two Lindblad resonances. Outward propagating waves are reflected at the inner Lindblad radius, $r_{\rm ilr}$, to then propagate inwards to $r_{\rm in}$ where they are again reflected. The eigen-frequencies are a set of frequencies for which the reflected outward propagating wave has the same phase as the original outward propagating wave (i.e. they correspond to spiral structures with an integer number of turns between $r_{\rm in}$ and $r_{\rm ilr}$). The fundamental eigen-frequency is therefore set by the mass and spin of the hole, disc parameters such as field strength and sound speed, and the inner disc radius. The observed changes in QPO frequency are again assumed to be caused by $r_{\rm in}$ moving towards the hole, with $\omega_i$ increasing as $r_{\rm in}$ reduces, except for when $r_{\rm in}$ is very close to the ISCO, where the relation reverses \cite{Varniere2002a}.

These spiral standing waves can create a further instability at the co-rotation radius: a \textit{Rossby vortex}. This is caused by an over-density forming at the co-rotation radius that spreads slightly in radius (red contours in Fig \ref{fig:aei}). Differential rotation of disc material means that this over-density is sheared, with the inner parts accelerating ahead of the outer parts to create a vortex at the co-rotation radius. This in turn twists the field lines threading the over-density to launch a vertical Alfven wave (yellow lines in the figure). If there is a low density corona above the disc, this Alfven wave can transfer energy and angular momentum from the accretion flow to the corona, contributing to the launching of a wind and/or jet (hence the name accretion ejection instability).

% More papers to maybe cite: Varniere \& Blackman (2005); Varniere \& Vincent (2017)

\subsection{Propagating oscillatory shock}

The propagating oscillatory shock model \cite{Molteni1996,Chakrabarti2008} assumes a two component accretion flow (TCAF) consisting of an outer Keplerian disc that is sandwiched above and below by a sub-Kelperian corona, and an inner sub-Keplerian \textit{halo} \cite{Titarchuk1995}, often referred to as a CENBOL (centrifugal pressure supported boundary layer). In general, the disc/corona and halo components have different accretion rates, and are separated by a shock at radius $r_{\rm shk}$ that marks a sharp transition in density from the outer (pre-shock) corona to the inner (post-shock) halo. According to the propagating oscillatory shock model, the shock location oscillates around a mean value $r_{\rm shk}$ if the cooling timescale of the halo is comparable with the infall timescale \cite{Molteni1996} or if the Rankine-Hugoniot condition for a stable shock is not satisfied \cite{Ryu1997}. It is this oscillation that is proposed to produce the QPO in the X-ray flux. The oscillation frequency is inversely proportional to the infall time in the halo (e.g. \cite{Chakrabarti2008})
\begin{equation}
    \nu_{\rm qpo} \sim \frac{c/R_g}{\mathcal{R} r_{\rm shk} \sqrt{r_{\rm shk}-2} } \sim \frac{\Omega_\phi(r=r_{\rm shk},a=0)}{\mathcal{R}},
\end{equation}
where $\mathcal{R} \equiv \rho_+/\rho_-$ is the ratio of the post-shock halo density to pre-shock corona density (the \textit{compression ratio}). Typical values of the compression ratio are $\sim 1-4$ \cite{Debnath2014}, meaning that the QPO frequency is, within a factor of a few, assumed to be the Keplerian orbital frequency at the shock location. The evolution of the QPO frequency is then reproduced by the shock location moving in as the QPO frequency increases. 

We see that this model requires $r_{\rm shk}$ to be very large for typical QPO frequencies measured in the hard and intermediate states. For $\nu_{\rm qpo}\sim 1$ Hz, typical values are $r_{\rm shk} \gtrsim 600$ \cite{Debnath2014}, which is very hard to reconcile with the relativistically broadened iron line routinely observed in the X-ray spectrum. Although measurements of the disc inner radius using the iron line are subject to modelling and calibration uncertainties, the observed broad iron line profiles are simply not compatible with a disc truncated at hundreds of $R_g$ (e.g. \cite{Garcia2015,Ingram2017}), which would instead produce a narrow line. This very large disc inner radius also means that the accretion rate through the disc must be implausibly high in the intermediate state (e.g. $\sim 7$ times the Eddington limit in \cite{Debnath2014}) in order to explain the observed disc temperatures. Whereas super-Eddington accretion rates are possible in nature \cite{Bachetti2014}, the Keplerian thin disc solution, which is used for this model, is not valid for such rates. Furthermore, the soft state still consists of a thin disc extending down to the ISCO in this model, meaning that as the source transitions from the intermediate to soft state, the disc accretion rate must sharply drop by more than an order of magnitude in order to reproduce the approximately constant X-ray luminosity observed over the transition, which seems rather fine-tuned.

\subsection{Pressure or accretion rate modes}

There are many QPO models in the literature that can be characterised as `intrinsic' variability models, in that the system geometry is assumed to stay constant with time, and the oscillation in flux is assumed to result from some resonant oscillation in a property such as pressure or accretion rate. One such model, proposed by \citet{Cabanac2010} in the context of a inner flow / truncated disc geometry, considers a white noise of perturbations to the pressure of the inner flow being generated at its outer radius, $r_{\rm tr}$. These perturbations set up pressure waves in the corona. Reflection of these waves at the flow inner radius leads to the growth of resonant modes with wavelengths that are an integer fraction of $(r_{\rm tr}-r_i)$. These oscillations in the pressure then modulate the X-ray flux through the Compton up-scattering process, leading to a predicted power spectrum consisting of band limited noise and a QPO with a fundamental frequency of $\nu_{\rm qpo} \sim {\rm few}~2\pi c_s / r_{\rm tr}$ and many higher harmonics. The QPO features become less sharp as transmission at $r_i$ is assumed to be more important. The increase in QPO frequency during the transition from hard to soft state is then explained by $r_{\rm tr}$ moving inwards, and to a lesser extent the sound speed increasing.

Other suggestions have been made based on MHD simulations. \citet{O'Neill2011} reported on large-scale, low-frequency dynamo cycles in a global MHD simulation. The cycles take the form of oscillations in azimuthal magnetic field in and above the disc (extending into a low density corona several scale heights above the disc). The oscillations in magnetic field at narrow radial ranges are quasi-periodic and occur at frequencies commensurate with the  observed LF QPO frequencies. However, the oscillation frequency is a function of radius, and so when the flux emitted form the entire disc is estimated, the oscillations are completely washed out and no QPO feature is seen in the prost-processed power spectrum. Similarly, \citet{Wang2012} proposed that LF QPOs may be caused by toroidal Alfv{\'e}n wave oscillations. \citet{Machida2008} noted the formation of a  constant angular momentum inner torus in a subset of their MHD simulations that deforms itself from a circle to a crescent quasi-periodically, in turn modulating the mass accretion rate and the mass outflow rate.

\section{Observational tests}
\label{sec:tests}

Here we summarise the current state-of-the-art in attempts to observationally constrain the QPO mechanism. The goal here is less to test individual theories, which in some cases overlap somewhat, but more to determine broad conclusions about the \textit{nature} of these oscillations. For instance, most models can be classed either as \textit{geometrical} -- whereby the shape and/or size of something varies quasi-periodically -- or \textit{intrinsic} -- whereby some fundamental property such as pressure or accretion rate oscillates in a stable geometry. In some models the disc emission is predicted to oscillate and in others the coronal emission is expected to oscillate. By testing these broad properties against the data, we can narrow down the allowed classes of models until we have a very specific picture of the physical properties that a successful model must have. In this manner, we need not even confine ourselves to existing theories proposed in the literature. We can instead imagine basic modulation mechanisms, such as a oscillations in the scale height or temperature of the corona.

\subsection{Inclination dependence}

\begin{figure}
	\includegraphics[width=0.48\textwidth,trim=0.8cm 1.3cm
        0.0cm 18.45cm,clip=true]{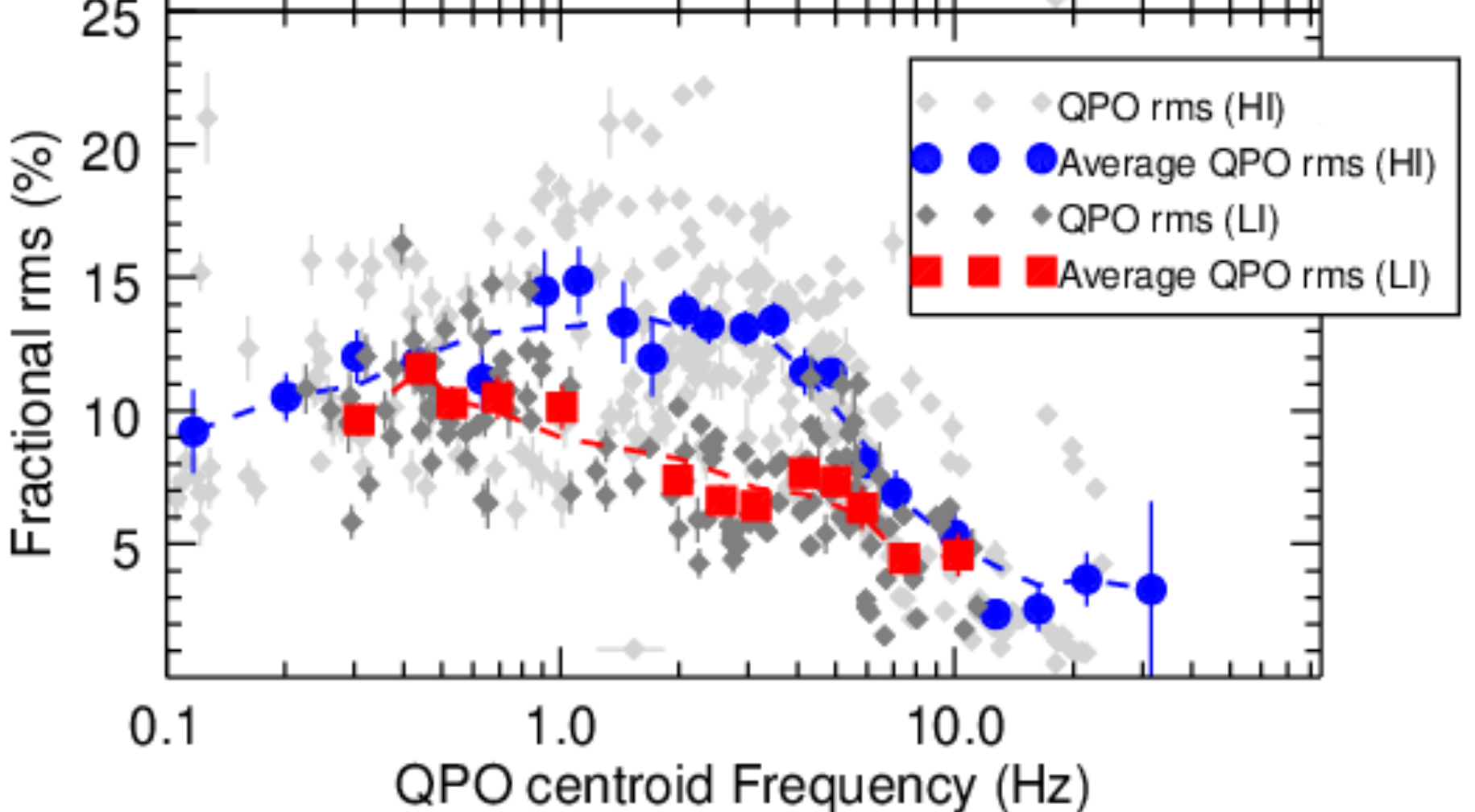} ~~
	\includegraphics[width=0.48\textwidth,trim=0.5cm 0.5cm
        0.0cm 0.5cm,clip=true]{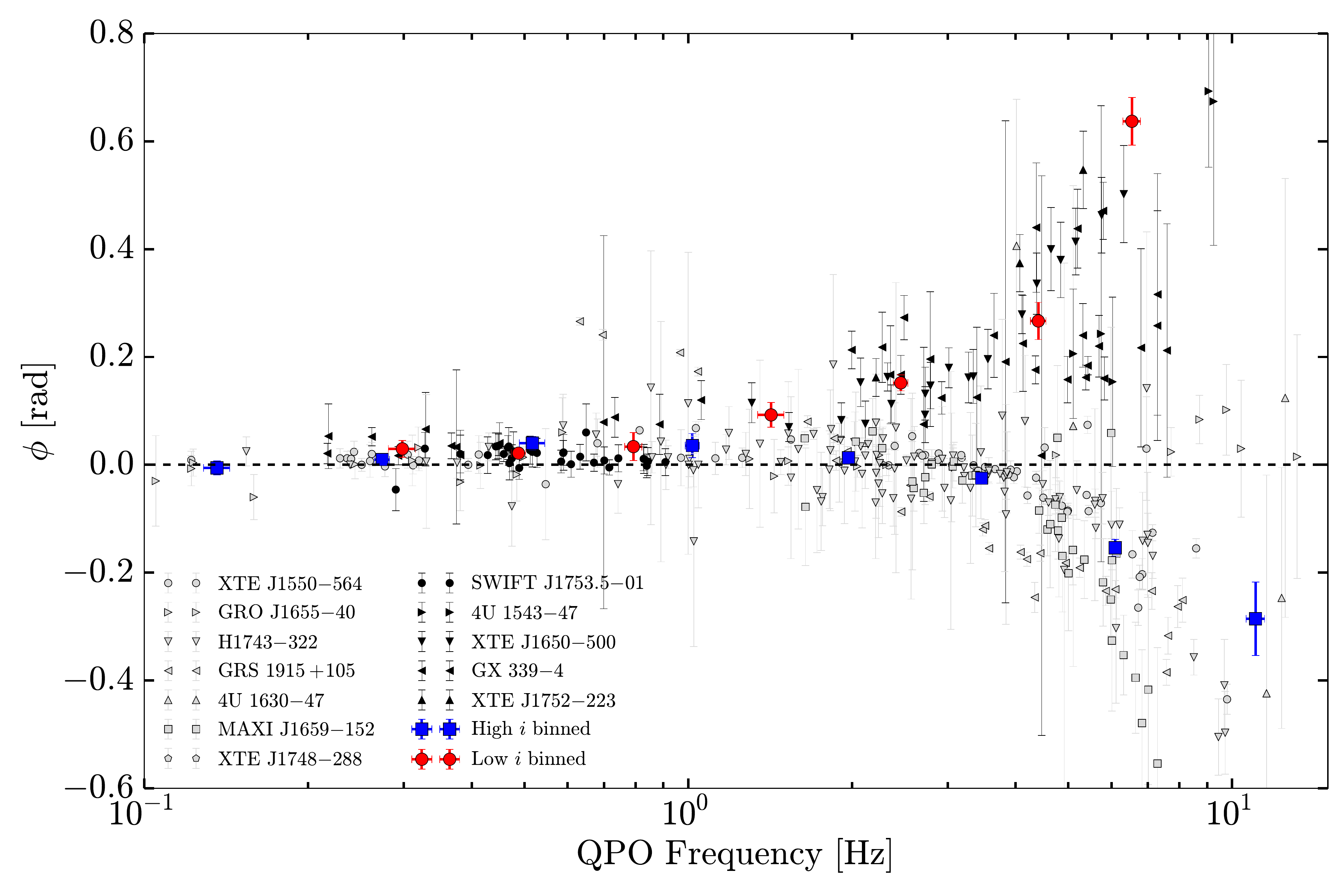}
    \centering
    \caption{Inclination dependence of Type-C QPO properties. \textit{Left:} Amplitude of the QPO fundamental versus the centroid frequency for many observations \cite{Motta2015}. \textit{Right:} Phase lag of the QPO fundamental between hard ($\sim 7-13$ keV) and soft ($\sim 2-7$ keV) energy bands (positive means hard photons lag soft) versus centroid frequency \cite{vandeneijnden2017}. Individual observations of low and high inclination sources are dark and light grey respectively. Red and blue points also correspond to low and high inclinations respectively, but for these points the population has been binned on QPO frequency.}
    \label{fig:inclination}
\end{figure}
%Left, Bottom, Right, Top

% As it has been known for a long time, the inclination with respect to the observer's line of sight strongly affects the observed properties of AGNs (see, e.g.,  \cite{Antonucci1993}, \cite{Urry1995}, \cite{Risaliti2002}, \cite{Bianchi2012}). Over the last decade it has become increasingly clear the same is true for Galactic accreting BH binaries. 
% \cite{Ponti2012} found strong evidence that the accretion disc winds observed in the soft states of BH binaries have an equatorial geometry with opening angles of a few degrees and therefore can only be observed in sources where the disc is inclined at a large angle to the line of sight (\textit{high-inclination sources}, as opposed to \textit{low-inclination sources}, where the orbital plane is closer to perpendicular to the line of sight). 
% Results by \cite{Munoz-Darias2013} indicated that the inclination modifies the q-shaped tracks that BHXBs in outburst display in a HID, which can be at least partially explained by considering inclination-dependent relativistic effects on the accretion disc. \cite{Corral-Santana2013} have found that self-obscuration - similar to that observed in AGN - can be relevant in very high inclination BHXBs. Similar findings were reported by \cite{Motta2017a}, which studied the dramatic effects of variable obscuration in the BH binary V404 Cyg. 

Given the size of the complete \textit{RXTE} archive, it is now feasible to conduct population studies of X-ray binaries that probe the dependence of QPO properties on the inclination angle between the binary rotation axis and our line-of-sight. \citet{Motta2015} used a sample of \textit{RXTE} observations to show that the Type-C QPOs are stronger in higher inclination (more edge-on) binary systems, whereas Type-B QPOs are stronger in \textit{lower} inclination (more face-on) systems (see Fig \ref{fig:inclination}, left). Such an inclination dependence of Type-C QPOs had previously been suggested in the literature \cite{Schnittman2006a}, but \citet{Motta2015} and \citet{Heil2015} were the first to use a large enough sample size to claim high statistical significance. This provides strong evidence that LF QPOs have a geometric origin. It also suggests that Type-B QPOs modulate the X-ray flux differently from Type-C QPOs. For instance, Type-C QPOs could be due to precession of the (oblate) inner flow, with Type-B QPOs instead caused by jet precession \cite{Motta2015,DeRuiter2019}. It is worth noting that the QPO amplitude in the precession model depends on the tilt angle and viewer azimuth as well as the (polar) inclination angle \cite{Veledina2013,Ingram2015a}, implying that there should be a rather large spread in the correlation between QPO amplitude and inclination. It is, however, difficult to test for such a spread because the inclination angle tends to be poorly constrained observationally beyond being categorised as `high' or `low'. 

\citet{vandeneijnden2017} used a sample slightly extended from the \citet{Motta2015} analysis (with the main difference being the inclusion of data from GRS 1915+105) to study the inclination dependence of the phase lag between hard ($\sim 7-13$ keV) and soft ($\sim 2-7$ keV) photons measured at the QPO fundamental frequency and its overtones. Fig \ref{fig:inclination} (right) shows their result for the Type-C QPO fundamental. We see that small hard lags (i.e. hard photons lag soft photons) are seen for all sources when the QPO frequency is low, whereas low / high (red / blue in the plot) inclination sources display hard / soft lags when the QPO frequency is high. This inclination dependence of the phase lag is more difficult to interpret than the amplitude inclination dependence, but is much more dramatic, and provides further evidence in favour of a geometric origin of the Type-C QPO. \citet{vandeneijnden2017} also found that soft lags for the Type-B fundamental appear to be preferentially associated with high-inclination sources, but not with high statistical significance because of the small sample size (15 sources in total, but only 9 of those sources have Type-B QPO detections and an inclination estimate). The sample size will continue to grow as hitherto unknown XRBs go into outburst, therefore there is scope to expand such population studies using X-ray missions such as \textit{NICER}.

\subsection{Frequency-resolved spectroscopy}

\begin{figure}
	\includegraphics[width=0.8\textwidth,trim=0.0cm 4.4cm
        0.0cm 0.3cm,clip=true]{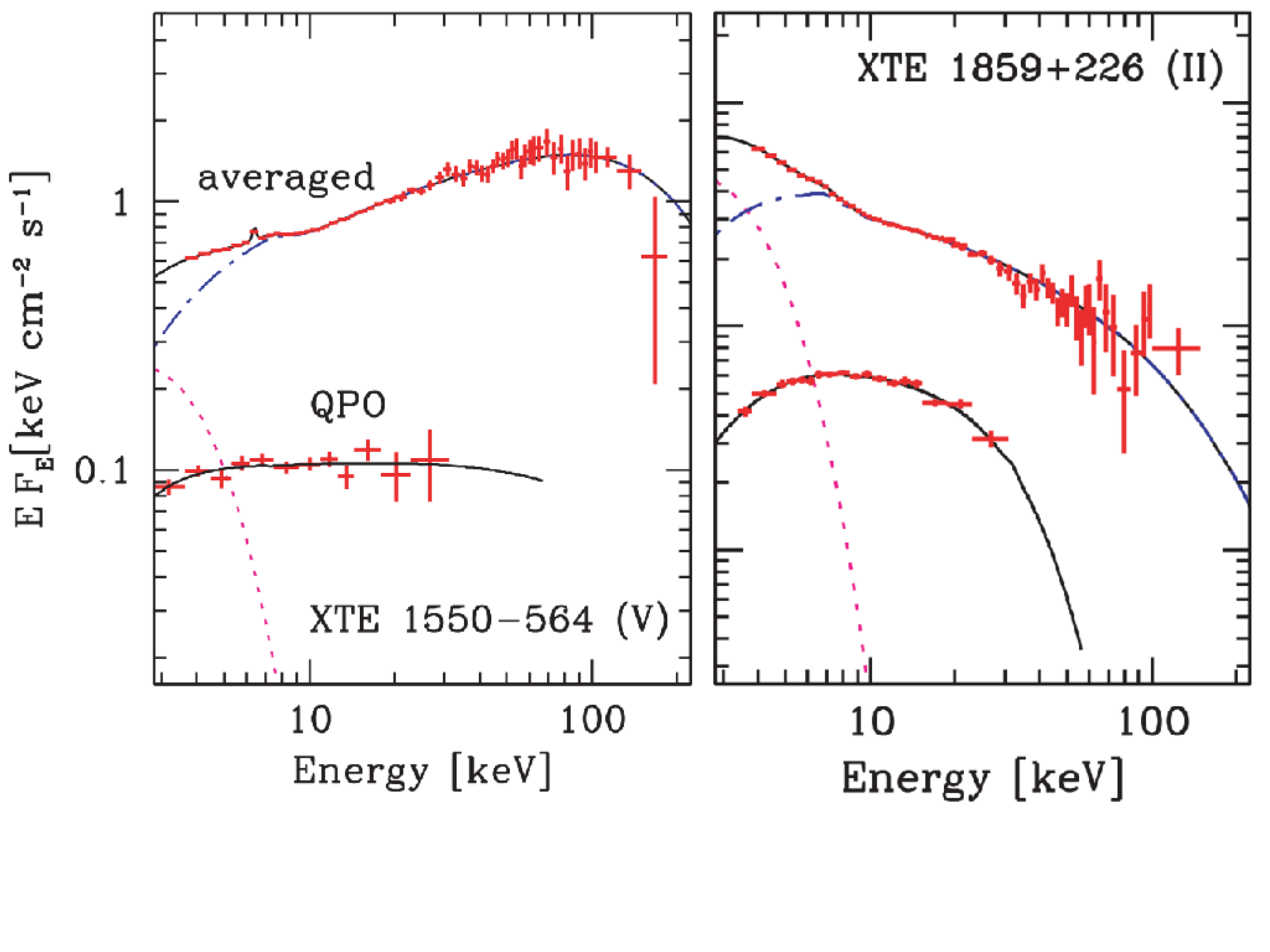}
    \centering
    \caption{Time-averaged (top) and frequency-resolved (bottom) spectra for the Type-C QPO fundamental component, unfolded around the best fitting spectral model (objects as labelled). The time-averaged spectrum includes disc blackbody (magenta dotted) and Comptonized (blue dot-dashed) radiation, whereas the QPO spectrum has no disc component. Adapted from \citet{Sobolewska2006}.}
    \label{fig:freqres}
\end{figure}
%Left, Bottom, Right, Top

Frequency-resolved spectroscopy provides a way to determine the energy spectrum of the QPO harmonics - in other words, it constrains which spectral components are oscillating. This is achieved by extracting light curves for many energy channels, and calculating the Poisson noise subtracted power spectrum of each channel in absolute rms normalization \cite{Revnivtsev1999} such that the integral of the power spectrum over all frequencies gives the variance of the corresponding light curve. The energy dependence of QPO harmonic components can be studied by fitting each power spectrum with a multi-Lorentzian model and plotting the square root of the integral of the Lorentzian component corresponding to  given harmonic against energy (the \textit{absolute rms spectrum}). The rms spectrum is in units of count rate and therefore the instrument energy response must be accounted for before it can be compared to physical models (see \cite{Mastroserio2018} for an extended discussion on this).

\citet{Sobolewska2006} studied the rms spectrum of the Type-C QPO fundamental component for a number of BH XRBs. Fig \ref{fig:freqres} shows a subset of their results. Perhaps the most striking conclusion is that the time-averaged spectrum includes a disc component (magenta), but the QPO rms spectrum does not. This indicates that the QPO fundamental does not modulate the directly observed disc emission, but strongly modulates the Comptonized emission. They also found that the QPO spectrum is softer than the time-averaged spectrum when the latter is hard, and becomes harder than the time-averaged spectrum as the latter becomes softer. \citet{Axelsson2014} studied the fundamental and second harmonic of Type-C QPOs in XTE J1550-564. They again found that the QPO fundamental rms spectrum contains no disc component, and noted that it is similar to the spectrum of high frequency ($\nu>10$ Hz) broad band noise variability, which displays weaker reflection components than the time-averaged spectrum (consistent with path length differences washing out the fastest variability in the reflection spectrum; \cite{Gilfanov2000}). The second harmonic, in contrast, is significantly softer than the time-averaged spectrum. A similar analysis of GX 339-4 \cite{Axelsson2016} additionally revealed that the rms spectrum of the second harmonic can be described by a low-temperature, optically thick Comptonized component that is not required to be present in the rms spectrum of the fundamental, which instead includes a higher temperature Comptonized component. The authors suggested that these results could be explained by precession of a stratified inner accretion flow, which is hotter closer to the BH. They could also potentially be explained by non-linear variability of a single Comptonized component \cite{Gierli'nski2005}. For instance, oscillations in the electron temperature will drive oscillations in the spectral index \cite{Zycki2003}. If these temperature oscillations have a strong second harmonic whereas the variations in total flux only have a strong first harmonic, this could lead to the second harmonic having a much softer rms spectrum than the first.

The lack of a QPO signal in the disc component is very troublesome for QPO models that are based on oscillations in the disc. For instance, it is simple to imagine how 
% the spiral disc waves of the AEI model or
disc corrugation modes can give rise to strong oscillations of the Comptonised flux (via e.g. variation of seed photons for Compton up-scattering), but much harder to imagine how the observed disc flux
% in either of these scenarios
is not modulated. We can speculate that the region of the disc where the oscillations take place is somewhat hidden from view by an X-ray corona located above the disc (but not entirely hidden, because the corona has optical depth $\tau \sim 1$), and therefore the disc spectrum we see is dominated by the stable region at larger radii that is not covered by the corona. \AI{The overdensities in the disc spiral waves of the AEI model radiate a Comptonized spectrum, and so this model can explain the spectrum of the fundamental, but it is not clear how it can explain the softer spectrum of the second harmonic.}

%We note that some approximations have been made in the frequency-resolved spectroscopy literature. First of all, the instrument response has often been accounted for simply by convolving models for the rms with the response matrix. This is only strictly appropriate if the instrument response is diagonal (in which case the convolution is the same as simply multiplying by the effective area curve), or if the phase lag between all energy bands is zero \cite{Mastroserio2018}. However, since the phase lags are generally fairly small for BH XRBs, the approximation is likely reasonable. It is also commonplace to use models consisting of a sum of additive spectral components, which is not appropriate in general since a time-dependent spectrum consisting of the sum of two components, $s(E,t)=a(E,t)+b(E,t)$, has an rms of $|S(E,\nu)| = \sqrt{ |A(E,\nu)|^2 + |B(E,\nu)|^2 + 2{\rm Re}[A(E,\nu)B^*(E,\nu)] }$. However, fits to the rms spectrum for which one spectral component dominates over the others (such as those shown in Fig XXX) do not suffer from this problem \cite{Kotov2001}. \citet{Mastroserio2018} note that all of these issues can be avoided entirely by instead fitting models to the real and imaginary parts of the cross-spectrum.

\subsection{QPO Phase-resolved spectroscopy}

\begin{figure}
	\includegraphics[width=0.48\textwidth,trim=2.0cm 1.5cm
        3.0cm 2.0cm,clip=true]{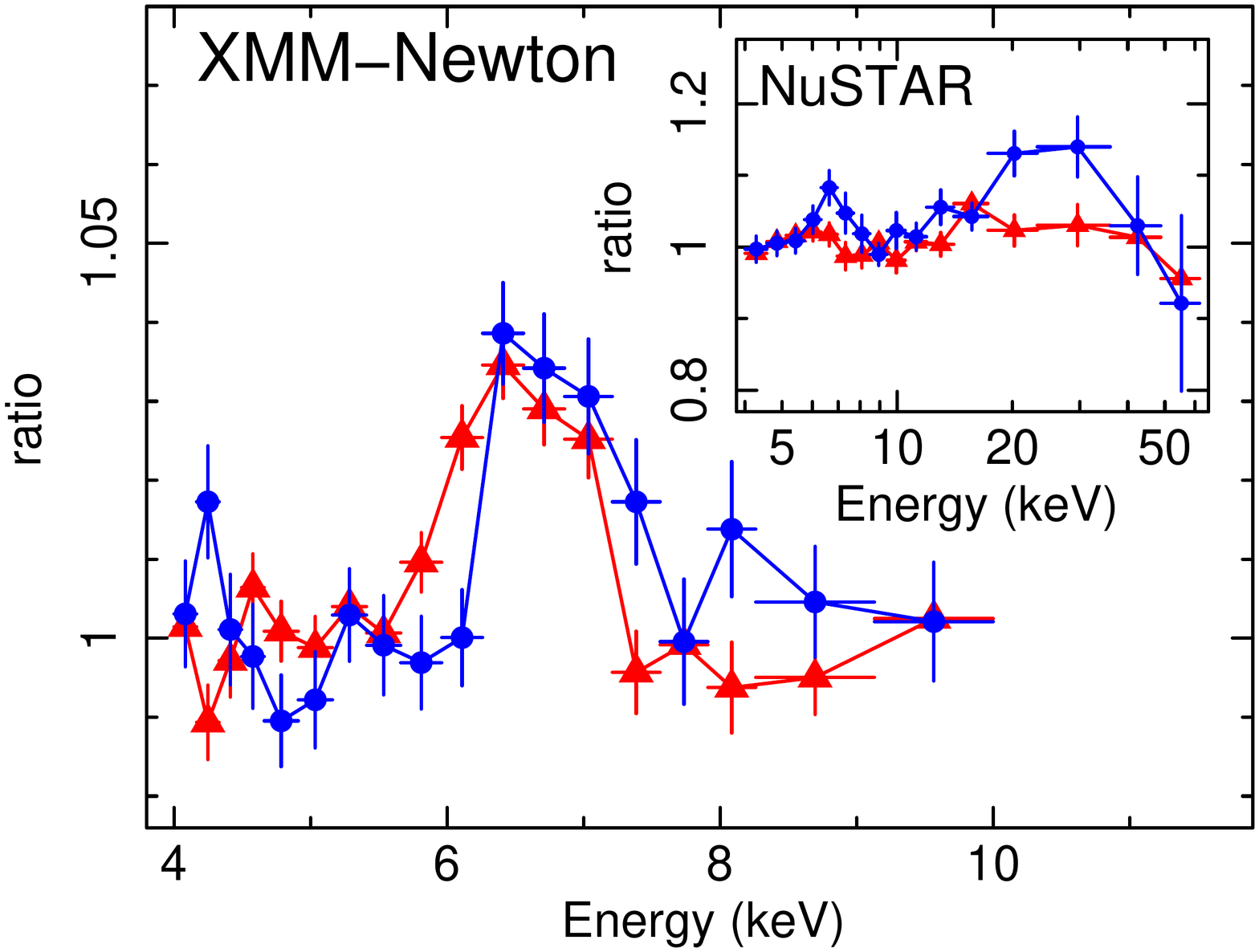} ~~
    \includegraphics[width=0.48\textwidth,trim=1.5cm 2.0cm
        2.0cm 11.0cm,clip=true]{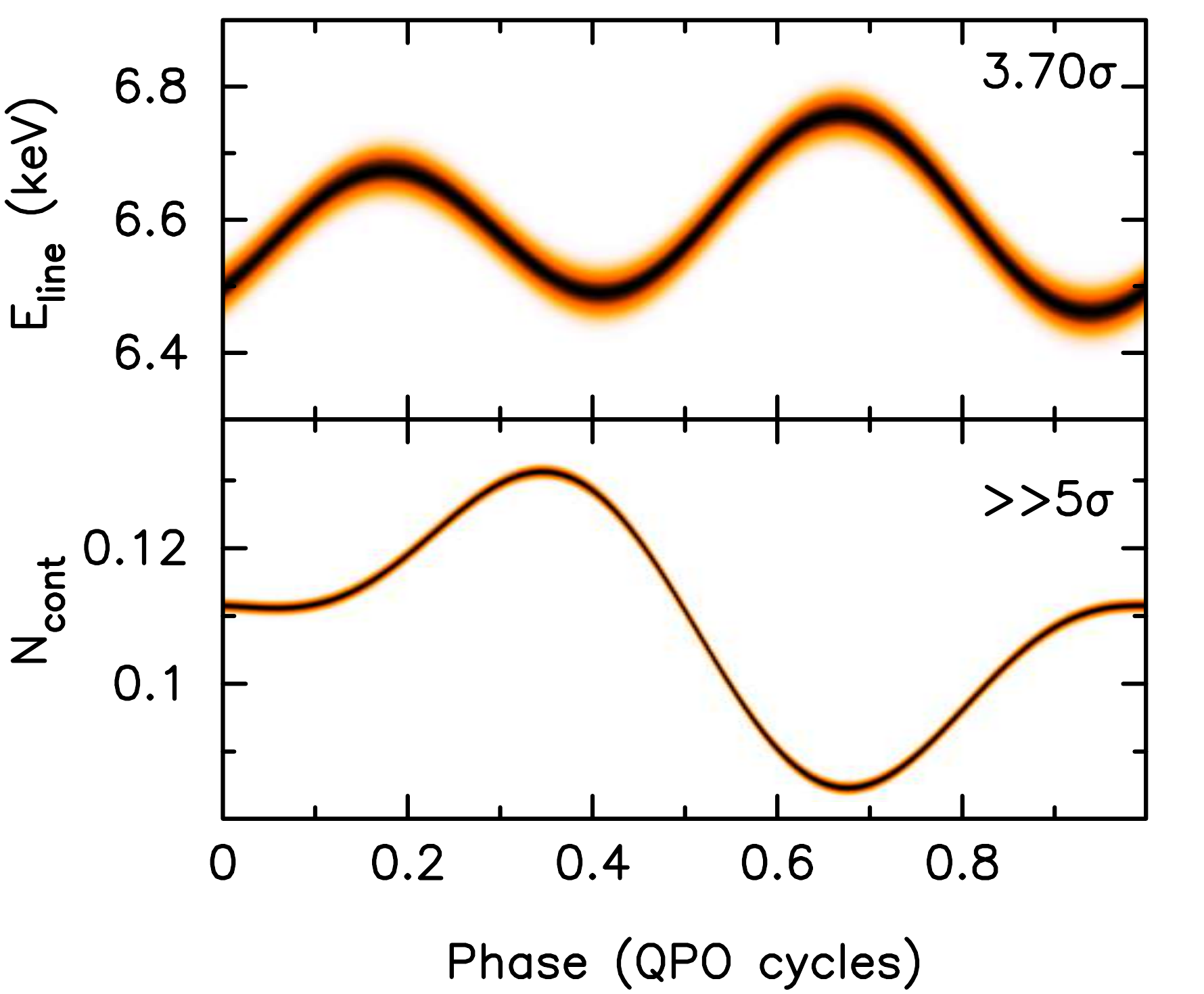}
    \centering
    \caption{\textit{Left:} Reconstructed spectra of H 1743-322 for two selected QPO phases (blue points are a quarter of a QPO cycle after red points) showing that the iron line profile changes with QPO phase. \textit{Right:} Centroid energy of the iron line (top) and continuum normalization (a proxy for X-ray flux; bottom) as a function of QPO phase, measured by fitting the iron line with a Gaussian function. Adapted from \citet{Ingram2016}.}
    \label{fig:centroid}
\end{figure}
%Left, Bottom, Right, Top

Whereas the rms spectrum gives the amplitude of different QPO harmonics for each energy channel, constraining the full QPO waveform for each energy channel provides a lot more information. We have already discussed in detail the QPO waveform for a single energy band in Section \ref{sec:qpowaveform}. The same techniques discussed therein can be applied to individual energy channels to obtain QPO phase-resolved spectra. This includes filtering and phase-folding such as the method of \citet{Tomsick2001} and the Hilbert-Huang transform explored by \citet{Su2015}. \citet{Ingram2015} argued that better signal to noise can be achieved by \textit{reconstructing} the waveforms from the harmonic amplitudes, the phase difference between harmonics in a broad (therefore high count rate) reference band and the phase difference for each harmonic between each energy channel and the broad reference band. \citet{Stevens2016} instead used the cross-correlation function between each energy channel and a broad reference band. The major advantage of this method is that it preserves the quasi-periodicity of the QPO instead of reducing it to a sum of harmonics. The major disadvantage is that the phase difference between harmonics is lost completely. Estimating parameter uncertainties is a subtle task for all reconstruction methods, since the error bars of the spectra are correlated across phases. \citet{Ingram2016} circumvented this problem by fitting the phase-resolved model in Fourier space - i.e. Fourier transforming the phase-resolved model and fitting it to the Fourier transform of the QPO. \citet{Stevens2016} instead used a bootstrapping scheme.

\citet{Ingram2016} used the phase-resolving method of \citet{Ingram2015} in order to study a Type-C QPO from the BH XRB H 1743-322. Fig \ref{fig:centroid} (left) shows reconstructed spectra at two QPO phases (blue circles are a quarter of a QPO cycle after red triangles). We see that the iron K$\alpha$ line profile changes between these two phases, both in the \textit{XMM-Newton} and \textit{NuSTAR} data (inset). Fig \ref{fig:centroid} (right) shows the results of fitting the phase-resolved spectra with a simple absorbed power-law plus Gaussian iron line model. Since the fitting is in the Fourier domain, the result is a probability map. This analysis confirms that the centroid energy of the line varies with QPO phase with a significance of $3.7\sigma$. This not only provides very strong evidence for a geometric QPO origin, but also confirms a specific prediction of the precession model \cite{Ingram2012a}. The only way to explain this in the context of a constant geometry is with variations in the disc ionization state, since the restframe iron line energy is $6.4$ keV for neutral iron and $6.9$ keV for the highest ionization state. This interpretation is ruled out by the \textit{NuSTAR} data since we see that the strongest Compton hump coincides with the bluest iron line, whereas increased disc ionization leads to the Compton hump being \textit{weaker} compared with the line \cite{Garcia2013}. Also, the variation of X-ray flux irradiating the disc would need to be $\sim 3$ orders of magnitude for the line energy modulation to be caused by ionization changes alone, when the variation in X-ray flux we see is only $\sim 15\%$. The modulation can therefore \textit{only} be explained by variations in Doppler shifts caused by changes in the accretion geometry.

\begin{figure}
	\includegraphics[width=\textwidth,trim=0.8cm 6.0cm
        1.5cm 5.5cm,clip=true]{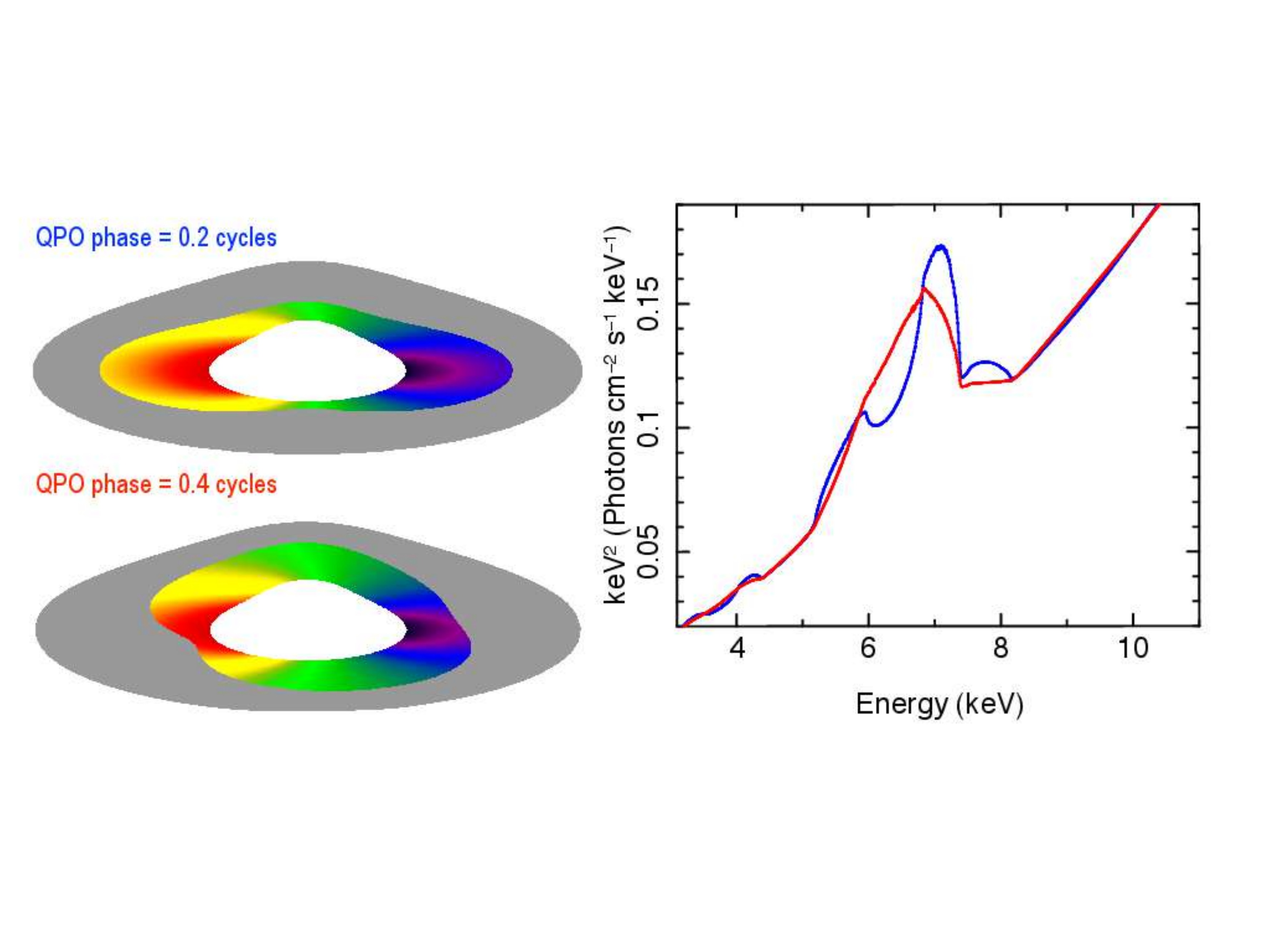}
    \centering
    \caption{Results of fitting a reflection model to the QPO phase-resolved spectra of H 1743-322, reproduced from \citet{Ingram2017}. The specific intensity of illumination on the disc surface (equation \ref{eqn:Ie}) is illustrated for two QPO phases (left) alongside the reflection spectrum corresponding to the same two phases (right). The rainbow colours on the disc represent red/blue shifts.}
    \label{fig:tomography}
\end{figure}
%Left, Bottom, Right, Top

\citet{Ingram2017} applied a more sophisticated phase-resolved spectral model to the same data set. They parameterized the restframe reflected specific intensity as a function of QPO phase, $\gamma$, and disc radius and azimuth with the analytic function
\begin{equation}
I_E(r,\phi) \propto r^{-q} \left\{ 1 + A_1 \cos^2[(\gamma-\phi+\phi_1)/2] + A_2 \cos^2[\gamma-\phi+\phi_2] \right\}\mathcal{R}(E),
\label{eqn:Ie}
\end{equation}
where the restframe reflection spectrum $\mathcal{R}(E)$ is calculated using the model \textsc{xillver} \cite{Garcia2013}. If the asymmetry parameters $A_1$ and $A_2$ are non-zero, this function corresponds to an asymmetric illumination pattern on the disc surface that rotates as the QPO phase increases. They then calculated the observed QPO phase-dependent reflection spectrum accounting for all relativistic effects. Fig \ref{fig:tomography} shows the best fitting illumination function (left) and the corresponding observed reflection spectra (right) for two selected QPO phases. We see that the there are two bright patches on the disc surface ($A_2 > 0$). This enables the model to reproduce the two maxima in line energy per QPO cycle seen in Fig \ref{fig:centroid} (right), since the bluest line occurs when the approaching and receding sides of the disc are both illuminated and the reddest line occurs when the front and back of the disc are illuminated. This is the case because blue shifted radiation is also Doppler boosted, and so the blue wing dominates over the red wing when approaching and receding sides of the disc are illuminated. \citet{Ingram2017} suggested that such an illumination could result from the top and bottom of an oblate precessing inner flow illuminating the disc (i.e. the misalignment angle between disc and flow is greater than the opening angle of the flow), or alternatively may occur if both the inner flow and the jet base are precessing. The best fitting model is preferred to a null-hypothesis of axisymmetric illumination ($A_1=A_2=0$) with $2.4\sigma$ significance. Confirmation of asymmetric illumination at $>3\sigma$ confidence levels is likely to be provided by the high count rate capabilities of \textit{NICER}.

Another important result of the \citet{Ingram2017} analysis is that the reflection fraction is required to vary with $3.5\sigma$ significance. Since the reflection fraction in this context is defined as the number of photons intercepted by the disc divided by the number of photons radiated by the corona (i.e. the covering fraction), this indicates once and for all that the geometry is changing over the course of each QPO cycle. \citet{Ingram2017} also tried alternative models with an axisymmetric illumination profile. They considered modulations in the ionization, disc inner radius and the radial emissivity parameter, $q$. The first of these has already been argued to be implausible, but qualitative testing is still worthwhile. The second is most similar to the propagating oscillatory shock model, although the truncation radius according to that model would be $>600~R_g$ for the observed QPO frequency ($\nu_{\rm qpo} \sim 0.25$ Hz), and therefore the rotational frequency is too low to cause the observed Doppler shifts. Still, an oscillation of the disc inner radius caused by some other physical mechanism could plausibly cause a line energy modulation. The third is intended to mimic oscillations in the shape or height of the corona. All three alternative models yield worse fits than the best fitting asymmetric model with $>3.5\sigma$ confidence.

The phase-resolved spectroscopy results strongly favour asymmetric geometrical models for the Type-C QPO. The line energy modulation could be caused by precession of the inner flow, jet or disc, or by the spiral density waves of the AEI \cite{Karas2001}. However, there is no strong QPO in the disc component. Therefore, although disc corrugation modes can explain the line energy modulation \cite{Tsang2013}, \AI{this requires the oscillating region of the disc to not be hidden by the corona and therefore predicts} a strong QPO in the disc spectrum.
% , which the AEI model can only explain if the \AI{oscillating} region of the disc containing the spiral density waves
% is largely hidden from our view, but on the other hand we must be able to see this region if the AEI is to explain the QPO phase dependence of the iron line profile. Disc corrugation modes suffer from a similar problem: they can explain the line energy modulation \cite{Tsang2013}, but not without predicting a strong QPO in the disc spectrum.
It is also unclear quite how the reflection fraction would change with QPO phase \AI{both in the corrugation modes and AEI models}, unless perhaps the disc is warped. These observational characteristics are instead naturally explained by precession of the corona or jet as long as the disc rotation axis is misaligned with the BH spin axis.

\citet{Stevens2016} studied a Type-B QPO in \textit{RXTE} data from GX 339-4. They we not able to constrain changes in the iron line due to signal to noise restraints, but were able to track the QPO phase dependence of the disc blackbody and Comptonised components. Their preferred model consisted of a constant multi-temperature disc, a variable blackbody component (representing thermal radiation from the inner disc), a variable power-law and a Gaussian iron line. They found that the blackbody temperature and power-law flux varied with $\sim 1.4\%$ and $\sim 25\%$ rms respectively, and that the temperature variations led the power-law flux variations by $\sim 0.3$ QPO cycles. The variations in bolometric flux of the blackbody component were $\propto T^4$. This is of course consistent with changes in the intrinsic disc flux, through either accretion rate fluctuations or variable irradiation, in a constant geometry (i.e. the Stefan-Boltzmann law). The $\sim 0.3$ cycles phase lag rules out irradiation as the cause, since in this case there should be no phase lag, but oscillations in the accretion rate of the inner disc are possible. An $F\propto T^4$ relation would also result if the blackbody component has a constant intrinsic flux, but the blue shift $g=E_{\rm obs}/E_{\rm em}$ (where $E_{\rm obs}$ and $E_{\rm em}$ are respectively observed and emitted photon energies) changes with QPO phase, as would be the case for precession. This is because the relation between the observed and emitted bolometric flux is $F_{\rm obs} \propto g^4 F_{\rm em}$ (see e.g. \cite{Luminet1979,Ingram2019}). Since the restframe flux follows the Stefan-Boltzmann law, $F_{\rm em} \propto T_{\rm em}^4$, this gives $F_{\rm obs} \propto (g T_{\rm em}^4)$. The effective blackbody temperature measured by the observer is $T_{\rm obs} = g T_{\rm em}$, and therefore $F_{\rm obs} \propto T_{\rm obs}^4$. The authors suggested a model whereby a precessing jet illuminates different disc azimuths. This fits the observations because the highest disc temperature would occur when the jet tilts towards the approaching disc material, and the peak in power-law flux would occur $\sim 1/4$ precession cycles later when the jet tilts towards the observer. The X-ray source must be reasonably far from the BH in order to explain the small variations in disc temperature.

\subsection{QPO triplets}\label{Sec: RPM_mass_spin}

Although the observational evidence strongly implies that LF QPOs are generated by precession, none of the tests discussed thus far give any indication that this is specifically \textit{Lense-Thirring} precession. If we knew the mass and spin of a given BH, and were able to track the disc inner radius over many observations, we could test if the QPO frequency evolves with disc inner radius in a manner predicted by the model. However, it is notoriously difficult to measure all three of these quantities. It is however possible to test the RPM whenever a Type-C LF QPO and a pair of HF QPOs are detected in the same observation, because the RPM associates these three features with Lense-Thirring precession, periastron prescession and orbital motion, all at the same characteristic radius. If a triplet of QPOs is detected, this leaves three equations and three unknowns: $a$, $M$ and $r$, which can be solved for analytically \cite{Ingram2014}. \citet{Motta2014} did just this for a QPO triplet detected from GRO J1655-40 (the pair of HF QPOs are pictured in Fig \ref{fig:HFQPOs}, left) to infer $r = 5.68 \pm 0.04$, $a=0.29 \pm 0.003$ and $M/M_\odot = 5.31 \pm 0.07$. Remarkably, this mass measurement agrees very well with the existing dynamical mass measurement of $M/M_\odot = 5.4 \pm 0.3$. Another validation of the model came when they found that the highest Type-C QPO frequency ever measured for this source is consistent with Lense-Thirring precession at the ISCO radius for these mass and spin values, as is predicted by the RPM.

\begin{figure}
	\includegraphics[width=0.6\textwidth]{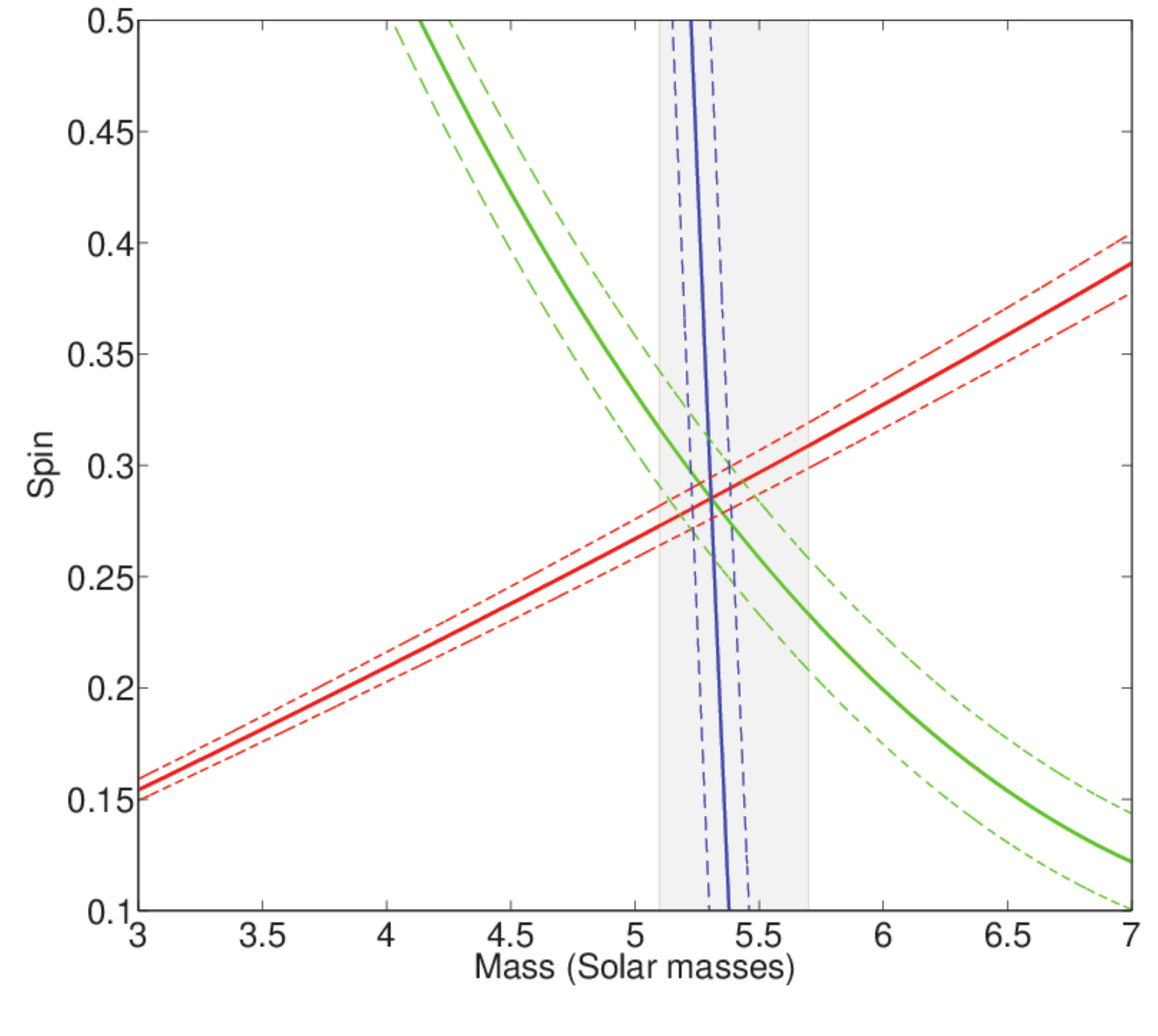}
    \centering
    \caption{\CH{Black hole spin as a function of the mass as predicted by the three equations of the relativistic precession model for the case of  GRO J1655-40. The derived black hole mass (m = 5.31$\pm$0.07 M$_{\odot}$ agrees with the black hole mass derived from 
    spectro-photometric optical observations \cite{Beer2002}, marked by the grey band in the plot (1-sigma confidence level). The green, red and blue lines show the spin as a function of the mass according to the functional form of the nodal precession frequency, the periastron precession frequency, and the orbital frequency, respectively. Taken from \cite{Motta2014}.}}
    \label{fig:RPMmass}
\end{figure}

Although the RPM is a rather simplistic model, it is the limiting case of the more physical precessing inner flow model when the radial extent of the precessing flow is small \cite{Motta2018}. The source is predicted to be in such a state for the observation containing the triplet, since the Type-C QPO frequency is fairly high ($\nu_{\rm qpo} \approx 17$ Hz). HF QPOs are only observed in softer states such as this one (i.e. with higher QPO frequencies) \cite{Belloni2012}. \citet{Psaltis1999} and \cite{Belloni2002} noted that broader power spectral components are observed in the hard state, whose characteristic frequencies smoothly evolve into the HF QPO frequencies as the source spectrum softens. They suggested that this happens because the flow is extended in the harder states, meaning that a range of periastron precession and orbital frequencies contribute to the broad components, but these features become narrower as the extent of the flow, and therefore the range of epicyclic frequencies, decreases. \citet{Fragile2016} showed that the GRO J1655-40 QPO triplet can also be explained by a precessing torus ($r_{\rm in} = 6.5~R_g$; $r_{\rm out}-r_{\rm in} \approx (0.2 - 2.9)~R_g$), with the lower and higher HF QPOs resulting respectively from breathing and vertical epicyclic modes of this torus. This model works for the dynamical mass measurement of GRO J1655-40 for a spin of $a\approx 0.63$, but there are $4$ parameters and therefore other values of BH mass also work.

The consistency of this triplet with the RPM and the more sophisticated model of \citet{Fragile2016} is very encouraging, but unfortunately HF QPO detections are sufficiently rare for this to be the only triplet in the entire \textit{RXTE} archive. It is therefore not possible to further test these models with more triplets. It is however possible to apply the RPM to doublets if a dynamical mass measurement is available. \citet{Motta2014a} did this for XTE J1550-564 and obtained a spin value consistent with that obtained via X-ray spectroscopic methods. It is also possible to place limits on the spin according to the RPM by combining the highest Type-C QPO frequency ever observed with a mass measurement \cite{Ingram2014,Franchini2017}. The very precise mass and spin values that can be inferred from the RPM make further observational validation of the model very desirable. Such tests will be made possible by future high throughput X-ray missions such as \textit{eXTP} and \textit{STROBE-X}, which will be very sensitive to detecting HF QPOs.

\subsection{Energy dependence of the QPO frequency}

One observational property of Type-C QPOs that is rather challenging for all models to reproduce is the energy dependence of QPO frequency observed for a number of sources \cite{Qu2010,vandeneijnden2016,vandeneijnden2017,Huang2018}. Most, if not all, QPO models invoke some sort of global mode, implying that there should be only one QPO frequency, regardless of the energy band we observe in. \citet{vandeneijnden2016} tested the hypothesis that there is no energy dependence of the QPO frequency at any given time, but that the QPO frequency changes with time (which it does) in a manner that correlates with QPO amplitude in one energy band and anti-correlates with amplitude in the other band. In this case, the power spectrum averaged over time would then contain a QPO with higher frequency for the correlated band than for the anti-correlated band. This hypothesis had to be ruled out because the phase of the higher frequency QPO was found to accelerate away from the phase of the lower frequency QPO (see the final paragraph of Section \ref{sec:freqamp}). This tells us that the \textit{instantaneous} QPO frequency depends on energy, and therefore if the QPO is driven by precession, there needs to be some \textit{differential} precession. This is not so difficult to imagine, since the Lense-Thirring precession frequency is a function of radius and so we can imagine the twist of the inner flow increasing to give a slightly higher precession frequency in the hotter inner region, in turn leading to the QPO frequency increasing with energy. However, \citet{vandeneijnden2016} found for GRS 1915+105 that the QPO frequency sightly \textit{decreases} with energy for low average $\nu_{\rm qpo}$ and increases with energy for higher average $\nu_{\rm qpo}$. This is very hard to explain with radial differential precession, since we always expect the precession frequency and spectral hardness to increase with proximity to the hole.

A possible answer to the riddle was provided by the GRMHD simulations of a moderately thick ($H/R=0.1$) tilted disc ran by \citet{Liska2019b}. They found that the disc precesses along with the low density corona above it and the jet. The jet precesses at a slightly slower rate than the corona, which precesses at a slightly slower rate than the disc. Such a `multi-region' setup could potentially also allow other models to reproduce the energy dependence of QPO frequency. Taking the AEI as an example, perhaps the spiral density wave in the disc mid-plane could have a slightly different frequency to the that in the disc upper atmosphere. 

\section{Discussion}
\label{sec:discussion}

\subsection{What are LF QPOs?}

From the previous Section, we can conclude a number of things about the nature of Type-C QPOs:

\noindent \textit{They modulate the Comptonised component, not the disc component:} We know this from frequency-resolved spectroscopy. \CH{Furthermore, QPOs are still visible - and matter of factly with a higher  fractional rms - at energies and in accretion states where the disc has no contribution.
 Note, however,} that the oscillation could still somehow originate in the disc but only modulate the X-ray flux emitted from the corona through, for example, variation of seed photons.

\noindent \textit{They are very likely a geometrical effect:} There are many strong lines of evidence that lead to this conclusion. There is the inclination dependence of QPO amplitude and phase lag, and the line energy and reflection fraction modulations observed in H 1743-322. A caveat is that the spectral modulations that form the strongest arguments are currently based entirely on one observation of one source. More observations therefore need to be analysed in order to gain confidence that the H 1743-322 observation is representative of the population.

\noindent \textit{They are likely an azymuthally asymmetric geometrical effect:} This conclusion is again driven by the line energy modulation in H 1743-322 and the QPO phase-resolved spectral modelling of \citet{Ingram2017}. We must therefore keep in mind that this conclusion is also essentially based on one observation of one source.

We can make similar statements about Type-B QPOs, perhaps with slightly less confidence. We also conclude that Type-B QPOs are different features to Type-C QPOs. This is evidenced by observations that contain both classes of QPO simultaneously \cite{Motta2012}, and also by the different inclination dependence of the two QPO classes. A model whereby Type-C and Type-B QPOs are respectively due to precession of the oblate inner accretion flow and the jet is consistent with the observational data, and Lense-Thirring precession is a theoretically appealing precession mechanism. There is still, however, a lot to learn about LF QPOs despite the many decades since they were first discovered, and it is of course important to carry on testing models other than the precession model such as the AEI.

\subsection{Are these systems misaligned?}
\label{sec:misaligned}

\begin{figure}
	\includegraphics[width=0.8\textwidth,trim=12.0cm 6.0cm
        11.0cm 9cm,clip=true]{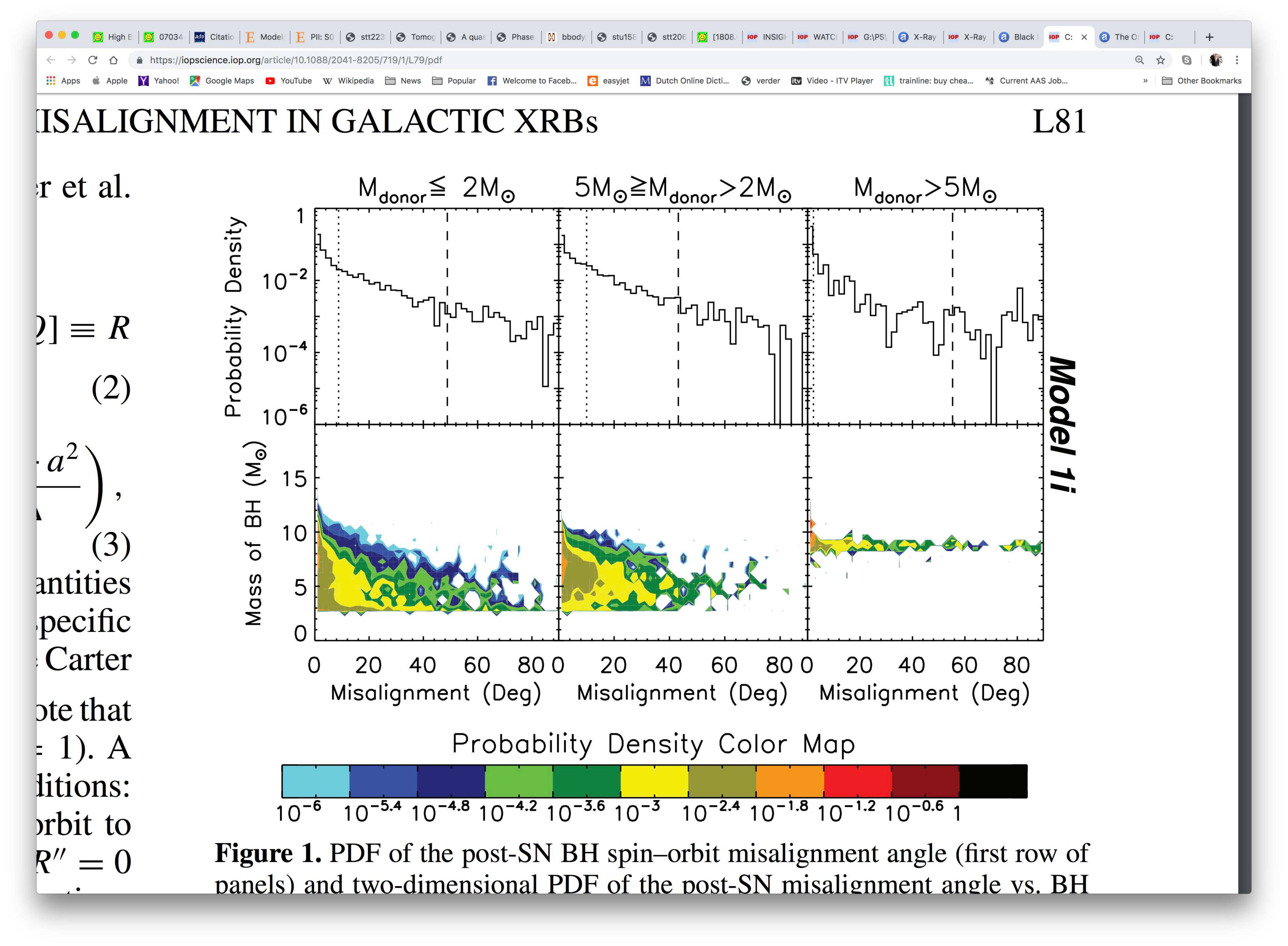}
    \centering
    \caption{Results of one of the supernova kick models ran by \citet{Fragos2010}: 1D (top) and 2D (bottom) probability density functions of the misalignment angle between the BH spin and the binary rotation axis immediately after the supernova that created the BH. Dotted and dashed lines represent $1 \sigma$ and $2\sigma$. Small misalignment is favoured because the largest supernova kicks destroy the binary system.}
    \label{fig:fragos}
\end{figure}
%Left, Bottom, Right, Top

For the Lense-Thirring precession model to apply to all Type-C QPOs, all sources that display them must have a misalignment between the BH spin axis and the binary orbital axis. This requirement can be relaxed a little if we assume that Type-C QPOs are driven by precession but remain agnostic about the physics behind this precession. That said, it is difficult to imagine
%a precession mechanism that could occur in a fully aligned system, and it is even harder to imagine 
how the reflection fraction could vary with QPO phase (as is observed) in an aligned system. Ray tracing calculations of a precessing ring \cite{Veledina2013} and torus \cite{Ingram2015a,Ingram2017} demonstrate that the observed QPO amplitudes can be reached if the half opening angle of the precession cone (i.e. the misalignment angle) is $\sim 10-15^\circ$. These calculations do not include the variation of seed photons that would result from the angle between flow and disc rotation axes varying between $0$ and $2\beta$, which provides an additional modulation mechanism and thus increase the QPO amplitude for a given misalignment angle \cite{Zycki2016,You2018}. The required range of misalignments can therefore be revised down slightly.

The first question we can ask is: is it theoretically reasonable for a large fraction of BH XRBs to have the $\gtrsim 5^\circ$ misalignment required by the precession model? For XRBs formed from existing binary star systems, misalignment is thought to be introduced by asymmetries in the supernova explosion that creates the BH (a \textit{supernova kick}). Fig \ref{fig:fragos} shows the result of one of the supernova kick models ran by \citet{Fragos2010} (Maxwellian distribution of kick velocities and uniform distrbution of kick directions) in the form of probability density functions for all of the binary systems that remain bound after the supernova (in order to become XRBs). We see that systems with smaller misalignments are more common, with $\sim 95\%$ in the $0-48^\circ$ range. This is because the largest kicks disrupt their binary systems. Accretion onto the hole will then drive the system towards alignment over time. \citet{King2016} show that this is a slow process, with the BH needing to accrete $\sim 10\%$ of its initial mass before the misalignment angle is significantly altered (also see \cite{Martin2008,Banerjee2019}), indicating that modest misalignments could be common amongst the BH XRB population. Even more mass would need to be accreted if the Bardeen-Petterson transition radius is smaller than analytic predictions, as recent GRMHD simulations imply \cite{Liska2019}.

It is therefore plausible for many systems to have a $\gtrsim 5^\circ$ misalignment, but it follows that there should also be many systems with a $\lesssim 5^\circ$ misalignment. A fraction of the BH XRB population should therefore display no, or at least very weak, Type-C QPOs according to the precesison model. The most famous example of such a source is perhaps Cygnus X-1. Although there are lumps and bumps in its power spectrum, the feature that has occasionally been speculated to be a Type-C QPO \cite{Axelsson2005} is very weak and broad ($Q\sim 2$) and is not seen in the bi-coherence \cite{Uttley2005}. Other sources with very weak and/or broad QPOs include GRO J0422+32 ($Q \lesssim 2$ \cite{vanderhooft1999}), LMC X--3 ($Q\sim 3$ \cite{Boyd2000}), 1E 1740.7--2942 (the great anhilator \cite{Smith1997}), GRS 1758--258 \cite{Smith1997}, Swift J1357.2--0933 \cite{ArmasPadilla2014} and GS 1354--64 \cite{Revnivtsev2000}. However, in some of these cases the QPO may have been smeared out by very long exposures, and in others the QPO may simply have been missed due to the only outburst being `hard state only' \cite{Tetarenko2016} or largely obstructed by sun constraints (e.g. Swift J1713.4--4219 \cite{Tetarenko2016}). Perhaps the best example for which these caveats do not apply is XTE J1652--453, which has no detectable Type-C QPO in its HIMS power spectrum \cite{Hiemstra2011}.
%Type-C QPOs were not detected for SAX J1819.3−2525, but this could be because it is thought to be embedded in an optically thick shroud (\cite{Wijnands2002,Revnivtsev2002}, similarly to V404 Cygni \cite{Huppenkothen2017,Motta2017a}.
Finally, there are a number of sources for which a power spectrum has never been published, despite having been observed by \textit{RXTE} or another telescope with similar timing capabilities. We speculate that there may be a bias at play amongst such sources, since Type-C QPOs are interesting features, and therefore a power spectrum displaying them is more likely to be published than one without.
%A handful of other sources have similar hard state / HIMS power spectra: GRO J0422+32 ($Q \lesssim 2$ \cite{vanderhooft1999}), LMC X-3 ($Q\sim 3$ \cite{Boyd2000}), 1E 1740.7-2942 (the great anhilator \cite{Smith1997}) and GRS 1758−258 \cite{Smith1997}. Type-C QPOs appear to be either very weak or missing entirely in a handful of other sources (although some of these may have been smeared by averaging over long observations). 
% Another good candidate is Swift J1357.2–0933, which has a typical hard state power spectrum except there is no (or possibly a very weak) QPO peak \cite{ArmasPadilla2014} at the frequency expected from the \citet{Wijnands1999} QPO-break relation ($\sim 0.3$ Hz).
%There are other sources with Type-C QPOs seemingly absent in their power spectra, but only for one or two hard state observations (e.g. GS 1354−64 \cite{Revnivtsev2000}), and therefore we can not rule out that there would have been QPOs later on in the outburst if it hadn't have either failed to transition (a so-called `failed' or `hard state only' outburst \cite{Tetarenko2016}) or been blocked from view by sun constraints (e.g. Swift J1713.4−4219 \cite{Tetarenko2016}). 

Although Type-C QPOs are therefore not completely ubiquitous, they are more commonly present than not, which is slightly in tension with the expectations of population synthesis models. One effect that has never been included in the binary evolution models, however, is radiative warping \cite{Pringle1996,Frank2002}. For an initially misaligned system, radiation pressure from the central X-ray source is capable of warping the outer disc. We speculate that this effect could further slow down the binary alignment process, and/or lead to the inner disc being more misaligned with the hole than the binary partner is. BH XRBs in dense environments such as globular clusters may have very different properties, since they are likely formed by binary capture for which there is no preferred misalignment angle \cite{Fabian1975}. Therefore, if a bright enough XRB in a globular cluster were ever detected, the precession model predicts that it would likely have very strong QPOs. Arbitrary misalignment angles are also expected for tidal disruption events (TDEs). It is somewhat encouraging for the precession model that the QPO recently discovered from a TDE reached a fractional rms as high as $\sim 50\%$ \cite{Pasham2019}.

There are a few (indirect) observational tests for binary misalignment. For instance, the proper motion of Cygnus X-1 is consistent with that of its natal association of O stars, implying that it felt no supernova kick \cite{Mirabel2017}. Before we get excited that this explains why Cygnus X-1 has no QPOs however, we must also note that GRS 1915+105, which has very strong QPOs, also has no peculiar velocity with respect to its association \cite{Mirabel2017}. Another test comes from super-luminal jet ejections, because the 3D jet trajectory can be reconstructed by tracking the proper motion and brightness of approaching and receding lobes. For GRO J1655--40, the inferred jet inclination angle is $\sim 85^\circ$ \cite{Hjellming1995}, whereas the binary inclination angle calculated from ellipsoidal modulations of absorption lines in the donor star's optical spectrum is $\sim 70^\circ$ \cite{Greene2001}. This means that the misalignment angle is \textit{at least} $15^\circ$, because we have no information about the projection of the binary rotation axis on the observer plane \cite{Fragile2001}. For SAX 1819--2525 (V4641 Sgr), a misalignment of $\gtrsim 50^\circ$ is inferred using the same method \cite{Maccarone2002a}. The misalignment for XTE J1550--564 could potentially be much smaller, but is still $\gtrsim 12^\circ$ \cite{Steiner2012b}. For H 1743--322, comparing the inferred jet orientation \cite{Steiner2012a} with the inner disc inclination measured via QPO phase-resolved reflection modelling \cite{Ingram2017} implies a misalignment angle $\gtrsim 5^\circ$, and an misalignment of $\sim 15^\circ$ is consistent within $0.5\sigma$ confidence.

Perhaps the most spectacular evidence of misalignment is the recent result of \citet{Miller-Jones2019}, who used very long baseline interferometry to discover that the jets of V404 Cygni are precessing. The half opening angle of the precession cone is $\sim 18^\circ$, and the precession period is in the range $\sim 1$ second to $2.6$ hours (the pattern of sampling makes it difficult to estimate the precession period, or even to conclusively confirm that that the changes in jet orientation are periodic). These observations are coincident with an intermittently detected X-ray QPO with a period of $\sim 56$ seconds \cite{Huppenkothen2017}, although it is difficult to reliably establish a connection between the two phenomena. The results of \citet{Miller-Jones2019} provide strong evidence that V404 is a misaligned system, although we note that this source cannot by any stretch be considered typical of the population. One argument in favour of alignment is that XRB jets are all observed to have a very small opening angle, which implies that the opening angle of the precession cone is equally small if the observed jet lobe is averaged over many precession cycles. However, resolved jet lobes are all from discrete, transient ejections and therefore there are no such constraints on the precession cone of the compact persistent jets coincident with Type-C QPOs.

\subsection{Outstanding Problems}

Although much progress has been made over the past decade or so, there are still some observational characteristics that challenge the precession hypothesis, and some that pose a challenge to most, if not all, proposed QPO models. Here, we summarise these problems.

\subsubsection{The sub-harmonic}

\begin{figure}
	\includegraphics[width=0.8\textwidth]{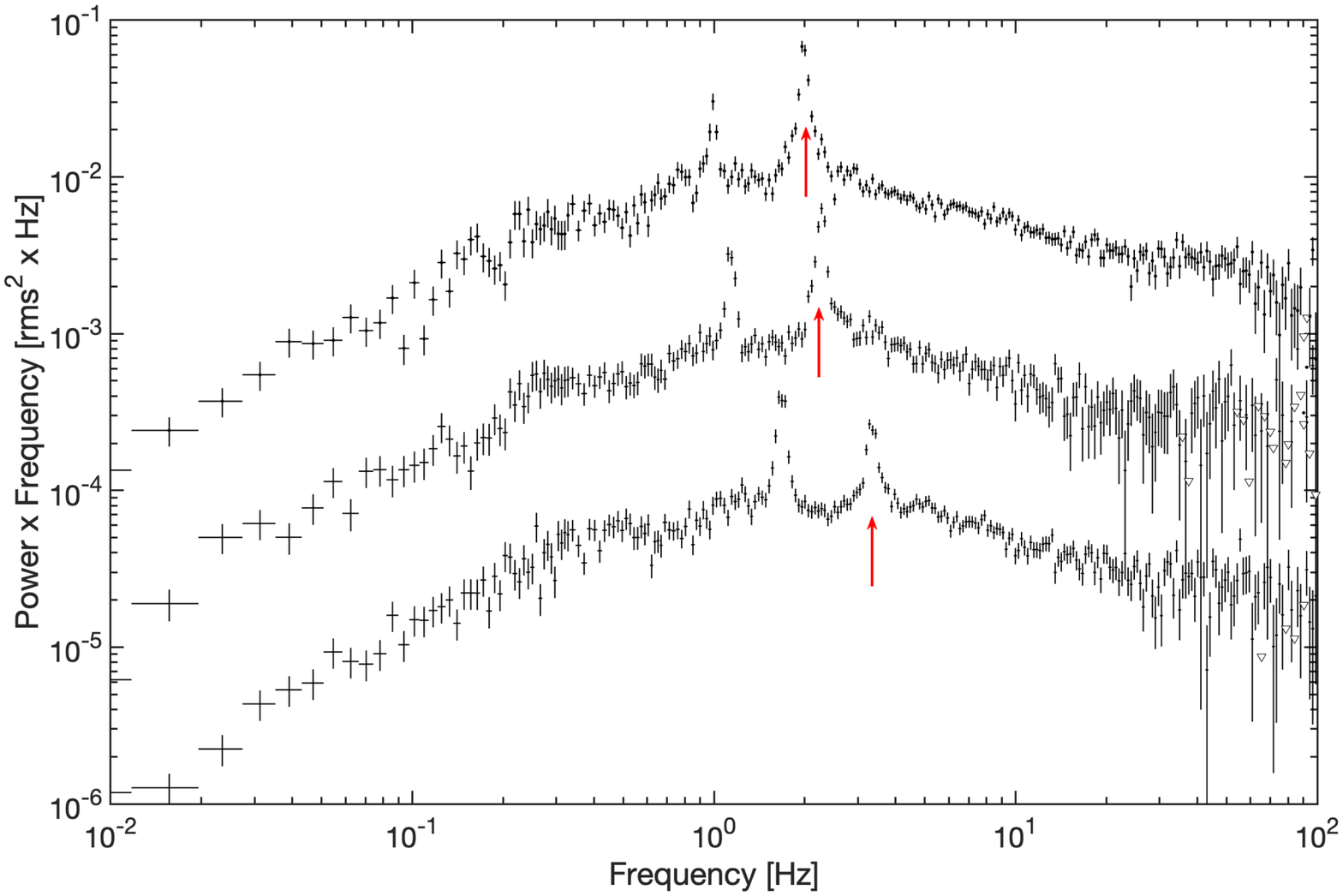}
    \centering
    \caption{Example of the sub-harmonic becoming stronger than the fundamental \CH{(marked by the red arrow)}. The observations are from three consecutive days of the 2007 outburst of GX339-4 (chronological order is top to bottom). The second and third power spectra from the top have been re-scaled by a factor of 1/10 and 1/100, respectively, to facilitate inspection.
    }
    \label{fig:swappy}
\end{figure}
%Left, Bottom, Right, Top

It is usual to classify the QPO fundamental as the highest amplitude peak in the power spectrum. However, a sub-harmonic -- at half the frequency of the fundamental -- is often observed. Sub-harmonics are seen for Type-C and Type-B QPOs, and they are perhaps the most puzzling property of QPOs. The most obvious explanation is that the sub-harmonic is the true fundamental, and somehow the second harmonic (which is what we call the fundamental) has more power than the fundamental. This is difficult to achieve in any model (e.g. \cite{Veledina2013}). Curiously, a sub-harmonic feature is often co-incident with peaks at $n$ times the frequency of the most powerful QPO peak, where $n$ is an integer $>1$, but \textit{not} with peaks at $(2n-1)/2$ times the frequency of the most powerful QPO peak. This means that, if the sub-harmonic is actually the true fundamental, then the QPO never has any strong odd harmonics, only even harmonics. The alternative of the sub-harmonic truly being a sub-harmonic is equally difficult to explain. The \textit{RXTE} archive also includes a few examples of the sub-harmonic amplitude overtaking that of the fundamental. Fig \ref{fig:swappy} shows an example of this over three observations on consecutive days from the 2007 outburst of GX339-4. The picture is further complicated by multi-wavelength observations, since the fundamental frequency of optical and IR QPOs are sometimes observed to be consistent with the X-ray fundamental (e.g. \cite{Hynes2003}) and sometimes with the X-ray sub-harmonic (e.g. \cite{Motch1983,Kalamkar2016}).

% We also note that \citet{Kalamkar2016} reported on the detection of a QPO in the infra-red and optical (U and V bands) data from GX 339-4, which is found at a frequency of 0.08 Hz. This QPO is found at half the X-ray QPO frequency, detected at 0.16 Hz, in correspondence to a low-amplitude sub-harmonic seen in the X-ray band. These authors strongly favoured a jet origin of the IR QPO, which cannot be explained in therms of Lense-Thirring precession of an hot flow, but could be originated by a jet (which dominates the emission in the Infra-Red band) if the latter is tied to the hot flow and precesses with it. Different explanations cannot be ruled out with the available data. Similarly,  the origin of the optical QPO remains unclear. These findings, however, once more cast doubts on the nature of the \textit{sub-harmonic}, that being the only feature significantly detected in both Optical and infra-red, might be in reality the fundamental frequency. 

% \subsubsection{The 11 Hz pulsar in Terzan 5: IGR J17480--2446}

\subsection{QPOs from accreting millisecond X-ray pulsars}

\AI{Since we know the spin frequency of accreting millisecond X-ray pulsars, we can estimate the spin parameter for a range of reasonable moment of inertia values and test the Lense-Thirring precession model against the observed LF QPO frequencies, under the assumption that the upper kHz QPO is due to orbital motion. Initial works to test the RPM in this manner concluded that the LF QPO frequency matches best to twice the Lense-Thirring precession frequency \cite{Morsink1999,Stella1999}. \citet{Ingram2010} later considered the precessing inner flow model, and found that the solid body precession frequency matched the LF QPO frequency if the surface density profile becomes more centrally peaked as the source gets brighter, as may result from material piling up on the MNS surface. More recently, \citet{vanDoesburgh2017} considered a sample of 13 sources and again found that the surface density profile is required to become more centrally peaked as the source flux increases in order to explain the QPO frequencies. They additionally find that the LF QPO frequencies are higher than the precession frequencies produced by assuming reasonable moment of inertial values, and suggest (again) that the LF QPO fundamental could actually be twice the precession frequency, or that the precession is modified by mechanisms other than the frame dragging effect, such as radiation pressure and magnetic precession.}

\AI{The opportunity for a stricter test of the Lense-Thirring precession model was provided by the discovery of QPOs in the $11$ Hz pulsar,} IGR J17480--2446, in the globular cluster Terzan 5. \AI{Shortly after the source was discovered,} \citet{Altamirano2012} reported on a QPO in its power spectrum with frequency evolving from 35-50 Hz, which is far too high to be driven by Lense-Thirring precession around a NS with a spin of only $11$ Hz. This QPO was also accompanied by a kHz QPO and broad band noise. Using well-known correlations between the frequencies of the different power spectral components \cite{Wijnands1999}, and by examining its general characteristics (quality factor, rms amplitude, overall power-spectral shape), \citet{Altamirano2012} classified the 35-50 Hz QPO as an HBO (thought to be the Z-source analogy of Type-C QPOs - see Section \ref{sec:NSQPOs}). This classification implies that HBOs cannot be due to Lense-Thirring precession, and the association between HBOs and Type-C QPOs also throws doubt on the Lense-Thirring model for BHs.

We note, however, that a search for QPOs in the frequency range expected for Lense-Thirring precession around an 11 Hz NS ($\lesssim 0.1$ Hz) has never been carried out for this source, due partially to the presence of a large number of type-I X-ray bursts in the data \cite{Motta2011b} and the relatively short length of the observations. The detection of such a very low frequency QPO in this source would cast doubt on the HBO classification of the 35-50 Hz QPO and thus re-establish the possibility that HBOs in general could be due to Lense-Thirring precession. It is also important to note that \AI{IGR J17480--2446 was behaving as a Z-source when these QPOs were observed, and therefore the Lense-Thirring precession frequency of particles close to the NS surface can be quite significantly modified by radiation pressure} \cite{Miller1999}.

% radiation pressure can quite significantly influence the Lense-Thirring precession frequency for a Z-source such as IGR J17480--2446 \cite{Miller1999}. Alternatively, it is possible that LF QPOs are caused by precession that is driven by a mechanism other than, or in addition to, the frame dragging effect \cite{vanDoesburgh2017}.

\AI{We can also consider whether it is reasonable for the NS spin axis to be misaligned with the binary rotation axis in accreting millisecond X-ray pulsars, given the standard paradigm in which the NS has been spun up} by accretion over its lifetime by a factor of $\sim$hundreds \cite{Radhakrishnan1982}.
% We have discussed in detail misalignment in BH XRBs and argued that it is plausible for such systems to be misaligned such that Type-C QPOs can be driven by Lense-Thirring precession. However, what of NS XRB systems that display QPOs with very similar properties? The standard paradigm is that accreting millisecond X-ray pulsars have been spun-up by accretion over the course of their lifetime by a factor of $\sim$hundreds \cite{Radhakrishnan1982}.
Is it possible for the spin magnitude to increase this much without the system aligning? We can adapt the calculations of \citet{King2016} to make a back of the envelope estimate, starting by expressing the magnitude of the spin-up torque on the NS as
\begin{equation}
    G_{\rm spinup} = \dot{M} ( G M_{\rm ns} R_{\rm ns} )^{1/2},
\end{equation}
where $M_{\rm ns}$ and $R_{\rm ns}$ are respectively the NS mass and radius, and we have assumed for simplicity that the accretion disc reaches the NS surface. Integrating over time for a constant mass accretion rate, ignoring any NS mass increase caused by accretion and assuming that today's spin frequency is $>>$ the birth frequency, the total mass that must be transferred to achieve a spin frequency of $\nu_{\rm s}$ is
\begin{equation}
    \frac{M_{\rm tr}}{M_{\rm ns}} \sim \frac{2\pi \delta }{(G M_{\rm ns})^{1/2}} R_{\rm ns}^{3/2} \nu_{\rm s},
\end{equation}
where the NS moment of inertia is $\delta~M_{\rm ns}R_{\rm ns}^2$. Assuming $\delta=2/5$ (solid sphere), $M_{\rm ns}=1.4~M_\odot$, $R_{\rm ns}=10$ km and $\nu_{\rm s}=500$ Hz, we find $M_{\rm tr}/M_{\rm ns} \sim 0.09$. We recall from Section \ref{sec:misaligned} that this is on the edge of how much mass can be transferred without significant alignment, implying that more NS XRBs should be aligned than BH XRBs, but not ruling out the possibility that some are still misaligned today.

\subsection{Future outlook}

New and upcoming X-ray missions promise to further increase our understanding of QPOs. Some of the best constraints to date have been obtained by QPO phase-resolved spectroscopy, which requires a large number of photons. High throughput X-ray detectors will therefore increase the quality and quantity of QPO phase-resolved spectroscopy studies. \textit{NICER}, which was recently installed on the International Space Station, is ideal for phase-resolved spectroscopy due to its good spectral resolution and ability to measure high count rates (unlike \textit{XMM-Newton}, it does not suffer from instrumental problems associated with high count rates such as pile-up or dead time). The bright BH XRB outbursts that have recently been extensively monitored by \textit{NICER} therefore promise to soon deliver insights into the physics of QPOs. In future, proposed missions with a large area X-ray instrument, such as \textit{eXTP} and \textit{STROBE-X} will further enhance our phase-resolved spectroscopy constraints, as well as enabling new tests such as QPO phase-resolved reverberation mapping. The hard X-ray capabilities of these missions will additionally enable the detection of more HF QPOs, and therefore more QPO triplets. Since the best current evidence that LF QPOs are specifically \textit{Lense-Thirring} precession comes from the QPO triplet in GRO J1655-40, this will be extremely valuable.

Perhaps the most distinctive test for a precession origin of LF QPOs is to search for QPOs in the X-ray polarization degree and angle. \citet{Ingram2015a} calculated the modulations in polarization degree and angle that would result from precession of a Comptonising torus. They found that the swings in polarization angle are greater when the torus is viewed more face-on. This is because an edge-on observer only sees the side of the precession cone, whereas a face-on observed sees the base of the precession cone, and so sees the polarization vector rotate through all angles. However, the mean polarization degree should be larger for a torus viewed more edge-on \cite{Sunyaev1985}, essentially because its image on the observer plane is less symmetric. Since polarization angle is easier to measure for a higher polarization degree, this introduces a trade-off. The \textit{Imaging X-ray Polarimetry Explorer} (\textit{IXPE}; \cite{Weisskopf2016}), due to be launched in 2021, will provide the first opportunity to measure the predicted polarization QPOs. This will, however, be a challenging measurement because many photons must be collected to make a significant detection of polarization. For the expected \textit{IXPE} count rate ($\lesssim 100$ cps) and BH XRB polarization degree ($\lesssim 8\%$), each polarization measurement will require a $\gtrsim 5$ minute exposure. It will therefore not be possible to probe the sub-second timescales of interest simply by making light curves of polarization degree and angle. \citet{Ingram2017a} introduced a statistical method to circumvent this problem, and showed that it should be possible for \textit{IXPE} to measure the predicted modulations with high statistical significance providing that the mean polarization degree for a high inclination source is $\gtrsim 4\%$. \textit{eXTP} is proposed to have an X-ray polarimeter of very similar design to the \textit{IXPE} instrument, but with higher effective area. This, combined with the possibility to enhance the signal by cross-correlating with the signal from the large area detector, will provide much more sensitive constraints on the variability of X-ray polarization \cite{Ingram2017a}.

\section{Conclusions}
\label{sec:conclusions}

Much progress has been recently  made in the physical interpretation of LF QPOs in BH XRBs. The observational evidence points strongly to a geometric origin of Type-C and Type-B QPOs, whereas Type-A QPOs are very rare and weak features that only have their own classification for historical reasons. Moreover, an azimuthally asymmetric geometry is required to explain Type-C QPOs, favouring a precession origin. Since the LF QPO is far stronger in the hard (i.e. Comptonised) emission than in the disc emission, it is likely that it is the X-ray corona that undergoes precession, as opposed to the disc. If the QPO mechanism is specifically Lense-Thirring precession (as implied by the QPO triplet in GRO J1655-40), this implies that the BH spin axis in most BH XRBs has to be misaligned with the binary rotation axis by $\gtrsim 5^\circ$. This could result from an asymmetric kick in the natal supernova and is supported by some observations of jet lobes.

An important caveat is that some of the strongest constraints come from phase-resolved spectroscopy of a single observation of H 1743-322, and so it is important to carry out the same analysis for more observations of more sources. In particular, tracking how the phase-resolved spectral properties of a given object evolve with QPO frequency will help diagnose what causes the QPO frequency to change. \textit{All} QPO models summarised here associate QPO frequency changes with a moving characteristic radius, usually the inner disc radius. However, observations in which the broad iron line profile appears to stay constant as the characteristic frequencies in the power spectrum increase imply that the disc inner radius may instead (at least at times) be fixed, or only move by a small amount, during the bright outburst phase \cite{Kara2019}. If such a paradigm is confirmed, some creativity will be required to determine how the observed $\gtrsim$two orders of magnitude increase in QPO frequency can be achieved without a moving radius. New, future and proposed X-ray missions promise further tests of the QPO mechanism through high count rate capabilities (e.g. \textit{NICER}, \textit{STROBE-X}) and polarimetry (\textit{IXPE}, \textit{eXTP}).
\\
\\
We thank Michiel van der Klis, Chris Fragile and Peggy Varniere for valuable and informative discussions. AI acknowledges support from the Royal Society. SME acknowledges support from the STFC and the Oxford Hintze Centre for Astrophysical Surveys. \AI{We thank the anonymous referee for insightful comments.}



%% References


\bibliographystyle{model2-names-astronomy.bst}
\bibliography{biblio.bib}






% \end{thebibliography}

\end{document}